\documentclass[12pt]{article}
\usepackage{rotating}
\usepackage{amsmath}

\usepackage{amsfonts}
\usepackage{chngcntr}

\usepackage{fancyhdr}
\usepackage[top=0.9in,bottom=0.9in,left=0.8in,right=0.8in]{geometry}
\usepackage{lscape}
\usepackage{amsthm}
\usepackage{enumerate}
\usepackage{threeparttable}
\usepackage{diagbox}
\usepackage{multirow}
\usepackage{makecell}
\usepackage{xcolor}
\usepackage{authblk}
\usepackage{amsmath}
\usepackage{graphicx}
\usepackage{bm}
\usepackage{enumerate}
\usepackage{subfig}
\usepackage[font=footnotesize,labelfont=bf]{caption}

\allowdisplaybreaks[4]
\usepackage[square,sort,comma,numbers,sectionbib,round]{natbib}

\usepackage{graphicx}

\usepackage{titlesec}
\titlespacing{\section}{0pt}{1ex}{0ex}
\titlespacing{\subsection}{0pt}{1ex}{0ex}
\titlespacing{\subsubsection}{0pt}{1ex}{0ex}

\usepackage{enumitem}
\setlist{nolistsep}

\usepackage{amssymb}
\def\be{\begin{equation}}
\def\ee{\end{equation}}
\def\ben{\begin{equation*}}
\def\een{\end{equation*}}
\def\bea{\begin{eqnarray}}
\def\eea{\end{eqnarray}}
\def\bd{\begin{displaymath}}
\def\ed{\end{displaymath}}
\def\bda{\begin{eqnarray*}}
\def\eda{\end{eqnarray*}}

\def\ha1{\hat \beta_1}

\def\bsc{\begin{scriptsize}}
\def\esc{\end{scriptsize}}

\def\diag{\mathrm{diag}}

\def\rd{\mathrm{d}}

\def\p{\mathrm{p}}
\def\P{\mathbb{P}}
\def\T{\scriptscriptstyle\top}
\def\cS{\mathcal{S}}

\newtheorem{la}{Lemma}

\renewcommand{\theequation}{\thesection.\arabic{equation}}
\newcommand{\E}{\rm E}

\renewcommand{\thetable}{\arabic{table}}

\numberwithin{equation}{section}

\usepackage[nodisplayskipstretch]{setspace}

\def\spacingset#1{\renewcommand{\baselinestretch}%
{#1}\small\normalsize} \spacingset{1}

\title{\bf Statistical Inferences for Complex Dependence of Multimodal Imaging Data\thanks{All coauthors contributed equally to the paper. We are grateful to the editor, the associate editor and three referees for their constructive comments and suggestions. Chang, He and Wu were supported in part by the National Natural Science Foundation of China (grant nos. 71991472, 72125008, 11871401 and 11701466). Chang was also supported by the Center of Statistical Research at Southwestern University of Finance and Economics. Kang was supported in part by NIH R01DA048993, NIH R01MH105561, NIH R01GM124061 and NSF IIS2123777.}}

\author[a,b]{Jinyuan Chang}
\author[a]{Jing He}
\author[c]{Jian Kang}
\author[a]{Mingcong Wu}
\affil[a]{\it \small Joint Laboratory of Data Science and Business
Intelligence, Southwestern University of Finance and Economics, Chengdu, China}
\affil[b]{\it \small Academy of Mathematics and Systems Science, Chinese Academy of Sciences, Beijing, China}
\affil[c]{\it \small Department of Biostatistics, University of Michigan, Ann Arbor, MI, U.S.A.}
\date{}                                           

\newtheorem{theorem}{Theorem}
\newtheorem{proposition}{Proposition}
\theoremstyle{definition}

\begin{document}

\bibliographystyle{agsm}
\bibpunct{(}{)}{,}{a}{}{;}

\maketitle
\begin{abstract}
\bigskip
Statistical analysis of multimodal imaging data is a challenging task, since the data involves high-dimensionality, strong spatial correlations and complex data structures. In this paper, we propose rigorous statistical testing procedures for making inferences on the complex dependence of multimodal imaging data. Motivated by the analysis of multi-task fMRI data in the Human Connectome Project (HCP) study, we particularly address three hypothesis testing problems: (a) testing independence among imaging modalities over brain regions, (b) testing independence between brain regions within imaging modalities, and (c) testing independence between brain regions across different modalities. Considering a general form for all the three tests, we develop a global testing procedure and a multiple testing procedure controlling the false discovery rate. We study theoretical properties of the proposed tests and develop a computationally efficient distributed algorithm. The proposed methods and theory are general and relevant for many statistical problems of testing independence structure among the components of high-dimensional random vectors with arbitrary dependence structures. We also illustrate our proposed methods via extensive simulations and analysis of five task fMRI contrast maps in the HCP study.

\end{abstract}

\noindent%

\vfill

\newpage



\section{Introduction} \label{sc:introduction}
\setlength{\baselineskip}{24pt}	
There is an exponential increase in neuroimaging studies during the past decades.
Recent advances in different neuroimaging techniques,
e.g., magnetic resonance imaging (MRI), positron emission
tomography (PET) and electroencephalography (EEG)
have generated large amount multimodal imaging data, which directly or indirectly measure the brain function and structure from different aspects in unprecedented detail.  It provides an important tool for researchers to have a better understanding of human brain pathology and etiology of neuropsychiatry diseases~\citep{liu2015multimodal}.  It is challenging to analyze the multimodal imaging data, which involve high-dimensionality, strong spatial correlations, and complex data structures, as the data type, range and scale can be quite different for different modalities.

In this paper, we aim at making statistical inferences on the complex dependence of multimodal imaging data.   Our motivation is the analysis of multiple task fMRI data from the Human Connectome Project~\citep[HCP]{van2012human}. The main goal of HCP is to understand the patterns of structural and functional connectivity in the healthy adult human brain.
In addition to collecting structural imaging data and the resting state fMRI data, HCP collects behavioral measures including cognitive, sensory and emotional processes and different types of task fMRI data to understand the relationships between individual differences in the neurobiological substrates of cognitive processing and functional connectivity~\citep{BARCH2013}.
The fMRI tasks include working memory tasks, language processing tasks and social cognition tasks. In the analysis of multimodal task fMRI data, we are interested in learning whether the brain activation patterns are independent between different fMRI tasks for each region of interests. For example, it has been shown that inferior frontal gyrus is related to cognitive control~\citep{drobyshevsky2006rapid}, language processing~\citep{binder2011mapping} and social cognition~\citep{castelli2000movement}, while it remains unknown whether these three types of brain activity are dependent with each other in inferior frontal gyrus.
Another interesting topic is to study  the brain activation dependence between regions for each type of task fMRI data. It is also of interest to learn the region pair dependence across different modalities~\citep{XiaLi2020}. To develop formal statistical inference procedures to address these questions, we focus on analyzing the five statistical parametric maps~\citep{penny2011statistical} or contrast maps that are derived from the task fMRI data.   However, our proposed methods and theory can be applied for general multimodal imaging data analysis.

Different statistical approaches have been proposed for analysis of the multi-task fMRI data such as the joint independent component analysis ~\citep{calhoun2006method} and the joint sparse representation analysis ~\citep{ramezani2014joint}, which focused on estimating the joint brain activation patterns rather than testing. Some statistical testing procedures have been developed for multimodal neuroimaging data.  For instance, a voxelwise testing procedure of multimodal MRI~\citep{naylor2014voxelwise} has been proposed for localizing brain activity rather than testing the association between modalities. A simultaneous covariance inferences procedure~\citep{XiaLi2020} has been developed for multimodal PET analysis. Although this method is general and can be applied for different types of neuromaging data analysis, it only focuses on region pair independence across two modalities and cannot be directly applied to address all the scientific questions of our interests.

Specifically, in the  imaging data, $J$ types of brain images with different modalities are collected for each subject. Let $j$ $(j=1,\ldots, J)$ index the image type. All the images are registered to the same standard brain template with $V$ voxels in total. Let $\mathcal{R}=\{1,\ldots,V\}$ denote the set of indices for the voxels in the whole brain template, and the voxels are indexed by $v \in \mathcal{R}$. The whole brain $\mathcal{R}$ is partitioned into $G$ predefined brain regions, denoted as $\mathcal{R}_1,\ldots, \mathcal{R}_G$, such that $\mathcal{R} = \cup_{g=1}^G \mathcal{R}_g $ and $\mathcal{R}_g \cap \mathcal{R}_{g'} = \emptyset$ for $g\neq g'$. Different imaging modalities may focus on measuring different types of functional brain activity, thus the voxels of interests within the same region is the image type specific.
Let $Y_{j}(v)$ represent the image signal of type $j$ at voxel $v \in \mathcal{R}$.  For each $g = 1,\ldots,G$,  let $Y_{j,g}$ be a vector consisting of $Y_j(v)$'s with the voxel $v$'s of interests in region $\mathcal{R}_g$ for image type $j$. Suppose that the dataset consists of $n$ independent subjects. 
We assume all the images are mutually independent over subjects.
In neuroimaging studies, different types of hypothesis testing problems have been considered according to the specific study aims. In particular, it is of great interest to test whether the different imaging modalities are independent within each brain region. The null hypothesis  can be formulated as
\begin{enumerate}
\item[{(a)}] For each region $g=1,\ldots,G$,
\begin{equation}\label{eq:region_test}
H_{0,g}^{(a)}: Y_{1,g}, \ldots,  Y_{J,g} \mbox{ are mutually independent}\,.
\end{equation}
\end{enumerate}
Another type of  problems has drawn much attention in neuroimaging studies is to test the independence in imaging measurements between different regions either within the same modalities or across different modalities. The corresponding null hypotheses are given as follows.
\begin{enumerate}
\item[{(b)}] For each pair $(g,g')$ with $g \neq g'$, 
\begin{equation}\label{eq:region_pair_test}
H_{0,g,g'}^{(b)}: Y_{j,g} \mbox{ and } Y_{j,{g'}} \mbox{ are independent for each modality } j = 1,\ldots, J\,.
\end{equation}
\item[{(c)}] For each pair $(g,g')$ with $g,g'=1,\ldots,G$,
\begin{equation}\label{eq:region_pair_cross_test}
H_{0,g,g'}^{(c)}: Y_{j,g} \mbox{ and } Y_{j',g'} \mbox{ are independent for all modality pairs } j \neq j'\,.
\end{equation}
\end{enumerate}
As we will mention later, 
testing for hypotheses (a)--(c) all can be formulated in a unified framework that testing independence structure among the components of a random vector, which is so fundamental that are relevant for many statistical problems.
In the application of the complex multimodal imaging data with large sample sizes, the dimension of random vectors can be extremely large.

Testing the independence between two random vectors $X\in \mathbb{R}^{p_x}$ and $Y\in\mathbb{R}^{p_y}$ can be viewed as a special case of the testing problems (a) with $J=2$, (b) with $J=1$ and (c) with $J=2$.  When $p_x=p_y=1$, the classical tests target on  testing whether the Pearson correlation coefficient~\citep{pearson1920}, Spearman's rho coefficient~\citep{spearman1904}, Kendall's tau coefficient~\citep{kendall1938}, Blum-Kiefer-Rosenblatt's R~\citep{blum1961} and the maximal information coefficient~\citep{reshef2011} equal to zero.  
When $p_x, p_y>1$, under the normality assumption, \cite{wilks1935} proposed the test based on the sample correlations and \cite{hotelling1936} proposed the test based on the canonical correlations. To get rid of the normality assumption, some nonparametric tests based on nonlinear dependence metrics have been proposed, for example, the distance covariance/correlation \citep{szekely2007}, the Hilbert-Schmidt independence criterion~\citep[HSIC]{gretton2008} and the sign covariance~\citep{bergsma2014}. See also \cite{heller2013} and \cite{gieser1997} for tests based on the ranks of the pairwise distances between the samples and quadrant statistic, respectively, and \cite{taskinen2005} for extensions of Kendall's tau and Spearman's rho statistics in multivariate cases. More recently, \cite{shi2020} proposed a distribution-free test combined the distance covariance with the center-outward ranks and signs~\citep{Hallin2021}, and \cite{deb2021} developed a more general framework for the multivariate rank-based test using optimal transportation. See also \cite{Berrett2021} for the permutation test based on $U$-statistic. When $p_x$ and $p_y$ diverge with the sample size $n$, \cite{gao2020} considered the test based on the bias-corrected distance correlation, \cite{zhu2020} proposed tests based on an aggregation of marginal sample distance covariance and HSIC, and \cite{chakraborty2019} suggested the test based on some newly proposed dependence metrics. \cite{zhu2017projcorr} used the squared projection covariance to characterize the dependence between two random vectors with arbitrary dimensions, where such measure equals to zero if and only if the two random vectors are independent.

A more general problem of interest is to test the mutually independence of $k$ random vectors $X^{(1)},\ldots,X^{(k)}$, which is identical to our testing problem (a) with fixed $g$. When $X^{(1)},\ldots,X^{(k)}$ are univariate, the methods in \cite{blum1961} and \cite{nagao1973} can be applied for the scenario with fixed $k$. For normally distributed $(X^{(1)},\ldots,X^{(k)})^{\T}$, \cite{bai2009}, \cite{ledoit2002} and \cite{schott2005} considered the cases with $k/n \to c $ for some constant $c>0$, where all these methods aim at testing whether the covariance matrix of $(X^{(1)},\ldots,X^{(k)})^{\T}$ is an identity matrix. \cite{han2017} considered two families of distribution-free test statistics, which allows $k$ to diverge exponentially fast in $n$.  \cite{leung2018} and \cite{yao2018} proposed tests based on pairwise rank correlations and pairwise distance covariance, respectively.  
When $X^{(1)},\ldots,X^{(k)}$ are multivariate, \cite{pfister2018} and \cite{jin2018} proposed tests based on $k$-variate HSIC and generalized distance covariance, respectively, with fixed $k$. See also \cite{ChakrabortyZhang2019} for the test based on the high order distance covariance.

Both global and multiple tests for hypotheses (a)--(c) are of great interest.
However, there is lack of formal statistical inference procedure to address these questions by jointly analyzing the complex multimodal imaging data. Our setting is more complicated than those considered in the aforementioned methods. For the three global null hypotheses (i) $H_{0,g}^{(a)}$ holds for all  $g=1,\ldots,G$, (ii) $H_{0,g,g'}^{(b)}$ holds for all pair $(g,g')$ with $g\neq g'$, and (iii) $H_{0,g,g'}^{(c)}$ holds for all pair $(g,g')$ with $g,g'=1,\ldots,G$, for (a)--(c), respectively, the aforementioned methods cannot be applied directly. Although some of these methods can be used to derive the p-value of each $H_{0,g}^{(a)}$ in the hypothesis (a), the validity of the associated multiple testing procedure is still unknown. Meanwhile, none of these methods can address the multiple tests for hypotheses (b) and (c). To fill this gap, in this work, we propose rigorous statistical testing procedures.

For a vector $X$ and an index set $\mathcal{S}$, denote by $X_{\mathcal{S}}$ the subvector of $X$ that includes all components of $X$ indexed by $\mathcal{S}$. As we will state in Section \ref{sc:preliminaries}, the testing problems (a)--(c) can be all covered by a general form
$
	H_{0,q}: X_{\cS_{1}} \mbox{ is independent of } X_{\cS_{2}} \mbox{ for any } (\mathcal{S}_{1}, \mathcal{S}_{2}) \in \mathcal{K}_q$,
where $q=1,\ldots,Q$, and $\mathcal{K}_q$ is a set consisting of all $(\mathcal{S}_{1}, \mathcal{S}_{2})$ of interest with $\mathcal{S}_1\cap\mathcal{S}_2=\emptyset$. For each $(\mathcal{S}_1,\mathcal{S}_2)\in\mathcal{K}_q$, we apply the squared projection covariance proposed in \cite{zhu2017projcorr} to characterize the pairwise dependency between $X_{\mathcal{S}_1}$ and $X_{\mathcal{S}_2}$, and propose a computationally efficient method to estimate the squared projection covariance rather than using the $U$-statistic type estimate considered in \cite{zhu2017projcorr}. For the global testing problem such that $H_{0,q}$ holds for all $q=1,\ldots,Q$, we first propose an $L$-statistic type test statistic based on the estimates of the squared projection covariances, and then approximate its null-distribution by the distribution of some functional of a normal random vector, which allows (i) arbitrary dependency among the components of $X$, and (ii) $d=\sum_{q=1}^Q|\mathcal{K}_q|$ diverging exponentially fast in the sample size $n$. For the multiple testing problem that identifies which $H_{0,q}$'s are not true, we also use an $L$-statistic as the marginal test statistic for each $H_{0,q}$. Since these marginal test statistics are not pivotal and the dependency among them are complicated,  to establish the theoretical guarantee of the associated FDR control procedure is quite challenging. In most FDR control works, the limiting distributions of the marginal test statistics usually have explicit forms (e.g., normal distribution, $t$ distribution, chi-square distribution) and thus the p-values of the marginal hypotheses can be obtained easily. However, in our setting, the limiting distributions of the proposed marginal test statistics do not have explicit forms (or even does not exist) in general. We need to approximate their null-distributions by the Gaussian approximation technique (\citealp{CCk2013}, \citealp{CCW2021}) and then obtain the associated p-values, where the approximation error due to Gaussian approximation plays a key role in our theoretical analysis of the  FDR procedure, for which we need to derive the uniform non-asymptotic error bounds for the approximations over $q=1,\ldots,Q$. Some recent development of the Gaussian approximation technique can be found in, for example, \cite{DengZhang2020}, \cite{CCKK2022} and \cite{Lopes2022}.

Our proposed method and theory  make several novel contributions.  Motivated by neuroimaging applications, we develop a general and unified approach for analyzing the independence structure among the components of an ultra high-dimensional random vector. Our method provides a powerful tool to study complex dependence structure in multimodal imaging data. It can construct the brain dependence networks by aggregating the local test results across region pairs or modality pairs. To implement the proposed test procedures, we develop a computationally efficient distributed algorithm which enables the scalability of our method to analyze large-scale high-resolution imaging data. To the best of our knowledge, our theoretical analysis is the first attempt in the literature of FDR control that uses Gaussian approximation technique to derive the p-values of the marginal hypotheses which rely on complicated test statistics, and also establish the theoretical guarantee of the associated FDR control procedure.




The rest of this paper is organized as follows. The methodology is given in Section \ref{sc:method} which includes the procedures for the global test and the multiple test with FDR control, and the distributed algorithm for the implementation of our proposed procedures. Section \ref{sc:simulation} studies the finite sample performance of our proposed methods. The analysis of HCP data is presented in Section \ref{sec:realdata}. Section \ref{sec:discussion} concludes this paper by some brief discussion. All technical proofs are relegated to the supplementary material.
At the end of this section, we introduce some notation that is used throughout the paper. For any positive integers $N$ and $a$, we write $[N]:=\{ 1, \ldots, N \}$ and $[N]+a := \{ 1+a, \ldots, N+a \}$. The notation $I(\cdot)$ denotes the indicator function and $1_{n}$ denotes the $n$-dimensional vector with all the elements being 1. For a $v_1\times v_2$ matrix $A=(a_{i,j})_{v_1\times v_2}$, define $|A|_\infty=\max_{i\in[v_1],j\in[v_2]}|a_{i,j}|$ and write $A_{\mathcal{V}_1,\mathcal{V}_2}=(a_{i,j})_{i\in\mathcal{V}_1,j\in\mathcal{V}_2}$ with $\mathcal{V}_{1}\subset[v_1]$ and $\mathcal{V}_{2}\subset[v_2]$. For a $v \times v$ matrix $B$, set $\diag(B)$ to be the $v \times v$ matrix only having the main diagonal of $B$.
For a $k$-dimensional vector $X=(X_{1},\ldots,X_{k})^{\T}$ and an index set $\mathcal{V}\subset [k]$, let $X_{\mathcal{V}}$ be the subvector of $X$ consisting of components indexed by $\mathcal{V}$. For any set $\mathcal{S}$, let $|\mathcal{S}|$ denote its cardinality. For two non-zero vectors $X,Y\in\mathbb{R}^k$, we write $\angle(X,Y)=\arccos\{X^{\T}Y/(|X|_2|Y|_2)\}$ for the angle between them, where $|\cdot|_2$ denotes the vector $l_2$-norm. If $X=0$ or $Y=0$, we define $\angle(X,Y)=0$. For two sequences of positive numbers $\{a_n\}$ and $\{b_n\}$, we write $a_n\lesssim b_n$ or $b_n\gtrsim a_n$ if $\limsup_{n\rightarrow\infty}a_n/b_n\leq c$ for some positive constant $c$, and $a_n\ll b_n$ or $b_n\gg a_n$ if $\lim_{n\rightarrow\infty}a_n/b_n=0$.

\section{Methodology}\label{sc:method}

\subsection{Preliminaries}\label{sc:preliminaries}

Let $X\sim F$ be a generic $p$-dimensional random vector. 
The null hypotheses (a)--(c) defined as \eqref{eq:region_test}--\eqref{eq:region_pair_cross_test} in Section \ref{sc:introduction} all can be formulated as the following general form:
\begin{align}\label{eq:general_test}
	H_{0,q}: X_{\cS_{1}} \mbox{ is independent of } X_{\cS_{2}} \mbox{ for any } (\mathcal{S}_{1}, \mathcal{S}_{2}) \in \mathcal{K}_q\,,
\end{align}
where $q\in[Q]$ and $\mathcal{K}_q$ is a set consisting of all $(\mathcal{S}_{1}, \mathcal{S}_{2})$ of interest with $\mathcal{S}_1\cap\mathcal{S}_2=\emptyset$. Here we allow that $\mathcal{K}_q$ may be different for different $q$.

Write $X = (Y_{1,1}^{\T}, \ldots, Y_{1,G}^{\T},\ldots, Y_{J,1}^{\T}, \ldots, Y_{J,G}^{\T})^{\T}$ with $Y_{j,g}$ defined above \eqref{eq:region_test}. The null hypothesis $H_{0,g}^{(a)}$ of the testing problem (a) given in \eqref{eq:region_test} can be reformulated as \eqref{eq:general_test} with $Q=G$, $q=g$ and $\mathcal{K}_q=\{(\mathcal{S}_{q,j},\mathcal{S}_{q,-j}):j\in[J]\}$ such that $X_{\mathcal{S}_{q,j}}=Y_{j,g}$, $X_{\mathcal{S}_{q,-j}}=(Y_{1,g}^{\T},\ldots,Y_{j-1,g}^{\T},Y_{j+1,g}^{\T},\ldots,Y_{J,g}^{\T})^{\T}$ and $|\mathcal{K}_q|=J$. Write $\mathcal{G}_1=\{(g,g'):g,g'\in[G]~\textrm{with}~g< g'\}$ and $Q=|\mathcal{G}_1|$. There exists a bijective mapping $\psi_1:\mathcal{G}_1\rightarrow[Q]$. For given $(g,g')\in\mathcal{G}_1$, let $q=\psi_1(g,g')$ and the associated null hypothesis $H_{0,g,g'}^{(b)}$ of the testing problem (b) given in \eqref{eq:region_pair_test} can be stated as \eqref{eq:general_test} with $\mathcal{K}_q=\{(\mathcal{S}_{q,j,1},\mathcal{S}_{q,j,2}):j\in[J]\}$ such that $X_{\mathcal{S}_{q,j,1}}=Y_{j,g}$, $X_{\mathcal{S}_{q,j,2}}=Y_{j,g'}$ and $|\mathcal{K}_q|=J$. Let $\mathcal{G}_2=\{(g,g'):g,g'\in[G]\mbox{ with } g \leq g'\}$ and $Q=|\mathcal{G}_2|$. We can also construct a bijective mapping $\psi_2:\mathcal{G}_2\rightarrow[Q]$. For given $(g,g')\in\mathcal{G}_2$, letting $q=\psi_2(g,g')$, the associated null hypothesis $H_{0,g,g'}^{(c)}$ of the testing problem (c) given in \eqref{eq:region_pair_cross_test} can be stated as \eqref{eq:general_test} with $\mathcal{K}_q=\{(\mathcal{S}_{q,j,1},\mathcal{S}_{q,j',2}):j\neq j'\}$ such that $X_{\mathcal{S}_{q,j,1}}=Y_{j,g}$, $X_{\mathcal{S}_{q,j',2}}=Y_{j',g'}$ and $|\mathcal{K}_q|=J(J-1)I(g\neq g')+2^{-1}J(J-1)I(g=g')$. 

To characterize the dependence between two random vectors $X_{\mathcal{S}_1}$ and $X_{\mathcal{S}_2}$, we consider using the squared projection covariance \citep{zhu2017projcorr} defined as
\begin{align}\label{eq:Pcov}
\mathrm{Pcov}^2(X_{\cS_{1}},X_{\cS_{2}}) = &\,\int_{|\alpha|_2=1}\int_{|\beta|_2=1}\int_{(u,v)\in\mathbb{R}^2} {\rm cov}^2\{I(\alpha^{\T}X_{\cS_{1}} \le u), I(\beta^{\T}X_{\cS_{2}} \le v) \}\,\notag\\
&~~~~~~~~~~~~~~~~~~~~~~~~~~~~~~~~~~~~~\times\rd F_{\alpha^{\T}X_{\cS_{1}},\beta^{\T}X_{\cS_{2}}}(u,v)\,\rd \alpha \, \rd \beta\,,
\end{align}
where $F_{\alpha^{\T}X_{\cS_1},\beta^{\T}X_{\cS_2}}(\cdot,\cdot)$ is the joint distribution function of $(\alpha^{\T}X_{\cS_1},\beta^{\T}X_{\cS_2})$. Such defined $\mathrm{Pcov}^2(X_{\cS_{1}},X_{\cS_{2}})$ has two advantages: (i) $\mathrm{Pcov}^2(X_{\cS_{1}},X_{\cS_{2}})\geq0$ and $\mathrm{Pcov}^2(X_{\cS_{1}},X_{\cS_{2}})=0$ if and only if $X_{\mathcal{S}_1}$ is independent of $X_{\mathcal{S}_2}$, and (ii) $\mathrm{Pcov}^2(X_{\cS_{1}},X_{\cS_{2}})$ does not require $X_{\mathcal{S}_1}$ and $X_{\mathcal{S}_2}$ to share the same dimensionality and allows arbitrary dimensions of $X_{\mathcal{S}_1}$ and $X_{\mathcal{S}_2}$. The second advantage is useful in high-dimensional problems where the dimensions of $X_{\mathcal{S}_1}$ and $X_{\mathcal{S}_2}$ can be quite large in practice. As shown in \cite{zhu2017projcorr}, $\mathrm{Pcov}^2(X_{\cS_{1}},X_{\cS_{2}})$ defined as \eqref{eq:Pcov} has the following explicit form:
\begin{align}\label{eq:pcovexp}
\mathrm{Pcov}^2(X_{\cS_{1}},X_{\cS_{2}})=&~\mathbb{E}\big\{\angle(X_{\cS_1}^{(1)}-X_{\cS_1}^{(3)},X_{\cS_1}^{(4)}-X_{\cS_1}^{(3)})\cdot\angle(X_{\cS_2}^{(1)}-X_{\cS_2}^{(3)},X_{\cS_2}^{(4)}-X_{\cS_2}^{(3)})\big\}\notag\\
&+\mathbb{E}\big\{\angle(X_{\cS_1}^{(1)}-X_{\cS_1}^{(3)},X_{\cS_1}^{(4)}-X_{\cS_1}^{(3)})\cdot\angle(X_{\cS_2}^{(2)}-X_{\cS_2}^{(3)},X_{\cS_2}^{(5)}-X_{\cS_2}^{(3)})\big\}\\
&-2\mathbb{E}\big\{\angle(X_{\cS_1}^{(1)}-X_{\cS_1}^{(3)},X_{\cS_1}^{(4)}-X_{\cS_1}^{(3)})\cdot\angle(X_{\cS_2}^{(2)}-X_{\cS_2}^{(3)},X_{\cS_2}^{(4)}-X_{\cS_2}^{(3)})\big\}\,, \notag
\end{align}
where $(X_{\cS_1}^{(1)},X_{\cS_2}^{(1)}), \ldots, (X_{\cS_1}^{(5)},X_{\cS_2}^{(5)})$ are five independent copies of $(X_{\cS_1},X_{\cS_2})$. Based on \eqref{eq:pcovexp}, for given independent and identically distributed observations $\{X_{1},\ldots,X_n\}\sim F$, \cite{zhu2017projcorr} suggest to estimate $\mathrm{Pcov}^2(X_{\cS_{1}},X_{\cS_{2}})$ by
\begin{align}\label{eq:pcovsample}
&\widehat{\mathrm{Pcov}^2}(X_{\cS_1},X_{\cS_2})\notag\\
&~~~~~~=\frac{1}{n^3}\sum_{i,k,l=1}^n\angle(X_{i,\cS_1}-X_{k,\cS_1},X_{l,\cS_1}-X_{k,\cS_1})\cdot\angle(X_{i,\cS_2}-X_{k,\cS_2},X_{l,\cS_2}-X_{k,\cS_2}) \notag\\
&~~~~~~~~~+\frac{1}{n^5}\sum_{i,j,k,l,r=1}^n\angle(X_{i,\cS_1}-X_{k,\cS_1},X_{l,\cS_1}-X_{k,\cS_1})\cdot\angle(X_{j,\cS_2}-X_{k,\cS_2},X_{r,\cS_2}-X_{k,\cS_2})  \\
&~~~~~~~~~-\frac{2}{n^4}\sum_{i,j,k,l=1}^n\angle(X_{i,\cS_1}-X_{k,\cS_1},X_{l,\cS_1}-X_{k,\cS_1})\cdot\angle(X_{j,\cS_2}-X_{k,\cS_2},X_{l,\cS_2}-X_{k,\cS_2})\,. \notag
\end{align}

When the dimensions of $X_{\cS_1}$ and $X_{\cS_2}$ are quite large, to calculate the $U$-statistic type estimate 
\eqref{eq:pcovsample} requires high computing cost and large storage space even with moderately large sample size $n$, since all observations need to be read from files at once and saved as a very large matrix. In our motivating HCP task fMRI data, the sample size $n=922$ and the number of voxels in the standard 2mm MNI brain template $V=185,405$. In this data set, we have five different fMRI contrast maps (2back-0back, task-rest, story-math, mental, random) and 116 predefined brain regions from the automated anatomical labelling (AAL)~\citep{tzourio2002automated}. If we are interested in the testing problem (a) given in \eqref{eq:region_test} for the first 90 brain regions and 2 modalities `story-math' and `mental',  we can formulate this problem as \eqref{eq:general_test} with $Q=90$ and $|\mathcal{K}_q|=1$ for each $q=1,\ldots,90$. On a server with two Intel(R) Xeon(R) Platinum 8160 CPU @ 2.10GHz, the computation time of the associated $\widehat{\mathrm{Pcov}^2}(X_{\cS_1},X_{\cS_2})$ given in \eqref{eq:pcovsample} for $H_{0,q}$ using all 922 observations ranges from 185.65 seconds to 1768.08 seconds for $q=1,\ldots,90$.

To improve the computational efficiency, we consider to divide the $n$ observations $\{X_1,\ldots,X_n\}$ into several moving subgroups and compute the associated squared projection covariance within these subgroups. Let $B\geq5$ denote the length of each subgroup, and write $M=n-B+1$. For any $m\in[M]$, let $\mathcal{B}_{m} = \{X_{m},\ldots, X_{m+B-1}\}$ be the $m$-th moving subgroup. For given $q\in[Q]$ and $(\mathcal{S}_1,\mathcal{S}_2)\in\mathcal{K}_q$, we estimate $\mathrm{Pcov}^2(X_{\cS_{1}},X_{\cS_{2}})$ based on $\mathcal{B}_m$ in the same manner as \eqref{eq:pcovsample} but with replacing $\{X_1,\ldots,X_n\}$ by $\mathcal{B}_m$, and then denote the associated estimate by $U_{m,q,\cS_1,\cS_2}$.
Gathering the estimates in all moving subgroups together, we finally estimate $\mathrm{Pcov}^2(X_{\cS_{1}},X_{\cS_{2}})$ by
\begin{align}\label{eq:ourestmethod}
\bar{U}_{q,\cS_1,\cS_2}=\frac{1}{M}\sum_{m=1}^{M}U_{m,q,\cS_1,\cS_2}
\end{align}
for any $q\in[Q]$ and $(\mathcal{S}_1,\mathcal{S}_2)\in\mathcal{K}_q$. By doing so, we just need to read $B$ observations of $\mathcal{B}_{m}$ at one time to calculate $U_{m,q,\cS_1,\cS_2}$, which can largely reduce the computing cost. The computation time of such method with $B=5$ using all 922 observations ranges between $2.72$ seconds and $33.30$ seconds for $q =1,\ldots,90$, which economizes $98\%$ computing cost in comparison to that of \cite{zhu2017projcorr}.  As we will discuss in Section \ref{sc:parallel} later, our method is convenient to develop a distributed algorithm for \eqref{eq:ourestmethod} to further enhance the computational efficiency. This is another advantage of our proposed method. Our theoretical analysis allows both fixed $B$ and diverging $B$.  Extensive numerical studies indicate that it works quite well even when we select $B=5$.


Our proposed estimate $\bar{U}_{q,\mathcal{S}_1,\mathcal{S}_2}$ in \eqref{eq:ourestmethod} also has another convenience in solving the hypothesis testing problem \eqref{eq:general_test}.  For the $U$-statistic type estimate $\widehat{\mathrm{Pcov}^2}(X_{\mathcal{S}_1},X_{\mathcal{S}_2})$ given in \eqref{eq:pcovsample}, \cite{zhu2017projcorr} shows that under null hypothesis $n\widehat{\mathrm{Pcov}^2}(X_{\mathcal{S}_1},X_{\mathcal{S}_2})$ converges in distribution to the weighted infinite  sum of independent chi-square distributions. If we construct the test statistic for the null hypothesis $H_{0,q}$ specified in \eqref{eq:general_test} based on $\{\widehat{\mathrm{Pcov}^2}(X_{\mathcal{S}_1},X_{\mathcal{S}_2})\}_{(\mathcal{S}_1,\mathcal{S}_2)\in\mathcal{K}_q}$, we will encounter two obstacles. First, if $|\mathcal{K}_q|>1$, how to characterize the dependency among different $\widehat{\mathrm{Pcov}^2}(X_{\mathcal{S}_1},X_{\mathcal{S}_2})$'s is extremely difficult (if not impossible)  since their limiting distributions are too complicated. Second, to do inference based on $\widehat{\mathrm{Pcov}^2}(X_{\mathcal{S}_1},X_{\mathcal{S}_2})$ usually needs to involve the random permutation procedure which requires very high computing cost. Notice that $\bar{U}_{q,\mathcal{S}_1,\mathcal{S}_2}$ is consistent to ${\rm Pcov}^2(X_{\mathcal{S}_1},X_{\mathcal{S}_2})$ for any $q\in[Q]$ and $(\mathcal{S}_1,\mathcal{S}_2)\in\mathcal{K}_q$. Then $\max_{(\mathcal{S}_1,\mathcal{S}_2)\in\mathcal{K}_q}\sqrt{M}\bar{U}_{q,\mathcal{S}_1,\mathcal{S}_2}/\hat{\sigma}_{q,\mathcal{S}_1,\mathcal{S}_2}$ provides a natural test statistic for the null hypothesis $H_{0,q}$ specified in \eqref{eq:general_test}, where $\hat{\sigma}_{q,\mathcal{S}_1,\mathcal{S}_2}$ is the estimation of the standard deviation of $\sqrt{M}\bar{U}_{q,\mathcal{S}_1,\mathcal{S}_2}$. Since $\sqrt{M}\bar{U}_{q,\mathcal{S}_1,\mathcal{S}_2}/\hat{\sigma}_{q,\mathcal{S}_1,\mathcal{S}_2}\rightarrow_d\mathcal{N}(0,1)$ for any $(\mathcal{S}_1,\mathcal{S}_2)\in\mathcal{K}_q$, using the Gaussian approximation technique, we know the null-distribution of $\max_{(\mathcal{S}_1,\mathcal{S}_2)\in\mathcal{K}_q}\sqrt{M}\bar{U}_{q,\mathcal{S}_1,\mathcal{S}_2}/\hat{\sigma}_{q,\mathcal{S}_1,\mathcal{S}_2}$ can be always approximated by the distribution of the largest component in a $|\mathcal{K}_q|$-dimensional normal distributed random vector with mean zero. 
Hence, the inference of the test statistic based on $\{\bar{U}_{q,\mathcal{S}_1,\mathcal{S}_2}\}_{(\mathcal{S}_1,\mathcal{S}_2)\in\mathcal{K}_q}$ can be implemented easily and efficiently. See Sections \ref{sc:test global}--\ref{sc:parallel} for details.

\subsection{A Global Testing Procedure}\label{sc:test global}
In this part, we first consider how to test the following global null hypothesis:
\begin{align}\label{eq:globaltest}
H_0: H_{0,q}\mbox{ given in \eqref{eq:general_test} holds for all }q\in[Q]\,.
\end{align}
Due to $\mathbb{E}(\bar{U}_{q,\cS_1,\cS_2})=\mathrm{Pcov}^2(X_{\cS_{1}},X_{\cS_{2}})$, large values of $\bar{U}_{q,\cS_1,\cS_2}$ provide evidence against the marginal null hypothesis that $X_{\cS_1}$ is independent of $X_{\cS_2}$ for given $(\cS_1,\cS_2)\in\mathcal{K}_q$. Denote by $\sigma_{q,\cS_1,\cS_2}$ the standard deviation of $\sqrt{M}\bar{U}_{q,\cS_1,\cS_2}$ and let $\hat{\sigma}_{q,\cS_1,\cS_2}$ be its consistent estimate which will be given below \eqref{eq:estsigma}. Then we define the standardized statistic
\begin{align}\label{eq:martest}
T_{q,\cS_1,\cS_2} = \frac{\sqrt{M}\bar{U}_{q,\cS_1,\cS_2}}{\hat{\sigma}_{q,\cS_1,\cS_2}}
\end{align}
for any $q\in[Q]$ and $(\cS_1,\cS_2)\in\mathcal{K}_q$. To test the global null hypothesis \eqref{eq:globaltest}, we use the information of the largest $L$ values among $\{T_{q,\cS_1,\cS_2}\}_{q\in[Q],(\cS_1,\cS_2)\in\mathcal{K}_q}$ to calculate the global test statistic. More specifically, for each $q\in[Q]$, we first stack $\{T_{q,\cS_1,\cS_2}\}_{(\cS_1,\cS_2)\in\mathcal{K}_q}$ into a $|\mathcal{K}_q|$-dimensional vector $T_{n}^{(q)}=(T_{n,1}^{(q)},\ldots,T_{n,|\mathcal{K}_q|}^{(q)})^{\T}$. We then obtain a $d$-dimensional vector $T_n= (T_{n,1},\ldots, T_{n,d})^{\T} =(T_{n}^{(1),\T},\ldots,T_{n}^{(Q),\T})^{\T} $ with $d =\sum_{q=1}^Q|\mathcal{K}_q|$. Given a pre-determined integer $L$, the global test statistic for \eqref{eq:globaltest} is given by
\begin{equation}
  W_{n,L}=\max_{1\le j_1 < \cdots < j_L \le d }\sum_{s=1}^L T_{n,j_s}\,,
  \label{eq:TS_MaxS}
\end{equation}
which is an $L$-statistic. Such kind of statistic that combines several largest signals together has been also used in \cite{Zhang2015} and \cite{FanShaoZhou2018} for solving other problems. Based on 
$W_{n,L}$, 
we formally reject the global null hypothesis $H_0$ in \eqref{eq:globaltest} at the significance level $\alpha\in(0,1)$ if
\begin{equation}\label{eq:cv}
W_{n,L}>{\rm cv}_{L,\alpha}\,,
\end{equation}
where ${\rm cv}_{L,\alpha}$ is the critical value satisfying $\mathbb{P}_{H_0}(W_{n,L}>{\rm cv}_{L,\alpha})\rightarrow\alpha$ as $n\rightarrow\infty$.

To determine ${\rm cv}_{L,\alpha}$, we need to investigate the null-distribution of $W_{n,L}$. When $L=1$, $W_{n,1}= \max_{j\in [d]}T_{n,j}$ is a maximum-type test statistic. Due to each $T_{n,j}\rightarrow_d\mathcal{N}(0,1)$ under $H_0$, we may consider to approximate the null-distribution of $W_{n,1}$ by some extreme-value distribution. However, such approximation relies heavily on the structural assumptions on the asymptotic covariance matrix of $T_n=(T_{n,1},\ldots,T_{n,d})^{\T}$ which are quite restrictive and difficult to be verified in practice. On the other hand, even such approximation works, its convergence rate is usually slow which will usually cause size distortion. Taking the extreme-value distribution of type I as an example, the convergence rate is just of order $O\{\log(\log n)/\log n\}$. In comparison to $W_{n,1}$, $W_{n,L}$ with $L>1$ gathers more information which will lead to power-enhancement under alternatives. 

To the best of our knowledge, the null-distribution of $W_{n,L}$ with $L>1$ and $d\gg n$ has been rarely studied in the literature. Note that $\bar{U}_{q,\mathcal{S}_1,\mathcal{S}_2}$ in \eqref{eq:martest} is the average of the dependent sequence $\{U_{m,q,\mathcal{S}_1,\mathcal{S}_2}\}_{m=1}^M$. Deriving the null-distribution of $W_{n,L}$ is challenging. In this paper, we propose a novel procedure to approximate the null-distribution of $W_{n,L}$ with general $L\geq1$ that can diverge with $n$. Our procedure does not need to impose any structural assumption on the asymptotic covariance matrix of $T_n$ and allows arbitrary dependency among the components of $T_n$.

For each $m\in[M]$, we stack $\{U_{m,q,\cS_1,\cS_2}\}_{q\in[Q],(\cS_1,\cS_2)\in\mathcal{K}_q}$ into a $d$-dimensional vector, denoted by $U_m=(U_{m,1},\ldots,U_{m,d})^{\T}$, with the same order as $T_n=(T_{n,1},\ldots,T_{n,d})^{\T}$. Write $\Sigma_n = \mathrm{Cov}(\sqrt{M}\bar{U})$ with $\bar{U}=M^{-1}\sum_{m=1}^{M}U_{m}$. Then $\{\sigma_{q,\cS_1,\cS_2}^2\}_{q\in[Q],(\cS_1,\cS_2)\in\mathcal{K}_q}$ are the $d$ elements in the main diagonal of $\Sigma_n$. Let $R_n=D_n^{-1/2}\Sigma_n D_n^{-1/2}$ with $D_n=\diag(\Sigma_n)$. For a $d$-dimensional vector $Z=(Z_1,\ldots,Z_d)^{\T}$, we sort its components in descending order as $Z_{(1)}\geq\cdots\geq Z_{(d)}$ and define a function
\begin{align*}
f_L(Z)=\sum_{j=1}^LZ_{(j)}\,.
\end{align*} Then $W_{n,L}=f_L(T_n)$. Proposition \ref{tm:GA}(i) shows that the critical value ${\rm cv}_{L,\alpha}$ in \eqref{eq:cv} can be selected as the $(1-\alpha)$-quantile of the distribution of $f_L(\xi)$, where $\xi\sim\mathcal{N}(0,R_{n})$.

In practice,  we need to estimate $R_n$. 
Since $\{U_{m}\}_{m=1}^{M}$ is a $d$-dimensional $(B-1)$-dependent process, then
$
\Sigma_n=M^{-1}\sum_{|m_1-m_2|< B}\mathrm{Cov}(U_{m_1},U_{m_2})$,
which indicates that
\begin{equation}\label{eq:estsigma}
\hat{\Sigma}_n = \sum_{j=-B+1}^{B-1}\hat{H}_j\,,
\end{equation} with $\hat{H}_j = M^{-1}\sum_{m=j+1}^{M}(U_{m}-\bar{U})(U_{m-j}-\bar{U})^{\T}$ for $j \geq 0$ and $\hat{H}_j = M^{-1}\sum_{m=-j+1}^{M}(U_{m+j}-\bar{U})(U_{m}-\bar{U})^{\T}$ for $j < 0$, provides an estimate of $\Sigma_n$. Let $\hat{D}_n=\diag(\hat{\Sigma}_n)$ which gives the estimates of $\{\sigma_{q,\cS_1,\cS_2}^2\}_{q\in[Q],(\cS_1,\cS_2)\in\mathcal{K}_q}$. Write  $\hat{R}_n=\hat{D}_n^{-1/2}\hat{\Sigma}_n\hat{D}_n^{-1/2}$ and $\hat{F}(t)=\mathbb{P}\{f_L(\hat{\xi})\leq t\,|\,X_1,\ldots,X_n\}$ with $\hat{\xi}\,|\,X_1,\ldots,X_n\sim\mathcal{N}(0,\hat{R}_n)$. Proposition \ref{tm:GA}(ii) shows that $\mathrm{cv}_{L,\alpha}$ can be approximated by
$$
\hat{\mathrm{cv}}_{L,\alpha}=\inf\big\{t\in\mathbb{R}: \hat{F}(t)\geq1-\alpha\big\}\,,
$$
the $(1-\alpha)$-quantile of the conditional distribution of $f_L(\hat{\xi})$ given $\{X_1,\ldots,X_n\}$. 
Practically, we can draw $d$-dimensional random vectors $\hat{\xi}_{1}, \ldots, \hat{\xi}_{N}$ independently from $\mathcal{N}(0,\hat{R}_n)$ for some large integer $N$, and use the $\lfloor N\alpha \rfloor$-th largest value of $f_L(\hat{\xi}_{1}), \ldots, f_L(\hat{\xi}_{N})$ to approximate $\hat{\mathrm{cv}}_{L,\alpha}$.
Such proposed test has several advantages: (i) it can hold the cases with $d\gg n$; 
(ii) it allows arbitrary dependence structure of $\Sigma_n$; and (iii) its implementation is quite simple which only needs to generate several $d$-dimensional normal random vectors. 
Write $\mathcal{F} = \{v=(v_1,\ldots,v_d)^{\T}: v_j \in \{ 0,1 \}~\textrm{for each}~j \in [d],~\textrm{and}~|v|_{0}= L \}$ and $D_n={\rm diag}(\sigma_1^2,\ldots,\sigma_d^2)$.


\begin{proposition}\label{tm:GA}
Assume there exists a universal constant $c_1>0$ such that $\min_{k\in[d]}\sigma_{k}^{2}\geq c_1$ and $\min_{v\in\mathcal{F}}v^{\T}R_nv\ge c_{1}$. Let $\xi\sim\mathcal{N}(0,R_n)$ and $\hat{\xi}\sim\mathcal{N}(0,\hat{R}_n)$. Under the global null hypothesis $H_{0}$ given in \eqref{eq:globaltest}, we have that
	{\rm(i)} $\sup_{z\in \mathbb{R}}|\mathbb{P}(W_{n,L}>z) - \mathbb{P} \{ f_L(\xi)>z\}|= o(1)$ provided that $B^4 L^{13}(\log d)^7=o(n)$ and $B^4L^{-2}(\log B)^3(\log d)^{-2}=o(n)$, and {\rm(ii)} $\sup_{z\in \mathbb{R}}|\mathbb{P}(W_{n,L}>z) - \mathbb{P} \{ f_L(\hat{\xi})>z\,|\,X_1,\ldots,X_n\}|= o_\p(1)$ provided that $B^4 L^{13}(\log d)^7=o(n)$ and $B^5L^8(\log d)^4\log(dB)=o(n)$.
\end{proposition}

The conditions $\min_{k\in[d]}\sigma_k^2\geq c_1$ and $\min_{v\in\mathcal{F}}v^{\T}R_nv\ge c_{1}$ require that the long-run variances of the sequences $\{U_{m,k}\}_{m=1}^M$ and $\{v^{\T}U_m\}_{m=1}^M$ are uniformly bounded away from zero over $k\in[d]$ and $v\in\mathcal{F}$, which are mild in practice and will be used when we apply Nazarov's inequality to bound the probability of a normal vector taking values in a small region. When we select $B$, the length of each subgroup, as a fixed constant, Proposition \ref{tm:GA} holds if $L^{13}(\log d)^7=o(n)$, which requires the pre-determined integer $L$ involved in the $L$-statistic $W_{n,L}$ defined as    \eqref{eq:TS_MaxS} should satisfy $L=o(n^{1/13})$. Furthermore, if $L$ is also selected as a fixed constant, Proposition \ref{tm:GA} holds provided that $\log d=o(n^{1/7})$, which allows $d=\sum_{q=1}^Q|\mathcal{K}_q|$ growing exponentially with the sample size $n$. 
Given Proposition \ref{tm:GA}(i), in order to prove Proposition \ref{tm:GA}(ii), it suffices to show $\delta_n:=\sup_{z\in\mathbb{R}}|\mathbb{P}\{f_L(\xi)>z\}-\mathbb{P}\{f_L(\hat{\xi})>z\,|\,X_1,\ldots,X_n\}|=o_{\p}(1)$. As shown in (S.15) of the supplementary material,  $\delta_n\lesssim \Delta_n^{1/3}\{1\vee \log(d^L/\Delta_n)\}^{2/3}$ with ${\Delta}_{n}= \max_{v_{1},v_{2}\in\mathcal{F}}|v_{1}^{\T}(R_n-\hat{R}_{n})v_{2}|$. Due to $\Delta_n\leq L^2|\hat{R}_n-R_n|_\infty$, then $\delta_n=o_\p(1)$ provided that $|\hat{R}_n-R_n|_\infty=o_\p\{L^{-4}(\log d)^{-2}\}$. For our suggested $\hat{R}_n$ specified below \eqref{eq:estsigma}, such requirement can be easily satisfied for any $d$-dimensional correlation matrix $R_n$ as long as $B^5L^8(\log d)^4\log(dB)=o(n)$ without imposing any structural assumption on $R_n$. This advantage of Gaussian approximation has also been found in, for example,  \cite{CCk2013}, \cite{CZZZ2017} and \cite{CCW2021}. The next theorems state the theoretical guarantee of the proposed global test.

\begin{theorem}\label{tm:size}
	Under the conditions of Proposition {\rm\ref{tm:GA}}, if $B^4 L^{13}(\log d)^7=o(n)$ and $B^5L^8(\log d)^4\log(dB)=o(n)$, then $\P_{H_0}(W_{n,L}>\hat{{\rm cv}}_{L,\alpha})\rightarrow \alpha$ as $n\rightarrow\infty$.
\end{theorem}

Theorem \ref{tm:GA} shows that the proposed global test has correct size control. The restrictions on $(B,L,d,n)$ for Theorem \ref{tm:size} are identical to those for Proposition \ref{tm:GA}(ii). If we select $B$ and $L$ as fixed constants, the proposed global test is valid provided that $\log d=o(n^{1/7})$. Write $\mathbb{E}(U_m)=(\mu_1,\ldots,\mu_d)^{\T}$. Theorem \ref{tm:power} indicates that the proposed global test is a consistent test under certain local alternatives.

\begin{theorem}\label{tm:power}
	Let $\rho=\max_{v\in\mathcal{F}}v^{\T}R_{n}v$ and $\lambda(|\mathcal{F}|,\alpha)=(2\log |\mathcal{F}|)^{1/2}+\{2\log(1/\alpha)\}^{1/2}$. Under the conditions of Proposition {\rm\ref{tm:GA}}, if $\max_{1\le j_{1}<\cdots<j_{L}\le d}\sum_{l=1}^{L}\sigma_{j_{l}}^{-1}\mu_{j_{l}}\ge \rho^{1/2}(1+\epsilon_{n})n^{-1/2}\lambda(|\mathcal{F}|,\alpha)$ under the alternative hypothesis $H_{1}$ for some $\epsilon_{n}>0$ satisfying $(\log |\mathcal{F}|)^{-1/2}\ll\epsilon_{n}\ll 1$,  then $\P_{H_1}(W_{n,L}>\hat{{\rm cv}}_{L,\alpha})\rightarrow 1$ as $n\rightarrow\infty$, provided that $B^5L^4\log(dB)(\log|\mathcal{F}|)^2 = o(n)$ and $B^3L^2\{\log(dB)\}^2\log|\mathcal{F}| = o(n)$.
\end{theorem}

Recall $d=\sum_{q=1}^Q|\mathcal{K}_q|$ and $U_{m,1},\ldots,U_{m,d}$, the $d$ components of $U_m$, are a permutation of the $d$ elements in the set $\{U_{m,q,\mathcal{S}_1,\mathcal{S}_2}:q\in[Q], (\mathcal{S}_1,\mathcal{S}_2)\in\mathcal{K}_q\}$. For any $j\in[d]$, there exists $q'\in[Q]$ and $(\mathcal{S}_1',\mathcal{S}_2')\in\mathcal{K}_{q'}$ such that $\mu_j=\mathbb{E}(U_{m,q',\mathcal{S}_1',\mathcal{S}_2'})={\rm Pcov}^2(X_{\mathcal{S}_1'},X_{\mathcal{S}_2'})$. Hence, the global null hypothesis $H_0$ in \eqref{eq:globaltest} is equivalent to $\mathbb{E}(U_m)=(\mu_1,\ldots,\mu_d)^{\T}=0$, and the  test statistic $W_{n,L}$ given in \eqref{eq:TS_MaxS} targets on testing whether such equivalence holds or not based on the obtained data $U_1,\ldots,U_M$, which essentially formulates the original hypothesis testing problem as a $d$-dimensional mean vector testing problem. The condition $\max_{1\le j_{1}<\cdots<j_{L}\le d}\sum_{l=1}^{L}\sigma_{j_{l}}^{-1}\mu_{j_{l}}\ge \rho^{1/2}(1+\epsilon_{n})n^{-1/2}\lambda(|\mathcal{F}|,\alpha)$ states the restriction on mean vector $(\mu_1,\ldots,\mu_d)^{\T}$ under which the proposed global test has power approaching one. Recall $\mathcal{F} = \{v=(v_1,\ldots,v_d)^{\T}: v_j \in \{ 0,1 \}~\textrm{for each}~j \in [d],~\textrm{and}~|v|_{0}= L \}$. For fixed $L$, the lower bound $\rho^{1/2}(1+\epsilon_{n})n^{-1/2}\lambda(|\mathcal{F}|,\alpha)$ has order $n^{-1/2}(\log d)^{1/2}$, which is the minimax optimal separation rate of any tests for $d$-dimensional mean vector \citep{CaiLiuXia2014}. In this case, our test shares the minimax optimal property.  Due to $|\mathcal{F}|={d\choose L}\leq d^L$, $B^5L^6\log (dB)(\log d)^2= o(n)$ provides a sufficient condition for the restrictions on $(B,L,d,n)$ for Theorem \ref{tm:power}. If we select $B$, the length of each subgroup, as a fixed constant, Theorem \ref{tm:power} holds if $L^6(\log d)^3= o(n)$, which requires the pre-determined integer $L$ involved in the $L$-statistic $W_{n,L}$ should satisfy $L=o(n^{1/6})$. Furthermore, if $L$ is also selected as a fixed constant, Theorem \ref{tm:power} holds provided that $\log d=o(n^{1/3})$, which allows $d$ growing exponentially with the sample size $n$. Together with the discussion below Theorem \ref{tm:size}, we conclude that when $\log d=o(n^{1/7})$ the proposed global test with fixed $B$ and $L$ can control Type I error at the prescribed significance level and also have power approaching one under certain local alternatives.

\subsection{A Multiple Testing Procedure with FDR Control}\label{sc:test multiple}

Section \ref{sc:test global} considers the global test whether all $H_{0,1},\ldots,H_{0,Q}$ given in \eqref{eq:general_test} hold simultaneously or not. When the global null hypothesis $H_0$ is rejected, we are interested in identifying which $H_{0,q}$'s are not true. Note that $T_{q,\cS_1,\cS_2}$ defined in \eqref{eq:martest} provides a valid test statistic for the marginal null hypothesis $H_{0,q,\cS_1,\cS_2}: X_{\cS_1}$ is independent of $X_{\cS_2}$ for given $(\cS_1,\cS_2)\in\mathcal{K}_q$. For $T_{n}^{(q)}=(T_{n,1}^{(q)},\ldots,T_{n,|\mathcal{K}_q|}^{(q)})^{\T}$ defined below \eqref{eq:martest}, we can formulate it as follows:
\begin{align*}
T_n^{(q)}=\sqrt{M}\{\hat{D}_{n}^{(q)}\}^{-1/2}\bar{U}^{(q)}\,,
\end{align*}
where $\bar{U}^{(q)}=M^{-1}\sum_{m=1}^{M}U_{m}^{(q)}$, $U_m^{(q)}$ and $\hat{D}_{n}^{(q)}$ are, respectively, the subvector of $U_m$ and the submatrix of $\hat{D}
_n$ including the associated components related to $\{T_{q,\cS_1,\cS_2}\}_{(\cS_1,\cS_2)\in\mathcal{K}_q}$. Let $\Sigma_n^{(q)}={\rm Cov}(\sqrt{M}\bar{U}^{(q)})$ and define its estimate  $\hat{\Sigma}_n^{(q)}$ in the same manner as $\hat{\Sigma}_n$ given in \eqref{eq:estsigma} but with replacing $\{U_m\}_{m=1}^M$ by $\{U_m^{(q)}\}_{m=1}^M$. The test statistic for $H_{0,q}$ can be selected as
\begin{align}\label{eq:WnLq}
 W_{n,L}^{(q)}=\max_{1\le j_1 < \cdots < j_L \le |\mathcal{K}_q|}\sum_{s=1}^L T_{n,j_s}^{(q)}\,.
\end{align}
If $|\mathcal{K}_q|=1$, we can only select $L=1$. Then $W_{n,L}^{(q)}$ can be also formulated as $W_{n,L}^{(q)}=f_L(T_n^{(q)})$ for $f_L(\cdot)$ defined in Section \ref{sc:test global}.
We reject $H_{0,q}$ when $W_{n,L}^{(q)}$ takes some large values.

Let $\mathcal{H}_0$ be the set of all true null hypotheses $H_{0,q}$ with $Q_0 = |\mathcal{H}_0|$. 
For each $q\in[Q]$, generate $\hat{\xi}_1^{(q)},\ldots,\hat{\xi}_N^{(q)}$ independently from $\mathcal{N}(0,\hat{R}_n^{(q)})$, where $\hat{R}_n^{(q)}=\{\hat{D}_n^{(q)}\}^{-1/2}\hat{\Sigma}_n^{(q)}\{\hat{D}_n^{(q)}\}^{-1/2}$. Identical to Proposition \ref{tm:GA}, we can also show that
\begin{align}\label{eq:gamult}
\max_{q\in\mathcal{H}_0}\sup_{z\in \mathbb{R}}\big|\mathbb{P}\{W_{n,L}^{(q)}>z\} - \mathbb{P}\{f_L(\hat{\xi}^{(q)})>z\,|\,X_1,\ldots,X_n\}\big|= o_\p(1)\,.
\end{align}
Then the p-value of the $q$-th hypothesis test $H_{0,q}$ is given by $\mathrm{pv}_{L}^{(q)} =  \P\{ f_{L}(\hat{\xi}^{(q)}) \ge W_{n,L}^{(q)} \,|\, X_1,\ldots,X_n \}$ with $\hat{\xi}^{(q)} \sim \mathcal{N}(0,\hat{R}_n^{(q)})$, and its normal quantile transformation has the form $V_{n,L}^{(q)} = \Phi^{-1}(1-\mathrm{pv}_{L}^{(q)})$. For implementation, $\mathrm{pv}_{L}^{(q)}$ can be approximated by $\hat{\mathrm{pv}}_{L}^{(q)} = N^{-1}\sum_{j=1}^{N}I\{f_L(\hat{\xi}_{j}^{(q)}) \ge W_{n,L}^{(q)} \}$ for some large integer $N$.
For any {$t\in\mathbb{R}$, the {\it False Discovery Proportion} (FDP) is defined as
\begin{align*}
\mathrm{FDP}(t) = \frac{ \sum_{q \in \mathcal{H}_0}I\{ V_{n,L}^{(q)} \ge t \} }{1\vee \sum_{q=1}^QI\{ V_{n,L}^{(q)}  \ge t \} }
\end{align*}
and the {\it False Discovery Rate} (FDR) is given by $\mathrm{FDR}(t) = \mathbb{E}\{ \mathrm{FDP}(t) \}$.

Given a prescribed level $\alpha\in(0,1)$, the key object here is to find the smallest $\hat{t}$ such that $\mathrm{FDR}(\hat{t})\leq \alpha$. To do this, we first consider $\mathrm{FDP}(t)$. Since the true null hypotheses set $\mathcal{H}_0$ is unknown, we need to estimate the numerator of $\mathrm{FDP}(t)$. Based on  \eqref{eq:gamult}, we have $\mathbb{P}\{V_{n,L}^{(q)} \ge t \} = 1-\Phi(t)+o(1)$ for any $q\in\mathcal{H}_0$. An ideal estimate for $\mathrm{FDP}(t)$ can be selected as
\begin{align*}
\widetilde{{\rm FDP}}(t)=\frac{ Q_0\{1-\Phi(t) \} }{1\vee\sum_{q=1}^{Q}I\{V_{n,L}^{(q)} \ge t\}}\,.
\end{align*}
Since $Q_0$ is unknown, $\widetilde{{\rm FDP}}(t)$ is infeasible in practice and we can only estimate $\mathrm{FDP}(t)$ via a more conservative way:
\begin{align}\label{eq:FDPt}
 \widehat{\mathrm{FDP}}(t) = \frac{ Q\{1-\Phi(t) \} }{1\vee\sum_{q=1}^{Q}I\{V_{n,L}^{(q)} \ge t\}}\,.
\end{align}
For given $\alpha\in(0,1)$, we choose
\begin{align}\label{eq:hatT}
\hat{t} = \inf \big\{ 0<t\leq (2\log Q - 2\log\log Q)^{1/2}: \widehat{\mathrm{FDP}}(t) \le \alpha \big\}\,.
\end{align}
If $\hat{t}$ defined in \eqref{eq:hatT} does not exist, let $\hat{t}=(2\log Q)^{1/2}$. We then reject all $H_{0,q}$'s with $V_{n,L}^{(q)}\ge \hat{t}$.

The dependency among the test statistics $\{W_{n,L}^{(q)}\}_{q\in[Q]}$ plays a key role in analyzing the theoretical property of the proposed multiple testing procedure. Since the limiting distribution of $W_{n,L}^{(q)}$ is not pivotal and does not admit explicit form, how to characterize the dependency among $\{W_{n,L}^{(q)}\}_{q\in[Q]}$ is nontrivial. Denote by $F_q(\cdot)$ the distribution function of $W_{n,L}^{(q)}$. Note that $\eta^{(q)}=\Phi^{-1}\{F_q(W_{n,L}^{(q)})\}$ follows the standard normal distribution. We can measure the dependency between $W_{n,L}^{(q)}$ and $W_{n,L}^{(q')}$ via the correlation between their transformations $\eta^{(q)}$ and $\eta^{(q')}$, where the latter one is essential in our theoretical analysis. It is obvious that the independence between $W_{n,L}^{(q)}$ and $W_{n,L}^{(q')}$ is equivalent to ${\rm Corr}\{\eta^{(q)},\eta^{(q')}\}=0$. Define the set
\begin{equation*}
 \mathcal{A}_{q}(\gamma) = \big\{ q' \in [Q]: q'\neq q, | \mathrm{Corr}\{\eta^{(q)}, \eta^{(q')}\} | \ge (\log Q)^{-2-\gamma}\big\}
\end{equation*}
for $\gamma>0$ and $q\in[Q]$. For the $q$-th test statistic $W_{n,L}^{(q)}$, we can divide the other $Q-1$ test statistics $\{W_{n,L}^{(q')}\}_{q'\in[Q]\setminus \{q\}}$ into two parts: (i) the ones  not belonging to $\mathcal{A}_q(\gamma)$ have quite weak dependence with $W_{n,L}^{(q)}$, and (ii) the ones belonging to $\mathcal{A}_q(\gamma)$ have relatively strong dependence with $W_{n,L}^{(q)}$. To construct the theoretical guarantee of the proposed multiple testing procedure, it is a common practice to analyze these two parts separately with different technical tools. See also \cite{Liu2013}, \cite{ChangShaoZhou2016} and \cite{XiaLi2020}. The next theorem indicates that our multiple testing procedure can control both FDR and FDP at the level $\alpha Q_0/Q$ asymptotically.

\begin{theorem}\label{tm:FDR}
	Let the conditions in Proposition {\rm\ref{tm:GA}} hold. Assume that $\max_{1\leq q\neq q' \leq Q}|\mathrm{Corr}\{\eta^{(q)}, \eta^{(q')}\}| \le r$ for some constant $r \in (0,1)$, and $\max_{q \in [Q]}|\mathcal{A}_q(\gamma)| = o(Q^{\nu})$ for some constants $\gamma>0$ and $0<\nu<(1-r)/(1+r)$. If $Q^{12}B^5L^4\log(nBK_{\max})\max\{(L\log K_{\max})^4,(\log n)^4\}=o(n)$ and $Q^{12}B^4L^{13}(\log K_{\max})^7=o(n)$ with $K_{\max}=\max_{q\in\mathcal{H}_0}|\mathcal{K}_q|$, then
	$\limsup_{n,Q \to \infty}\mathrm{FDR}(\hat{t}) \le \alpha Q_0/Q$ and $\lim_{n,Q \to \infty}\P\{ \mathrm{FDP}(\hat{t})\le \alpha Q_0/Q + \varepsilon \} = 1$ for any $\varepsilon > 0$. 
\end{theorem}

The conditions on ${\rm Corr}\{\eta^{(q)},\eta^{(q')}\}$ and $\mathcal{A}_q(\gamma)$ are commonly used in the literature, which are   applied to bound the variance of $\sum_{q\in\mathcal{H}_0}I\{V_{n,L}^{(q)}\geq t\}$ appeared in the numerator of ${\rm FDP}(t)$. See, for example, \cite{Liu2013}, \cite{ChangShaoZhou2016} and \cite{XiaLi2020}. Specifically, the condition $\max_{1\leq q\neq q'\leq Q}|{\rm Corr}\{\eta^{(q)},\eta^{(q')}\}|\leq r$ for some constant $r\in(0,1)$ imposes a constraint on the magnitudes of the correlations between different $\eta^{(q)}$ and $\eta^{(q')}$. The condition  $\max_{q \in [Q]}|\mathcal{A}_q(\gamma)| = o(Q^{\nu})$ controls the number of pairs $\{\eta^{(q)},\eta^{(q')}\}$ those are highly correlated. Recall $B$ and $L$ are, respectively, the length of each subgroup and the pre-determined integer involved in the $L$-statistic $W_{n,L}^{(q)}$ defined as  \eqref{eq:WnLq}. When $B$ and $L$ are selected as fixed constants, Theorem \ref{tm:FDR} holds if $Q^{12}(\log K_{\max})^7=o(n)$ and $Q^{12}\log(nK_{\max})\max\{(\log K_{\max})^4,(\log n)^4\}=o(n)$. Under the mild condition $K_{\max}\lesssim n^{\kappa}$ for some sufficiently large constant $\kappa>0$ in practice, the restrictions can be simplified as $Q^{12}(\log n)^7=o(n)$, which requires $Q$, the number of multiple tests, to satisfy $Q=o\{n^{1/12}(\log n)^{-7/12}\}$.

\subsection{A Distributed Algorithm}\label{sc:parallel}

Generating multivariate normal random vectors is a key step in the implementation of our proposed tests. For example, in the global test \eqref{eq:globaltest}, we need to generate $\hat{\xi}_1,\ldots, \hat{\xi}_N$ independently from $\mathcal{N}(0,\hat{R}_n)$ to determine the critical value $\hat{\mathrm{cv}}_{L,\alpha}$, where $\hat{R}_n$ is a $d \times d$ matrix with $d=\sum_{q=1}^Q|\mathcal{K}_q|$.  The standard approach for generating an $s$-dimensional random vector $\zeta\sim\mathcal{N}(0,\Xi)$ includes three steps: (i) perform the Cholesky decomposition for the $s\times s$ matrix $\Xi=A^{\T}A$; (ii) generate $s$ independent $\mathcal{N}(0,1)$ random variables $z=(z_1,\ldots,z_s)^{\T}$; (iii) perform the transformation $\zeta=A^{\T}z$. Computationally this is an $s^3$-hard problem. In practice, $Q$ and $|\mathcal{K}_q|$'s can be large, which causes the generation of a $d$-dimensional random vector followed $\mathcal{N}(0,\hat{R}_n)$ requires very high computing cost. To overcome this challenge and further enhance the computational efficiency of our proposed methods, we propose in this section a distributed algorithm that can avoid generating high-dimensional normal distributed random vectors directly.

We use the global test \eqref{eq:globaltest} as the example to illustrate our idea. We first divide the full dataset into $K$ blocks with $n_k$ observations in the $k$-th block $\mathcal{I}_k$. Given $B\geq 5$, we can obtain $M_k=n_k-B+1$ moving subgroups in the $k$-th block $\mathcal{I}_k$ and calculate $U_{m,q,\cS_1,\cS_2}^{(k),{\rm dist}}$ in the same manner as $U_{m,q,\mathcal{S}_1,\mathcal{S}_2}$ appeared in \eqref{eq:ourestmethod} but with replacing $\{X_m,\ldots,X_{m+B-1}\}$ by the $m$-th subgroup in the $k$-th block $\mathcal{I}_k$ for $m \in [M_k]$, $(\cS_1,\cS_2) \in \mathcal{K}_q$ and $q \in [Q]$. For given $k\in[K]$ and $m\in [M_k]$, we stack $\{U_{m,q,\cS_1,\cS_2}^{(k),{\rm dist}}\}_{q\in[Q],(\cS_1,\cS_2)\in\mathcal{K}_q}$ into a $d$-dimensional vector $U_m^{(k),{\rm dist}}=(U_{m,1}^{(k),{\rm dist}},\ldots,U_{m,d}^{(k),{\rm dist}})^{\T}$. Write $\bar{U}^{(k),{\rm dist}}=(\bar{U}_1^{(k),{\rm dist}},\ldots,\bar{U}_d^{(k),{\rm dist}})^{\T}=M_k^{-1}\sum_{m=1}^{M_k}U_m^{(k),{\rm dist}}$. Then we define the distributed analogue of the global test statistic $W_{n,L}$ specified in \eqref{eq:TS_MaxS} as follows:
\begin{align}\label{eq:diststat}
W_{n,L}^{{\rm dist}}=\max_{1\le j_1 < \cdots < j_L \le d }\sum_{s=1}^L \frac{n_{\rm dist}^{-1/2}\sum_{k=1}^{K}M_k\bar{U}_{j_s}^{(k),{\rm dist}}}{[n_{\rm dist}^{-1}\sum_{k=1}^KM_k\{\hat{\sigma}_{j_s}^{(k),{\rm dist}}\}^2]^{1/2}}\,,
\end{align}
where $n_{{\rm dist}}=\sum_{k=1}^{K}M_{k}$ and $\hat{\sigma}_j^{(k),{\rm dist}}$ is the estimate for the standard deviation of $\sqrt{M_k}\bar{U}_j^{(k),{\rm dist}}$. Define $\hat{\Sigma}^{(k),{\rm dist}}$ in the same manner as $\hat{\Sigma}_n$ given in \eqref{eq:estsigma} but with replacing $\{U_m\}_{m=1}^M$ by $\{U_m^{(k),{\rm dist}}\}_{m=1}^{M_k}$. We can select $\{\hat{\sigma}_j^{(k),{\rm dist}}\}^2$ as the associated element in the main diagonal of $\hat{\Sigma}^{(k),{\rm dist}}$.

Identical to Proposition \ref{tm:GA}, we can also show that the null-distribution of $W_{n,L}^{{\rm dist}}$ can be approximated by the distribution of $f_L(\check{\xi})$ with $f_L(\cdot)$ defined in Section \ref{sc:test global} and $\check{\xi}\sim\mathcal{N}(0,\check{R}_n)$ for some $d\times d$ matrix $\check{R}_n$. To avoid generating $d$-dimensional normal distributed random vector $\check{\xi}$, we consider to generate $d$-dimensional $\mathring{\xi}$ as follows:
\begin{align}\label{eq:mathringxi}
\mathring{\xi}=(\tilde{D}_n^{\rm dist})^{-1/2}
\sum_{k=1}^K\varepsilon_kn_{\rm dist}^{-1/2}M_k\{\bar{U}^{(k),{\rm dist}}-\bar{U}^{{\rm dist}}\}
\end{align}
with $\tilde{D}_n^{{\rm dist}}={\rm diag}[n_{\rm dist}^{-1}\sum_{k=1}^KM_k\{\tilde{\sigma}_1^{(k),{\rm dist}}\}^2,\ldots,n_{\rm dist}^{-1}\sum_{k=1}^KM_k\{\tilde{\sigma}_d^{(k),{\rm dist}}\}^2]$, where $\{\varepsilon_k\}_{k=1}^K$ are independent Rademacher random variables with $\P(\varepsilon_{k} = 1) = \P(\varepsilon_{k} = -1) = 1/2$, $\bar{U}^{{\rm dist}}=(\bar{U}_1^{\rm dist},\ldots,\bar{U}_d^{\rm dist})^{\T}=n_{\rm dist}^{-1}\sum_{k=1}^KM_k\bar{U}^{(k),{\rm dist}}$, and $\tilde{\sigma}_j^{(k),{\rm dist}}$ is the estimate for the standard deviation of $\sqrt{M_k}\{\bar{U}_j^{(k),{\rm dist}}-\bar{U}_j^{\rm dist}\}$ that can be determined in the same manner as $\hat{\sigma}_j^{(k),{\rm dist}}$ in \eqref{eq:diststat} but with replacing the data $\{U_m^{(k),{\rm dist}}\}_{m=1}^{M_k}$ by $\{\varepsilon_kU_m^{(k),{\rm dist}}-\varepsilon_k\bar{U}^{{\rm dist}}\}_{m=1}^{M_k}$. In comparison to generating $\check{\xi}\sim\mathcal{N}(0,\check{R}_n)$, the newly method \eqref{eq:mathringxi} only requires to generate $K$ independent Rademacher random  variables $\varepsilon_1,\ldots,\varepsilon_K$ which can largely reduce the computational cost. Let $R_{n}^{{\rm dist}}=(D_{n}^{{\rm dist}})^{-1/2}\Sigma_{n}^{{\rm dist}}(D_{n}^{{\rm dist}})^{-1/2}$ with $\Sigma_{n}^{{\rm dist}}=\mathrm{Cov}(n_{{\rm dist}}^{1/2}\bar{U}^{{\rm dist}})$ and $D_{n}^{{\rm dist}}=\diag(\Sigma_{n}^{{\rm dist}})=\diag\{(\sigma_{1}^{{\rm dist}})^2,\ldots,(\sigma_{d}^{{\rm dist}})^2\}$. Proposition \ref{pn:rade} provides the theoretical guarantee for such method.

\begin{proposition}\label{pn:rade}
	Write $\sigma_{j}^{(k),{\rm dist}}={\rm Var}^{1/2}(M_k^{-1/2}\sum_{m=1}^{M_k}U_{m,j}^{(k),{\rm dist}})$ for any $k\in [K]$ and $j\in [d]$. Let $\min_{k\in[K]}n_k\asymp \max_{k\in[K]}n_k$. Assume there exists a universal constant $c_{2}>0$ such that $\min_{v\in\mathcal{F}}v^{\T}R_{n}^{{\rm dist}}v\ge c_{2}$ and $\min_{k\in [K]}\min_{j\in [d]}\sigma_{j}^{(k),{\rm dist}}\ge c_{2}$. Under the global null hypothesis $H_{0}$ given in \eqref{eq:globaltest}, we have $\sup_{z\in\mathbb{R}}|\mathbb{P}(W_{n,L}^{{\rm dist}}>z)-\mathbb{P}\{f_L(\mathring{\xi})>z\,|\,X_1,\ldots,X_n\}|=o_{\p}(1)$ provided that $B^5L^8K\log (dBK)(\log d)^4=o(n)$, $B^3L^{13}\{\log (dK)\}^{3}(\log d)^7=o(K)$ and $B^4K^{3/2}L^{-2}\{\log (dBK)\}^3(\log d)^{-2}=o(n)$.
\end{proposition}

When we select $B$ as a fixed constant, Proposition \ref{pn:rade} holds provided that $L^8K\log (dK)(\log d)^4=o(n)$, $L^{13}\{\log (dK)\}^{3}(\log d)^7=o(K)$ and $K^{3/2}L^{-2}\{\log (dK)\}^3(\log d)^{-2}=o(n)$, which requires the pre-determined integer $L$ should satisfy $L\ll \min\{n^{1/8},K^{1/13}\}$. Furthermore, if $L$ is also selected as a fixed constant, Proposition \ref{pn:rade} holds provided that $\{\log(dK)\}^3(\log d)^7=o(K)$ and $K^{3/2}\{\log(dK)\}^3(\log d)^{-2}=o(n)$. Additionally, with restricting $K$, the number of dataset blocks involved in the distributed algorithm, satisfying $K=o(d)$, Proposition \ref{pn:rade} is valid for any $K$ such that $(\log d)^{10}\ll K\ll \min\{d, n^{2/3}(\log d)^{-2/3}\}$, which allows $\log d=o(n^{1/16})$ that $d=\sum_{q=1}^Q|\mathcal{K}_q|$ growing exponentially with the sample size $n$.
To obtain a guideline for the choices of $K$ in practice, we have performed simulations to assess the statistical properties of the proposed tests for different $K$. The simulation settings in Section \ref{sc:simulation} have similar sample size and image resolutions as the real applications. For example, $V = 200^2$ and $V = 500^2$ correspond to the number of voxels in the 3mm and 2mm MNI standard brain templates respectively. For the data set with a large sample size, e.g., $n\geq 900$, the proposed testing method performs very well for a wide range of $K$ $(K = 20, 30, 40)$. For the data set with a moderate or small sample size, i.e. $n \leq 600$,  a relatively small $K$ (e.g. $K = 20$) may lead to a slightly better power compared to other choices while the empirical size still can be controlled at around 0.05. Please refer to Section~\ref{sc:simulation} for more details.  Besides considering the statistical properties,  we may choose $K$ taking into account the available computational resources  including the number of processors,  memory storage and computing speed. A smaller $K$ may require less processors but more local memory per processors leading to a longer computational time. A larger $K$ may need more processors but less local memory which can reduce the total computational time.

Let $\hat{F}^{{\rm dist}}(t)=\mathbb{P}\{f_L(\mathring{\xi})\leq t\,|\,X_1,\ldots,X_n\}$ for $\mathring{\xi}$ defined as \eqref{eq:mathringxi}. We can reject the  global null hypothesis $H_0$ in \eqref{eq:globaltest} at the significance level $\alpha\in(0,1)$ if $W_{n,L}^{{\rm dist}}>\hat{{\rm cv}}_{L,\alpha}^{\rm dist}$ with
$
\hat{{\rm cv}}_{L,\alpha}^{\rm dist}=\inf\{t\in\mathbb{R}:\hat{F}^{{\rm dist}}(t)\geq1-\alpha\}
$.
In practice, for some sufficiently large $N$, letting $\mathring{\xi}_1,\ldots,\mathring{\xi}_N$ be independent random vectors followed \eqref{eq:mathringxi}, we can approximate $\hat{{\rm cv}}_{L,\alpha}^{\rm dist}$ by the $\lfloor N\alpha \rfloor$th largest value of $f_L(\mathring{\xi}_{1}), \ldots, f_L(\mathring{\xi}_{N})$.

In the multiple test, the distributed analogue of the p-value $\mathrm{pv}_{L}^{(q)}$ specified below \eqref{eq:gamult} can be similarly defined. For given $q\in [Q]$, $k\in[K]$ and $m\in [M_k]$, we stack $\{U_{m,q,\cS_1,\cS_2}^{(k),{\rm dist}}\}_{(\cS_1,\cS_2)\in\mathcal{K}_q}$ into a $|\mathcal{K}_q|$-dimensional vector $U_{m,(q)}^{(k),{\rm dist}}=(U_{m,q,1}^{(k),{\rm dist}},\ldots,U_{m,q,|\mathcal{K}_q|}^{(k),{\rm dist}})^{\T}$. For the $q$-th hypothesis $H_{0,q}$, the test statistic $W_{n,L}^{(q),{\rm dist}}$ can be defined in the same manner as $W_{n,L}^{{\rm dist}}$ in \eqref{eq:diststat} but with replacing $\{U_{m}^{(k),{\rm dist}}\}_{m=1}^{M_k}$ by $\{U_{m,(q)}^{(k),{\rm dist}}\}_{m=1}^{M_k}$. Write $\bar{U}_{(q)}^{(k),{\rm dist}}=M_k^{-1}\sum_{m=1}^{M_k}U_{m,(q)}^{(k),{\rm dist}}$ and $\bar{U}_{(q)}^{{\rm dist}}=n_{\rm dist}^{-1}\sum_{k=1}^KM_k\bar{U}_{(q)}^{(k),{\rm dist}}$. Then the $|\mathcal{K}_q|$-dimensional $\mathring{\xi}^{(q)}$ can be generated in the same manner as $\mathring{\xi}$ in \eqref{eq:mathringxi} but with replacing $\bar{U}^{(k),{\rm dist}}$ and $\bar{U}^{{\rm dist}}$ by $\bar{U}_{(q)}^{(k),{\rm dist}}$ and $\bar{U}_{(q)}^{{\rm dist}}$, respectively. The distributed analogue of the p-value of $H_{0,q}$ is then defined as $\mathrm{pv}_{L}^{(q),{\rm dist}}=\P\{ f_{L}(\mathring{\xi}^{(q)}) \ge W_{n,L}^{(q),{\rm dist}} \,|\, X_1,\ldots,X_n \}$. For implementation, we generate independent random vectors $\mathring{\xi}_1^{(q)},\ldots,\mathring{\xi}_N^{(q)}$ for some sufficiently large $N$, and $\mathrm{pv}_{L}^{(q),{\rm dist}}$ can be approximated by $\hat{\mathrm{pv}}_{L}^{(q),{\rm dist}} = N^{-1}\sum_{j=1}^{N}I\{f_L(\mathring{\xi}_{j}^{(q)}) \ge W_{n,L}^{(q),{\rm dist}} \}$.

\section{Simulations} \label{sc:simulation}

In this section, we perform simulations to evaluate the finite-sample performance of the proposed tests. We consider the following data generating process for the multimodal imaging data. Assume there are $J$ different imaging modalities for each subject. Let $\mathcal{R}=\{1,\ldots,V\}$ denote the set of indices of the two-dimensional (2D) image, which is divided as $G$ predefined brain regions $\mathcal{R}_1, \ldots, \mathcal{R}_G$ by the $k$-means clustering method, and $v \in \mathcal{R}$ denote the index of the voxels. The coordinates of the voxel $v$ are denoted by $h_{v}\in \mathbb{R}^2$, and the coordinates of the center in the region $\mathcal{R}_g$ are denoted by $\bar{h}_g$. For each $i\in[n]$, let $Y_{j}^{(i)}(v)$ be the image signal of type $j$ at voxel $v$ for the subject $i$.  Assume
$Y^{(i)}_{j}(v) = b^{(i)}_{j}(v) + \sum_{j'=1}^J d_{j,j'}(v) \alpha^{(i)}_{j,j'}(v)
+ \epsilon^{(i)}_{j}(v)$,
where $
b^{(i)}_{j}(v) = \sum_{g=1}^G \beta^{(i)}_{j,g} I(v \in \mathcal{R}_g)$ and $d_{j,j'}(v) =\sum_{g=1}^G \delta_{j,j',g} I(v \in \mathcal{R}_g)$ with $\delta_{j, j',g}=  \delta_{j', j,g}\geq 0$ for $j<j'$. Here $d_{j,j'}(v)$'s are regression coefficients, and $b_j^{(i)}(v)$ and $\alpha_{j,j'}^{(i)}(v)$'s are some random variables. For each $i\in[n]$, write $\beta^{(i)} =\{\beta_{1}^{(i),\T}, \ldots, \beta_{J}^{(i),\T}\}^{\T}$ with $\beta_{j}^{(i)} =\{\beta_{j,1}^{(i)},\ldots, \beta_{j,G}^{(i)}\}^{\T}$. We generate $\beta^{(i)}$ independently from the normal distribution $\mathcal{N}(0, E)$, where $E = (E^{j,j'})_{j,j' \in [J]}$ with $E^{j,j'} = (e^{j,j'}_{g,g'})_{g,g' \in [G]}$. For any $i\in[n]$ and $(j,j')$ such that $j\leq j'$, $\{\alpha^{(i)}_{j,j'}(v)\}_{v \in \mathcal{R}}$ is independently generated from a Gaussian process $\mathcal{GP}(0,\kappa_{\alpha})$ with the covariance function  $\kappa_{\alpha}(v,v') =
  \sum_{g=1}^G\exp\{-0.001(|h_v-\bar{h}_g|_2^2+|h_{v'}-\bar{h}_g|_2^2)-10|h_v-h_{v'}|_2^2\}I(v, v'\in \mathcal{R}_g)$. For $j > j'$, let $\alpha_{j,j'}^{(i)}(v) = \alpha_{j',j}^{(i)}(v)$. We will consider several scenarios for the regression coefficients $d_{j,j'}(v)$'s and the covariance matrix $E$ of $\beta^{(i)}$ to minic the null hypothesis and the alternative hypothesis, which will be specified later. Given the regression coefficients $d_{j,j'}(v)$'s and the generated $b_{j}^{(i)}(v)$'s and $\alpha_{j,j'}^{(i)}(v)$'s, $\epsilon_{j}^{(i)}(v)$'s are independently generated from $\mathcal{N}(0,\sigma^2)$, where $\sigma^2$ is chosen such that the R-squared is 0.95.



In our simulations, we set $J=3$, $G\in\{16,25\}$ and $V\in\{100^2, 200^2, 500^2\}$. We consider the sample size $n\in\{300, 600, 900\}$, and divide the observations equally to $K=30$ blocks and put each block on a processor to implement the distributed computing procedures presented in Section \ref{sc:parallel} with $N=5000$. All simulation results are based on $1000$ replications. We have also tried $K=20$ and $40$ in the simulation which lead to similar results as $K=30$. We only present the results based on $K=30$. The results based on $K=20$ and $40$ can be found in the supplementary material.

We consider both the global and multiple tests for the null hypotheses (a)--(c) given in \eqref{eq:region_test}--\eqref{eq:region_pair_cross_test}, respectively. 
 To evaluate the empirical sizes, we set $\delta_{j,j',g} = 0$ and $e_{g,g'}^{j,j'}=I(j=j')I(g=g')$ for any $j,j'\in [J]$ and $g,g' \in [G]$. To evaluate the empirical powers, we consider the following three scenarios (M1)--(M3), which corresponds to the testing problems (a)--(c), respectively.
\begin{enumerate}
    \item[(M1)] Let $e_{g,g'}^{j,j'}=I(j=j')I(g=g')$ and $\delta_{j,j',g} =\{I(j=j')+6I(|j-j'|=1)\}I(1\leq g\leq 4)$.
        \item[(M2)]  Let $e_{g,g'}^{j,j'}=I(j=j')\{I(g=g')+0.4I(|g-g'|=1)-0.6I(|g-g'|=1)I(j=2)\}$ and $\delta_{j,j',g}=0$.
    \item[(M3)] Let $e_{g,g'}^{j,j'}=I(j=j')I(g=g')+I(g'=g+1)\{0.4I(j=1,j'=2)-0.4I(j=1,j'=3)+0.2I(j=2,j'=3)\}$ for $j\leq j'$, and $e_{g,g'}^{j,j'}=e^{j',j}_{g',g}$ for $j>j'$, and also let $\delta_{j,j',g} = 0$.
\end{enumerate}

For the global tests, we set the significance level $\alpha = 5\%$. When $G=16$, $d=48$, $360$ and $768$ in the global tests of (a)--(c), respectively. When $G=25$, $d=75$, $900$ and $1875$ in the global tests of (a)--(c), respectively. See the definition of $d$ above \eqref{eq:TS_MaxS}. The empirical sizes and powers of the proposed global tests are reported in Table \ref{tb:global_mb_pair_sc_size} with $L=1,$ 3 and 5. For $L=1$, the proposed tests have good size control around $5\%$ and the power of the proposed tests increases quickly to 1 in all the three testing problems as the sample size $n$  grows from $300$ to $900$.  In comparison to the proposed tests with $L=1$, we find that the proposed tests with $L>1$ still have good size control around $5\%$ when $d$ is small (testing problem (a)) and they are a little bit conservative  when $d$ is large (testing problems (b) and (c)), but the associated powers grow as $L$ increases.

The empirical FDRs and powers of the proposed multiple tests are reported in Table \ref{tb:efdr_mb_pair}. Due to the number of modalities $J=3$ in our setting, we know $|\mathcal{K}_q|=3$ in \eqref{eq:general_test} for all $q \in [Q]$ in the multiple tests for (a)--(b) and we can only consider the proposed test statistics with $L\leq 3$. For the testing problems (a)--(c), $(Q,Q_0)=(16,12), (120,105)$ and $(136,121)$ when $G=16$, and $(Q,Q_0)=(25,21), (300,276)$ and $(325,301)$ when $G=25$, respectively. Given the significance level $\alpha=5\%$, we can obtain $\hat{t}_r$ defined as \eqref{eq:hatT} in the $r$-th simulation replication. For each $q\in[Q]$, denote by $V_{n,L,r}^{(q)}$ the normal quantile transformation of the p-value for $H_{0,q}$ in the $r$-th simulation replication. See its definition below \eqref{eq:gamult}. We define the empirical power of the proposed multiple test as
 $\frac{1}{1000} \sum_{r=1}^{1000}\frac{1}{Q-Q_0}\sum_{q\notin \mathcal{H}_{0}}I\{ V_{n,L,r}^{(q)} > \hat{t}_{r} \}$. Theorem \ref{tm:FDR} implies that the FDR should be controlled by $\alpha Q_{0}/Q$, which equals to $3.75\%$, $4.38\%$ and $4.45\%$ when $G=16$, and $4.20\%$, $4.60\%$ and $4.63\%$ when $G=25$ in the testing problems (a)--(c), respectively.
 Table \ref{tb:efdr_mb_pair} shows that the proposed multiple tests have good performances for FDR control, and the empirical powers improve considerably when the sample size $n$ increases from $300$ to $900$. 

As we mentioned in Section \ref{sc:introduction}, some existing works can be also used to derive the p-value of each $H_{0,g}^{(a)}$ in testing problem (a). Based on the p-values obtained by these methods, we can consider the associated multiple testing procedure in the same manner as ours given in Section \ref{sc:test multiple}. Here we consider three methods: (i) the $k$-variate HSIC-based test (dHSIC) in \cite{pfister2018}, (ii) the generalized distance covariance based test (GdCov) in \cite{jin2018}, and (iii) the test based on the high order distance covariance in \cite{ChakrabortyZhang2019}. The dHSIC test and GdCov test are implemented by calling the R-functions {\tt dhsic.test} and {\tt mdm\_test}, respectively, in the R-packages {\tt dHSIC} and {\tt EDMeasure}. There are the three test statistics in \cite{ChakrabortyZhang2019}. Here, we choose their bias-corrected Studentized test statistic (JdCov) with $c=2$ as suggested by the authors. The code for JdCov test is  available in the supplementary material of \cite{ChakrabortyZhang2019}. 
Table \ref{tb:efdr_mb_pair} also includes the empirical FDRs and powers of the multiple testing procedures based on these three competitors in testing problem (a), which shows that our proposed method has similar performance as these three methods. However, these three competitors cannot be applied to address the multiple tests in the testing problems (b) and (c). Since the implementation of the JdCov test relies on a bootstrap procedure which requires high computing cost, the results of JdCov in Table \ref{tb:efdr_mb_pair} are  only based on 100 replications, and its associated results for $V=500^2$ are omitted. Different from the linear dependency considered here among $Y_j^{(i)}(v)$'s, we have also considered a simulation setting with nonlinear dependency imposed among $Y_j^{(i)}$'s for the testing problem (a). See Section A.2 in the supplementary material.  It shows that (i) the proposed method significantly outperforms the three competitors; (ii) the proposed method is robust to $V$; and (iii) the three competitors suffer obvious power
loss with the increase of $V$.

\begin{table}[t]
\renewcommand\arraystretch{1.2}
\centering
\begin{threeparttable}
\tiny
\caption{Empirical sizes and powers of the proposed global tests with $K=30$ for the testing problems (a)--(c) given in \eqref{eq:region_test}--\eqref{eq:region_pair_cross_test}, respectively. All numbers reported below are multiplied by 100.}\label{tb:global_mb_pair_sc_size}
\begin{tabular}{cccc|ccc|ccc|ccc}
\hline\hline
&&&&\multicolumn{3}{c|}{Problem (a)}&\multicolumn{3}{c|}{Problem (b)}&\multicolumn{3}{c}{Problem (c)}\tabularnewline
&$G$&$n$&$V$&$L=1$&$L=3$&$L=5$&$L=1$&$L=3$&$L=5$&$L=1$&$L=3$&$L=5$\tabularnewline\hline
       & 16 & 300 & $100^2$ & 5.70   & 5.70   & 3.90   & 6.00   & 3.50   & 2.10   & 5.10   & 4.00   & 2.90   \\
       &    &     & $200^2$ & 6.30   & 4.80   & 3.80   & 6.50   & 4.40   & 2.20   & 5.20   & 4.00   & 3.10   \\
       &    &     & $500^2$ & 5.20   & 4.50   & 4.40   & 6.20   & 3.50   & 1.70   & 7.60   & 5.70   & 3.70   \\
       &    & 600 & $100^2$ & 4.40   & 4.40   & 4.00   & 3.90   & 3.30   & 2.30   & 3.90   & 2.90   & 2.30   \\
       &    &     & $200^2$ & 4.70   & 5.10   & 4.40   & 4.30   & 3.00   & 1.90   & 4.80   & 2.50   & 1.80   \\
       &    &     & $500^2$ & 5.20   & 4.90   & 3.70   & 5.10   & 3.70   & 2.30   & 5.10   & 2.90   & 2.50   \\
       &    & 900 & $100^2$ & 4.90   & 4.40   & 4.60   & 2.90   & 2.90   & 2.60   & 4.70   & 2.80   & 2.30   \\
       &    &     & $200^2$ & 6.00   & 5.30   & 5.20   & 4.10   & 2.30   & 2.00   & 2.90   & 2.40   & 2.00   \\
Global &    &     & $500^2$ & 5.20   & 5.10   & 4.90   & 4.40   & 2.30   & 2.50   & 3.70   & 2.30   & 1.90   \\ \cline{2-13}
Sizes  & 25 & 300 & $100^2$ & 6.20   & 4.80   & 3.20   & 5.40   & 3.10   & 1.50   & 6.20   & 3.50   & 2.10   \\
       &    &     & $200^2$ & 7.40   & 5.50   & 4.40   & 7.20   & 3.50   & 1.60   & 7.50   & 4.30   & 3.10   \\
       &    &     & $500^2$ & 7.50   & 6.10   & 5.60   & 6.20   & 3.00   & 1.90   & 7.20   & 5.10   & 3.20   \\
       &    & 600 & $100^2$ & 7.20   & 5.30   & 5.80   & 4.70   & 3.10   & 2.20   & 3.90   & 3.00   & 2.50   \\
       &    &     & $200^2$ & 4.90   & 4.70   & 3.80   & 4.60   & 2.80   & 1.50   & 3.60   & 2.40   & 1.20   \\
       &    &     & $500^2$ & 4.50   & 4.70   & 4.40   & 4.70   & 2.80   & 2.30   & 4.90   & 2.80   & 1.00   \\
       &    & 900 & $100^2$ & 2.90   & 3.80   & 4.20   & 3.40   & 2.50   & 1.70   & 3.30   & 1.90   & 1.00   \\
       &    &     & $200^2$ & 3.90   & 3.90   & 3.70   & 3.00   & 2.10   & 1.50   & 3.70   & 2.20   & 1.30   \\
       &    &     & $500^2$ & 4.50   & 4.40   & 3.80   & 6.10   & 2.70   & 2.10   & 2.40   & 1.80   & 1.50   \\ \hline
       & 16 & 300 & $100^2$ & 100.00 & 100.00 & 100.00 & 93.30  & 99.40  & 99.90  & 88.90  & 98.20  & 99.30  \\
       &    &     & $200^2$ & 100.00 & 100.00 & 100.00 & 92.70  & 99.10  & 99.40  & 86.90  & 96.30  & 97.70  \\
       &    &     & $500^2$ & 100.00 & 100.00 & 100.00 & 94.40  & 99.00  & 99.60  & 86.20  & 97.70  & 99.40  \\
       &    & 600 & $100^2$ & 100.00 & 100.00 & 100.00 & 100.00 & 100.00 & 100.00 & 100.00 & 100.00 & 100.00 \\
       &    &     & $200^2$ & 100.00 & 100.00 & 100.00 & 100.00 & 100.00 & 100.00 & 100.00 & 100.00 & 100.00 \\
       &    &     & $500^2$ & 100.00 & 100.00 & 100.00 & 100.00 & 100.00 & 100.00 & 100.00 & 100.00 & 100.00 \\
       &    & 900 & $100^2$ & 100.00 & 100.00 & 100.00 & 100.00 & 100.00 & 100.00 & 100.00 & 100.00 & 100.00 \\
       &    &     & $200^2$ & 100.00 & 100.00 & 100.00 & 100.00 & 100.00 & 100.00 & 100.00 & 100.00 & 100.00 \\
Global &    &     & $500^2$ & 100.00 & 100.00 & 100.00 & 100.00 & 100.00 & 100.00 & 100.00 & 100.00 & 100.00 \\ \cline{2-13}
Powers & 25 & 300 & $100^2$ & 100.00 & 100.00 & 100.00 & 93.50  & 99.50  & 99.90  & 85.00  & 98.40  & 99.50  \\
       &    &     & $200^2$ & 100.00 & 100.00 & 100.00 & 94.60  & 99.40  & 99.90  & 85.50  & 98.40  & 99.50  \\
       &    &     & $500^2$ & 100.00 & 100.00 & 100.00 & 93.00  & 99.20  & 99.70  & 82.50  & 94.80  & 96.10  \\
       &    & 600 & $100^2$ & 100.00 & 100.00 & 100.00 & 100.00 & 100.00 & 100.00 & 100.00 & 100.00 & 100.00 \\
       &    &     & $200^2$ & 100.00 & 100.00 & 100.00 & 100.00 & 100.00 & 100.00 & 100.00 & 100.00 & 100.00 \\
       &    &     & $500^2$ & 100.00 & 100.00 & 100.00 & 100.00 & 100.00 & 100.00 & 100.00 & 100.00 & 100.00 \\
       &    & 900 & $100^2$ & 100.00 & 100.00 & 100.00 & 100.00 & 100.00 & 100.00 & 100.00 & 100.00 & 100.00 \\
       &    &     & $200^2$ & 100.00 & 100.00 & 100.00 & 100.00 & 100.00 & 100.00 & 100.00 & 100.00 & 100.00 \\
       &    &     & $500^2$ & 100.00 & 100.00 & 100.00 & 100.00 & 100.00 & 100.00 & 100.00 & 100.00 & 100.00 \\
\hline\hline
\end{tabular}

\end{threeparttable}
\end{table}

\begin{table}[t]
\renewcommand\arraystretch{1.2}
\centering
\begin{threeparttable}
\tiny
\caption{Empirical FDRs and powers of the proposed multiple tests with $K=30$, and three competitors (JdCov, dHSIC, GdCov) for the testing problems (a)--(c) given in \eqref{eq:region_test}--\eqref{eq:region_pair_cross_test}, respectively. All numbers reported below are multiplied by 100.}\label{tb:efdr_mb_pair}
\begin{tabular}{cccc|ccccc|cc|cc}
\hline\hline
   &     &    & & \multicolumn{5}{c|}{Problem  (a)}                & \multicolumn{2}{c|}{Problem (b)}     & \multicolumn{2}{c}{Problem  (c)}      \\
& $G$  & $n$   & $V$   & $L=1$ & $L=3$ & JdCov     & dHSIC  & GdCov    & $L=1$  &  $L=3$ &  $L=1$ & $L=3$\\
 \hline
&16 & 300 & $100^2$ & 1.96          & 2.19         & 2.80   & 1.83   & 3.83   & 4.09          & 2.43         & 4.11          & 2.32         \\
&   &     & $200^2$ & 2.44          & 2.09         & 3.53   & 1.89   & 3.26   & 3.75          & 2.55         & 4.50          & 2.57         \\
&   &     & $500^2$ & 1.93          & 2.46         & NA     & 1.81   & 3.22   & 3.10          & 2.52         & 4.08          & 1.46         \\
&   & 600 & $100^2$ & 2.37          & 2.42         & 2.73   & 1.76   & 2.23   & 3.75          & 5.11         & 2.99          & 3.32         \\
&   &     & $200^2$ & 2.77          & 2.67         & 2.33   & 2.04   & 1.75   & 3.44          & 4.98         & 2.12          & 2.73         \\
&   &     & $500^2$ & 2.31          & 2.57         & NA     & 2.09   & 1.96   & 3.58          & 4.95         & 2.35          & 2.98         \\
&   & 900 & $100^2$ & 2.27          & 2.86         & 1.80   & 1.60   & 1.95   & 4.00          & 5.47         & 3.16          & 3.39         \\
&   &     & $200^2$ & 2.43          & 2.67         & 2.13   & 2.04   & 2.88   & 3.88          & 5.28         & 3.21          & 3.33         \\
Empirical &  & & $500^2$ & 2.37     & 2.52         & NA     & 1.93   & 2.49   & 3.83          & 5.39         & 3.59          & 3.59         \\ \cline{2-13}
FDRs & 25 & 300 & $100^2$ & 2.67    & 2.31         & 2.20   & 1.54   & 3.88   & 4.31          & 3.88         & 4.29          & 1.79         \\
&   &     & $200^2$ & 2.55          & 2.49         & 1.60   & 1.58   & 3.38   & 3.83          & 4.08         & 4.55          & 2.00         \\
&   &     & $500^2$ & 2.40          & 2.65         & NA     & 1.53   & 3.81   & 3.91          & 4.02         & 5.70          & 3.38         \\
&   & 600 & $100^2$ & 2.69          & 2.79         & 1.60   & 1.56   & 3.97   & 4.47          & 5.74         & 3.72          & 3.92         \\
&   &     & $200^2$ & 2.59          & 2.71         & 3.80   & 1.69   & 3.74   & 4.00          & 5.53         & 3.65          & 3.83         \\
&   &     & $500^2$ & 2.37          & 2.62         & NA     & 1.74   & 3.89   & 4.13          & 5.79         & 3.69          & 4.14         \\
&   & 900 & $100^2$ & 2.72          & 3.30         & 2.13   & 1.80   & 3.43   & 4.40          & 6.47         & 3.13          & 4.16         \\
&   &     & $200^2$ & 3.07          & 3.09         & 3.13   & 1.83   & 4.18   & 4.33          & 6.53         & 3.25          & 4.22         \\
&   &     & $500^2$ & 2.45          & 2.83         & NA     & 1.87   & 3.77   & 4.29          & 6.37         & 3.38          & 4.30         \\ \hline
& 16 & 300 & $100^2$& 99.78         & 98.80        & 100.00 & 100.00 & 100.00 & 18.57         & 41.80        & 14.63         & 22.96        \\
&   &     & $200^2$ & 99.88         & 99.70        & 100.00 & 100.00 & 100.00 & 19.10         & 42.93        & 14.07         & 22.03        \\
&   &     & $500^2$ & 99.93         & 99.73        & NA     & 100.00 & 100.00 & 18.87         & 42.48        & 14.12         & 23.73        \\
&   & 600 & $100^2$ & 100.00        & 100.00       & 100.00 & 100.00 & 100.00 & 87.21         & 96.95        & 72.99         & 91.04        \\
&   &     & $200^2$ & 100.00        & 100.00       & 100.00 & 100.00 & 100.00 & 87.00         & 96.82        & 68.83         & 88.44        \\
&   &     & $500^2$ & 100.00        & 100.00       & NA     & 100.00 & 100.00 & 86.83         & 96.99        & 68.66         & 89.43        \\
&   & 900 & $100^2$ & 100.00        & 100.00       & 100.00 & 100.00 & 100.00 & 99.44         & 99.89        & 97.63         & 99.20        \\
&   &     & $200^2$ & 100.00        & 100.00       & 100.00 & 100.00 & 100.00 & 99.36         & 99.85        & 97.63         & 99.25        \\
Empirical&  & & $500^2$ & 100.00    & 100.00       & NA     & 100.00 & 100.00 & 99.45         & 99.91        & 97.90         & 99.36        \\ \cline{2-13}
Powers& 25 & 300 & $100^2$ & 99.95  & 99.78        & 100.00 & 100.00 & 100.00 & 12.67         & 41.88        & 8.93          & 15.28        \\
&   &     & $200^2$ & 99.93         & 99.58        & 100.00 & 100.00 & 100.00 & 12.80         & 40.80        & 8.92          & 15.30        \\
&   &     & $500^2$ & 99.90         & 99.50        & NA     & 100.00 & 100.00 & 12.38         & 40.90        & 9.03          & 15.23        \\
&   & 600 & $100^2$ & 100.00        & 100.00       & 100.00 & 100.00 & 100.00 & 86.85         & 96.00        & 78.22         & 91.86        \\
&   &     & $200^2$ & 100.00        & 100.00       & 100.00 & 100.00 & 100.00 & 87.39         & 95.77        & 78.43         & 91.77        \\
&   &     & $500^2$ & 100.00        & 100.00       & NA     & 100.00 & 100.00 & 87.20         & 95.76        & 78.99         & 90.88        \\
&   & 900 & $100^2$ & 100.00        & 100.00       & 100.00 & 100.00 & 100.00 & 99.00         & 99.79        & 98.34         & 99.42        \\
&   &     & $200^2$ & 100.00        & 100.00       & 100.00 & 100.00 & 100.00 & 99.07         & 99.76        & 98.51         & 99.47        \\
&   &     & $500^2$ & 100.00        & 100.00       & NA     & 100.00 & 100.00 & 99.05         & 99.81        & 97.85         & 99.32        \\
\hline\hline
\end{tabular}
\end{threeparttable}
\end{table}

\section{Analysis of HCP Data}\label{sec:realdata}

We performed the analysis of the motivating HCP task fMRI data using the proposed testing procedures. All the data were collected from the HCP-1200 release \citep{van2013wu}. The HCP minimally preprocessed pipeline~\citep{glasser2013minimal} was adopted to preprocess the fMRI time series data. In particular, the pipeline includes motion correction, distortion correction, brain-boundary based linear registration, nonlinear registration to the standard MNI 2mm space and the intensity normalization. Our analysis focused on  922 subjects with three fMRI task domains: working memory (n-back), language processing (story and math) and social cognition. For the details of the tasks, please refer to~\cite{BARCH2013}.  To generate the subject-specific contrast maps from the fMRI time series data, the fixed-effects models were fitted to estimate the average brain activation over time using FSL~\citep{jenkinson2012fsl}.  For some tasks, multiple contrast maps were created beyond the standard experimental versus control condition. We analyzed five contrast maps with two (2back-0back, task-rest) for the working memory task, one (story-math) for language processing task and two (mental, random) for the social cognition task. Our analysis indicates that the language processing, working memory and social cognition brain activities are strongly dependent with statistical significance in most parts of the inferior frontal gyrus. For more details, please refer to Section B in the supplementary material.

We considered three testing problems (a)--(c), defined as \eqref{eq:region_test}--\eqref{eq:region_pair_cross_test} respectively, in the analysis of the HCP fMRI contrast maps.
We performed the distributed computing procedure by splitting the first 900 subjects equally on $30$ processors and putting the rest 22 observations on the 31st processor, and calculated the critical values by the method given in Section \ref{sc:parallel}.

\noindent\textbf{Testing independence among modalities over regions:}~
In the testing problem (a), we tested the independence of the brain activation from the three fMRI tasks, i.e., working memory (2back-0back), language processing (story-math) and social cognition (mental) over the 90 AAL regions.
 The global test, in which $d=270$, rejected the null hypothesis of the three types of brain activation being mutually independent in all 90 regions with a p-value $<0.0001$.  Of note, we took different values for $L$: $1, 5, 10$ and $15$; and the associated p-values were all less than $0.0001$. The multiple tests were conducted for the same null hypothesis in each of the 90 AAL regions  with the FDR controlled at $0.001$ and $L=1$. Each test was rejected if the corresponding p-value $\le 1.35 \times 10^{-3}$.
 We identified 30 out of 90 AAL regions where the three types of fMRI task  activities are significantly dependent. This result provides some new insights about relationships between brain functions and human behaviors.




\begin{figure}[htbp]
\centering
\subfloat[{Working memory}]{
\label{fig2_a}
\begin{minipage}[c]{.33\linewidth}
\centering
\includegraphics[width=3.5cm]{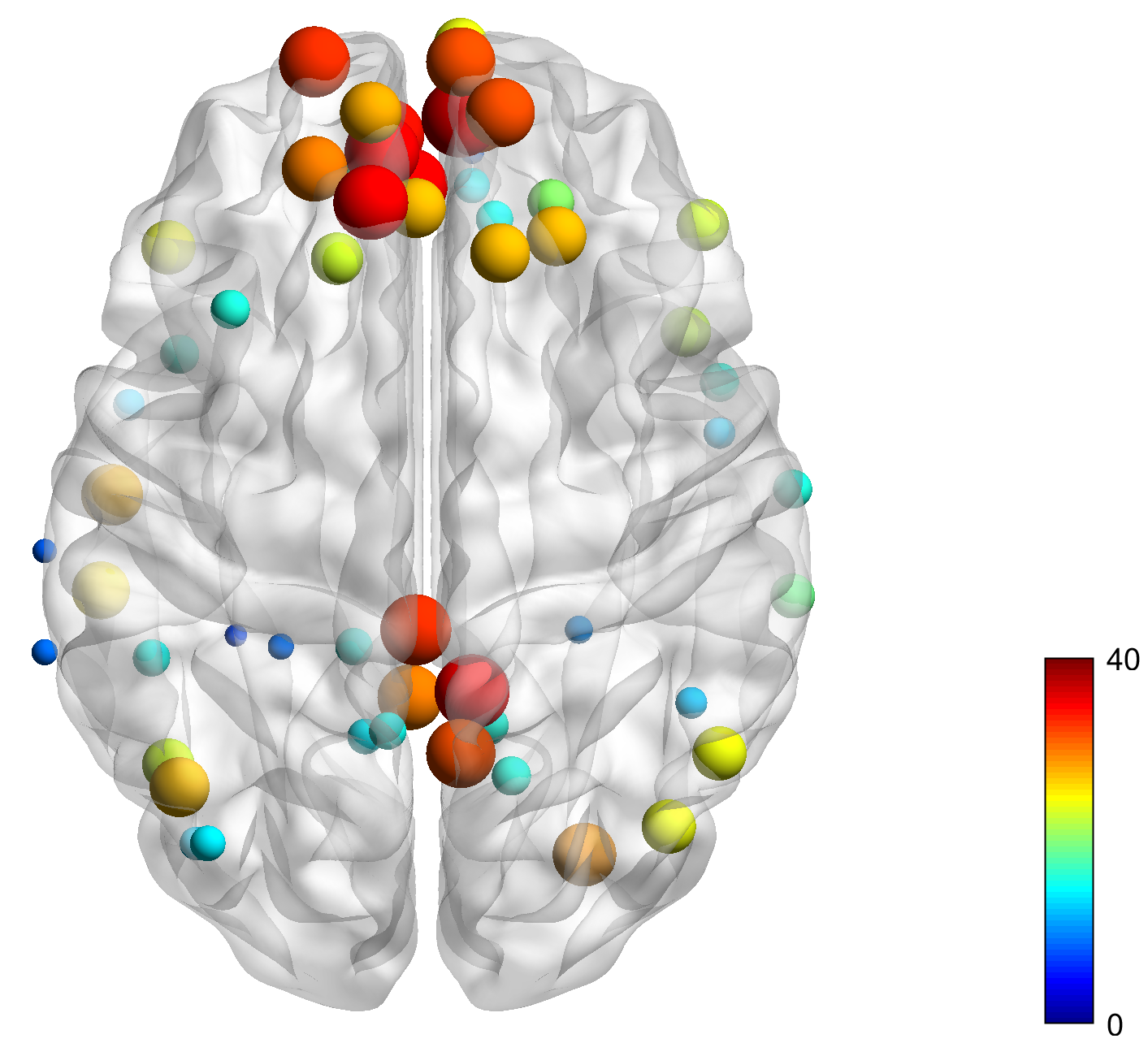}
\end{minipage}
}
\subfloat[Language processing]{
\label{fig2_b}
\begin{minipage}[c]{.33\linewidth}
\centering
\includegraphics[width=3.5cm]{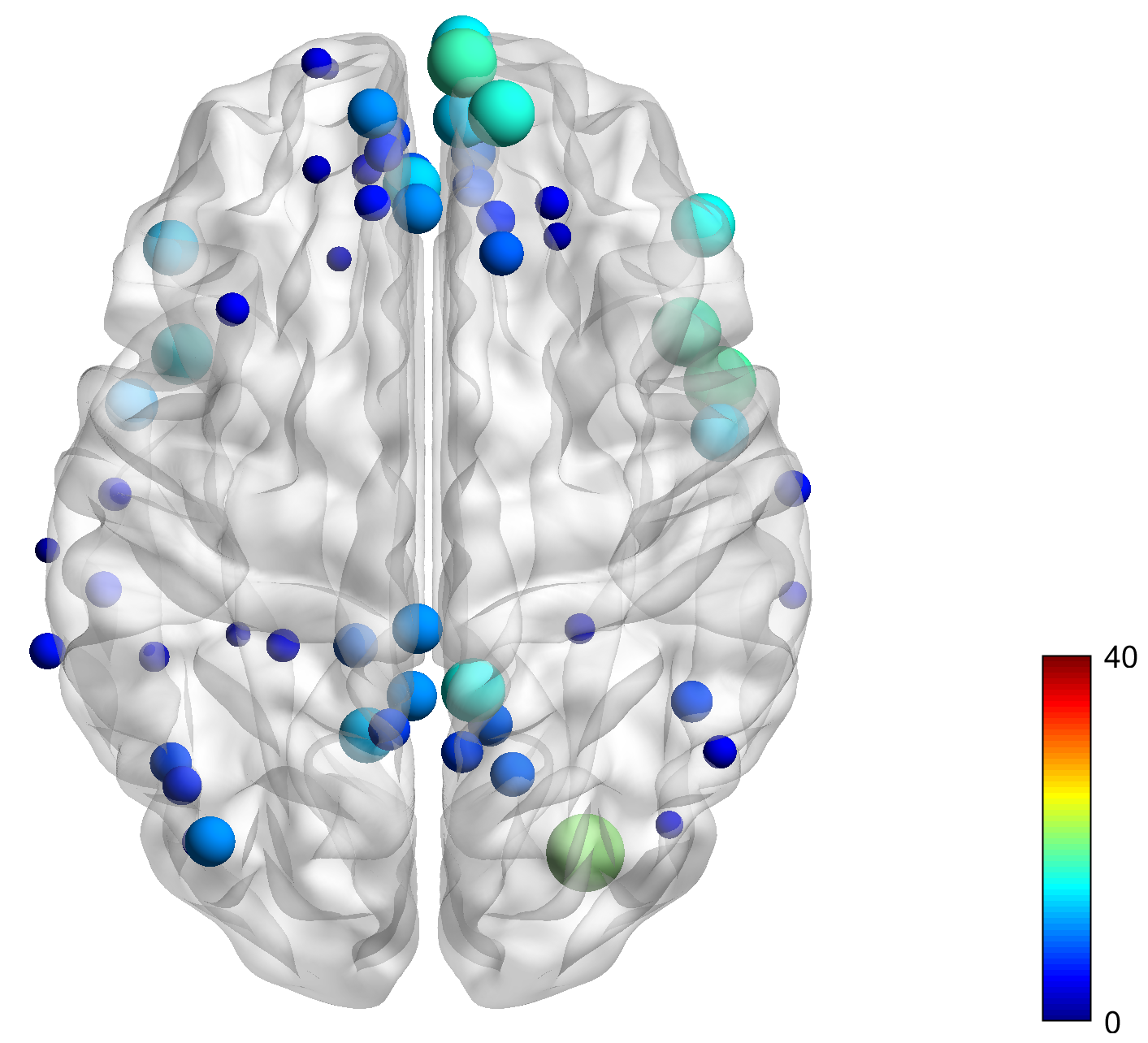}
\end{minipage}
}
\subfloat[Social cognition]{
\label{fig2_c}
\begin{minipage}[c]{.33\linewidth}
\centering
\includegraphics[width=3.5cm]{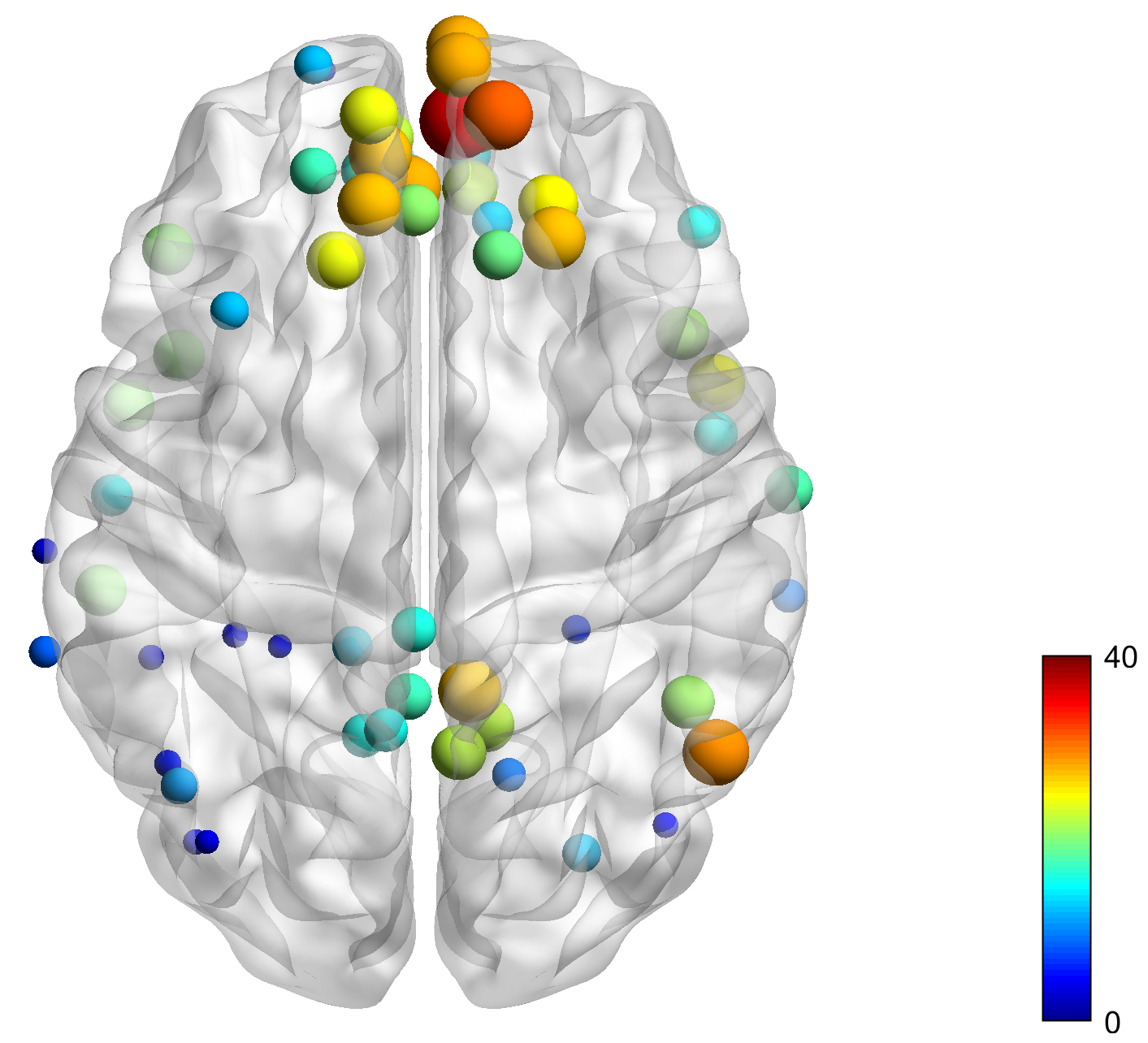}
\end{minipage}
}
\caption{\label{fig2} The degrees of nodes (regions) in the default mode network. The sizes of the nodes are proportional to the degrees of nodes.} 
\end{figure}

\noindent\textbf{Testing independence between regions within modalities:}~In the testing problem (b), we mainly concentrated on assessing the independence of fMRI task activity between regions in the default mode network~\citep[DMN]{raichle2015brain}. In our analysis, we adopted the parcellation of brain by~\cite{power2011functional} consisting of 264 regions among which the DMN has 58 regions.
The DMN was developed for the brain activity in experimental participants at wakeful rest, and the DMN also can be active in social cognition and working memory tasks~\citep{spreng2012fallacy}. In this problem, we tested the independence of brain regions in the three fMRI tasks, i.e., working memory (2back-0back), language processing (story-math) and social cognition (mental) over the 58 DMN regions. The global test, in which $d=4959$, rejected the null hypothesis of independence of brain activities between all the region pairs in DMN for the three fMRI task contrasts with a p-value $<0.0001$ for $L=1,5,10$ and 15. Furthermore, we are interested in identifying brain regions which are statistically significantly correlated in the working memory task ($j=1$), the language processing task ($j=2$) and the social cognition task ($j=3$), respectively. For each $j \in \{1,2,3\}$, we consider the multiple tests for null hypotheses $H_{0,j,g,g'}: Y_{j,g}$ and $Y_{j,g'}$ are independent. Due to $g<g'$, there are ${58\choose 2} = 1653$ null hypotheses. The multiple tests were conducted with the FDR controlled at 0.001 and $L=1$.
Each test was rejected if the corresponding p-value $\le 5.91 \times 10^{-5}$ for the working memory task (2back-0back), the language processing (story-math) task and the social cognition (mental) task, respectively. There are 627, 227 and 494 region pairs identified for strong dependence of brain activity in the working memory task (2back-0back), language processing (story-math) task and the social cognition (mental) task, respectively. Each testing result can be represented as an undirected graph, where the regions in the DMN denote the nodes and each significant region pair has an edge.  Given the graphs are very dense, in Figure \ref{fig2}, we present the degree for each region,  i.e., the number of other regions in the DMN that are significantly dependent on this region.  The brain activation dependence among the default mode network on the working memory task is much larger than that on the language processing task. The medial prefrontal cortex has  relatively strong co-activation interactions with other regions on all three tasks.  On the social cognition task, the left angular gyrus has weak co-activation interactions with other regions compared to the posterior cingulate cortex and the precuneus.



\begin{figure}[htbp]
\centering
\subfloat[Coronal view]{
\label{fig3_b}
\begin{minipage}[c]{.3\linewidth}
\centering
\includegraphics[height=2cm]{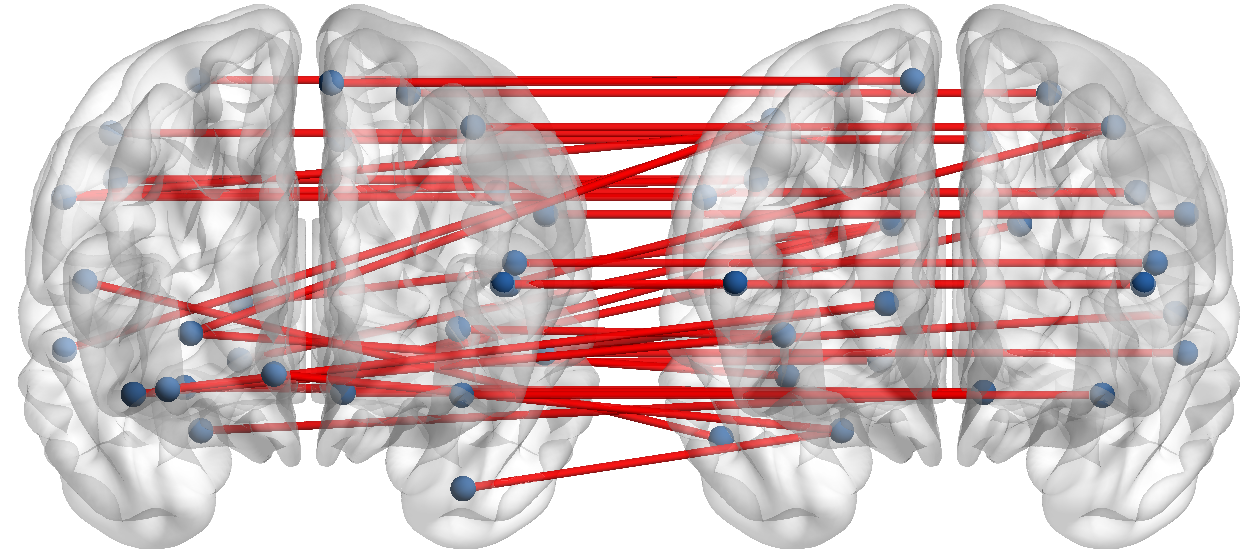}
\end{minipage}
}
\subfloat[Sagittal view]{
\label{fig3_c}
\begin{minipage}[c]{.3\linewidth}
\includegraphics[height=2cm]{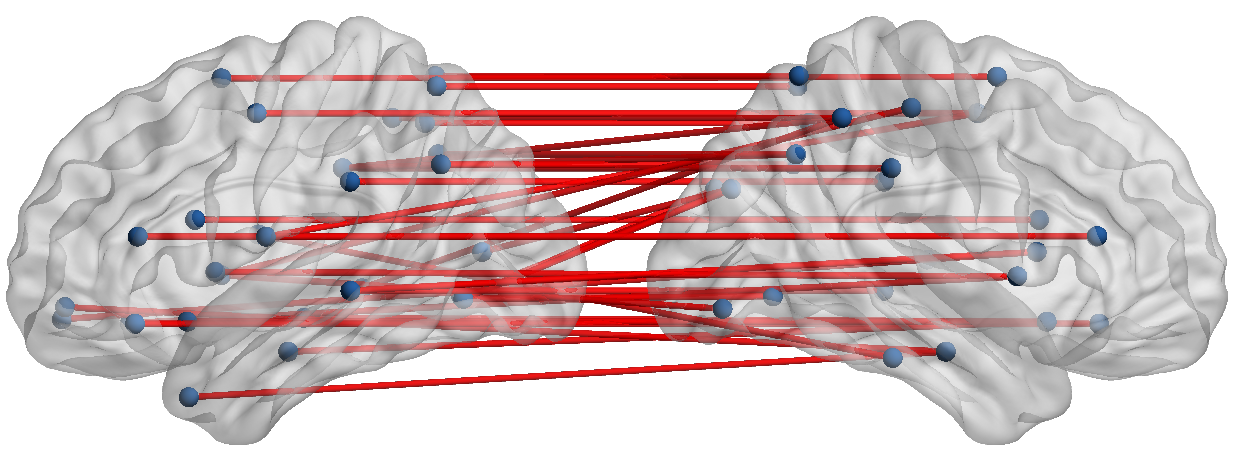}
\end{minipage}
}
\subfloat[Axial view]{
\label{fig3_a}
\begin{minipage}[l]{.3\linewidth}
\centering
\includegraphics[width=3.5cm]{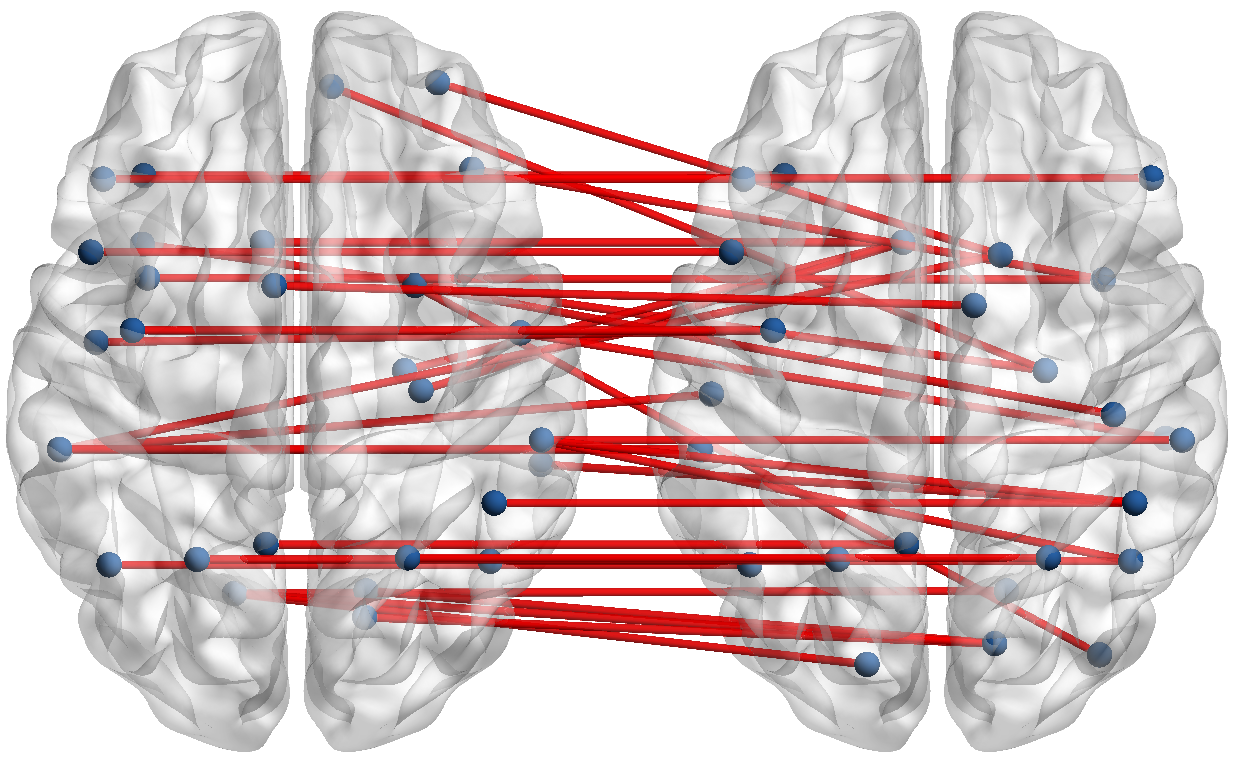}
\end{minipage}
}
\caption{Bipartite graph for the testing result on the independence between regions across the working memory (2back-0back) task (left) and the language processing (story-math) task (right). The edge between two nodes represents significantly correlation and the corresponding p-values are all $<0.0001$.}\label{fig3}
\end{figure}


\noindent\textbf{Testing independence between regions across different modalities:}~
In the testing problem (c), we consider to test the independence of brain activity between the 90 AAL regions across the working memory (2back-0back) task and the language processing (story-math) task. The global test, in which $d=8100$, rejected the null hypothesis of the two task fMRI activations being independent for across all 90 regions under the significance level $\alpha=0.05$ with p-values $< 0.0001$ for $L = 1, 5, 10$ and 15. To identify brain regions where the working memory activity ($j=1$) and the language processing activity ($j'=2$) are statistically significantly correlated with each other, we consider the multiple tests for null hypotheses $H_{0,g,g'}$: $Y_{1,g}$ and $Y_{2,g'}$ are independent. Due to $g,g'=1,\ldots,90$, there are $90\times90=8100$ null hypotheses. The multiple tests were conducted with the FDR controlled at 0.001 and $L=1$. Each test was rejected if the corresponding p-value $\le 1.10 \times 10^{-5}$. Our testing results indicate that the two types of task fMRI activations are significantly dependent with each other in 14 brain regions (p-value $<0.0001$)  and across 23 different region pairs (p-value $<0.0001$). The results imply some new findings for the multi task fMRI activity.  For example, the inferior frontal gyrus is well known for language comprehension and production~\citep{tyler2011left}, but also related to working memory~\citep{nissim2017frontal}. We found that working memory activity is strongly associated with the language processing activity in the left opercular part and the left triangular part of the inferior frontal gyrus. In addition,  working memory activity in the left triangular part of the inferior frontal gyrus is strongly associated with the language processing activity in the right triangular part of Inferior frontal gyrus.   Figure \ref{fig3} shows the whole bipartite graph indicating the dependence in brain activation between the working memory task (2back-0back) and the language processing (story-math) task.

\section{Discussion}\label{sec:discussion}
In this paper, we propose a novel  method for testing independence structure among the components of the ultra high-dimensional random vector. For a general form of null hypothesis, we develop a global testing procedure and a multiple testing procedure with FDR control using Gaussian approximations.   To substantially improve the computational efficiency, we develop a distributed algorithm that is scalable to ultra high-dimensional data. We perform rigorous theoretical analyses and extensive simulations to demonstrate that the tests have good properties for a wide range of dimensions and sample sizes. In the analysis of task fMRI data in the HCP study,  we apply the proposed method to test three types of complex dependences for five different contrast maps at group level. We obtain significant evidences for some scientifically meaningful findings,  providing insights on associations among brain activities in different regions under different cognitive tasks. We highlight several advantages of our methods. Compared to existing tests that may be sensitive to the dependence structure of data, our proposed test is particularly powerful for analysis of multimodal imaging data which appear to be noisy, spatially dependent, high-dimensional and complexly associated with anatomical structures and modalities.  The scalable computation algorithm is another key contribution of the proposed test compared to other methods. To the best of our knowledge, we are among the first to propose the computational feasible independence tests with theoretical guarantees on high-dimensional voxel-level data ($\approx$200,000 voxels) in neuroimaging studies.  In addition to imaging applications, the proposed test can be considered as  a general tool for testing independence for many other statistical problems.  For example, a multi-omics study may collect SNP genotypes,  gene expressions, DNA copy number alternations, DNA methylation changes, along with other genetic information. It is also of great interest to test the independence among multimodal genomic data.  Our proposed method provides a useful tool to address this problem.


{
\spacingset{1}
 \bibliography{ref5}
 }













\newpage

\setcounter{page}{1}
\rhead{\bfseries\thepage}
\lhead{\bfseries SUPPLEMENTARY MATERIAL}

\setcounter{page}{1}
\begin{center}
{\bf\Large Supplementary Material for ``Statistical inferences for complex dependence of multimodal imaging data'' }
\end{center}

\appendix

\counterwithout{equation}{section}

\renewcommand{\theequation}{S.\arabic{equation}}

\setcounter{equation}{0}

 In this supplementary material, we present some additional simulation results, additional details in analysis of HCP data and proofs of the main results in the paper. Throughout the supplementary material, we use $C$ to denote a generic positive constant that may be different in different uses.

\section{Additional Simulation Results}\label{se:simu}

\subsection{Simulation Results for $K=20$ and $40$}
We first present the empirical sizes and powers of the global tests in Table~\ref{tb:global_K20} ($K=20$) and Table~\ref{tb:global_K40} ($K=40$) and report  the empirical FDRs and powers of the multiple tests in Table~\ref{tb:efdr_K20} ($K = 20$) and Table~\ref{tb:efdr_K40} ($K = 40$). 

\setcounter{table}{0}
\renewcommand{\thetable}{S\arabic{table}}

\begin{table}[htbp]
\renewcommand\arraystretch{1.2}
\centering
\begin{threeparttable}
\tiny
\caption{Empirical sizes and powers of the proposed global tests with $K=20$ for the testing problems (a)--(c) given in (1.1)--(1.3), respectively. All numbers reported below are multiplied by 100.}\label{tb:global_K20}
\begin{tabular}{cccc|ccc|ccc|ccc}
\hline\hline
&&&&\multicolumn{3}{c|}{Problem (a)}&\multicolumn{3}{c|}{Problem (b)}&\multicolumn{3}{c}{Problem (c)}\tabularnewline
&$G$&$n$&$V$&$L=1$&$L=3$&$L=5$&$L=1$&$L=3$&$L=5$&$L=1$&$L=3$&$L=5$\tabularnewline\hline
       & 16 & 300 & $100^2$ & 6.30   & 5.30   & 5.20   & 6.90   & 4.60   & 4.20   & 7.80   & 5.50   & 3.30   \\
       &    &     & $200^2$ & 7.50   & 7.10   & 5.60   & 10.20  & 5.90   & 3.80   & 8.30   & 6.10   & 4.90   \\
       &    &     & $500^2$ & 8.60   & 7.50   & 7.30   & 11.10  & 7.10   & 4.40   & 9.40   & 6.00   & 3.40   \\
       &    & 600 & $100^2$ & 6.30   & 6.20   & 5.40   & 6.80   & 4.50   & 3.40   & 2.90   & 2.30   & 1.60   \\
       &    &     & $200^2$ & 5.80   & 6.30   & 6.30   & 4.30   & 2.30   & 1.70   & 4.00   & 2.60   & 1.90   \\
       &    &     & $500^2$ & 6.50   & 5.90   & 5.40   & 5.60   & 3.90   & 2.40   & 4.30   & 2.50   & 1.50   \\
       &    & 900 & $100^2$ & 5.90   & 4.80   & 5.50   & 3.30   & 2.40   & 2.00   & 4.60   & 2.60   & 2.50   \\
       &    &     & $200^2$ & 3.70   & 4.30   & 4.40   & 3.10   & 2.10   & 2.00   & 3.70   & 2.60   & 2.00   \\
Global &    &     & $500^2$ & 3.70   & 4.10   & 3.70   & 4.20   & 3.00   & 2.40   & 3.90   & 1.90   & 1.50   \\ \cline{2-13}
Sizes  & 25 & 300 & $100^2$ & 9.60   & 8.50   & 7.10   & 8.20   & 5.00   & 3.00   & 9.00   & 5.80   & 3.70   \\
       &    &     & $200^2$ & 8.90   & 7.30   & 5.70   & 7.90   & 5.30   & 3.70   & 11.60  & 7.60   & 4.90   \\
       &    &     & $500^2$ & 9.40   & 6.20   & 5.40   & 8.70   & 6.70   & 4.70   & 11.90  & 10.60  & 8.80   \\
       &    & 600 & $100^2$ & 4.90   & 4.40   & 3.80   & 4.80   & 4.00   & 2.10   & 6.90   & 3.20   & 2.20   \\
       &    &     & $200^2$ & 5.50   & 5.60   & 5.90   & 4.40   & 2.60   & 1.90   & 5.90   & 3.50   & 2.20   \\
       &    &     & $500^2$ & 6.00   & 5.00   & 4.60   & 3.70   & 3.20   & 2.20   & 5.10   & 2.70   & 1.90   \\
       &    & 900 & $100^2$ & 5.50   & 4.60   & 4.70   & 3.20   & 2.20   & 1.70   & 4.60   & 2.80   & 2.20   \\
       &    &     & $200^2$ & 3.90   & 3.40   & 3.30   & 3.90   & 2.10   & 2.30   & 3.90   & 2.80   & 2.20   \\
       &    &     & $500^2$ & 5.10   & 5.10   & 4.60   & 4.20   & 2.20   & 1.80   & 3.80   & 3.20   & 2.50   \\ \hline
       & 16 & 300 & $100^2$ & 100.00 & 100.00 & 100.00 & 98.50  & 99.80  & 99.90  & 94.00  & 99.10  & 99.70 \\
       &    &     & $200^2$ & 100.00 & 100.00 & 100.00 & 98.40  & 100.00 & 100.00 & 93.70  & 99.10  & 99.60 \\
       &    &     & $500^2$ & 100.00 & 100.00 & 100.00 & 98.40  & 100.00 & 100.00 & 99.70  & 99.60  & 99.80  \\
       &    & 600 & $100^2$ & 100.00 & 100.00 & 100.00 & 100.00 & 100.00 & 100.00 & 100.00 & 100.00 & 100.00 \\
       &    &     & $200^2$ & 100.00 & 100.00 & 100.00 & 100.00 & 100.00 & 100.00 & 100.00 & 100.00 & 100.00 \\
       &    &     & $500^2$ & 100.00 & 100.00 & 100.00 & 100.00 & 100.00 & 100.00 & 100.00 & 100.00 & 100.00 \\
       &    & 900 & $100^2$ & 100.00 & 100.00 & 100.00 & 100.00 & 100.00 & 100.00 & 100.00 & 100.00 & 100.00 \\
       &    &     & $200^2$ & 100.00 & 100.00 & 100.00 & 100.00 & 100.00 & 100.00 & 100.00 & 100.00 & 100.00 \\
Global &    &     & $500^2$ & 100.00 & 100.00 & 100.00 & 100.00 & 100.00 & 100.00 & 100.00 & 100.00 & 100.00 \\ \cline{2-13}
Powers & 25 & 300 & $100^2$ & 100.00 & 100.00 & 100.00 & 99.10  & 100.00 & 100.00 & 99.10  & 99.80  & 99.90 \\
       &    &     & $200^2$ & 100.00 & 100.00 & 100.00 & 99.70  & 99.90  & 100.00 & 96.80  & 99.10  & 99.20  \\
       &    &     & $500^2$ & 100.00 & 100.00 & 100.00 & 99.20  & 99.90  & 100.00 & 97.20  & 99.20  & 99.30  \\
       &    & 600 & $100^2$ & 100.00 & 100.00 & 100.00 & 100.00 & 100.00 & 100.00 & 100.00 & 100.00 & 100.00 \\
       &    &     & $200^2$ & 100.00 & 100.00 & 100.00 & 100.00 & 100.00 & 100.00 & 100.00 & 100.00 & 100.00 \\
       &    &     & $500^2$ & 100.00 & 100.00 & 100.00 & 100.00 & 100.00 & 100.00 & 100.00 & 100.00 & 100.00 \\
       &    & 900 & $100^2$ & 100.00 & 100.00 & 100.00 & 100.00 & 100.00 & 100.00 & 100.00 & 100.00 & 100.00 \\
       &    &     & $200^2$ & 100.00 & 100.00 & 100.00 & 100.00 & 100.00 & 100.00 & 100.00 & 100.00 & 100.00 \\
       &    &     & $500^2$ & 100.00 & 100.00 & 100.00 & 100.00 & 100.00 & 100.00 & 100.00 & 100.00 & 100.00 \\
\hline\hline
\end{tabular}
\end{threeparttable}
\end{table}

\begin{table}[htbp]
\renewcommand\arraystretch{1.2}
\centering
\begin{threeparttable}
\tiny
\caption{Empirical FDRs and powers of the proposed multiple tests with $K=20$ for the testing problems (a)--(c) given in (1.1)--(1.3), respectively. All numbers reported below are multiplied by 100.}\label{tb:efdr_K20}
\begin{tabular}{ccc|cc|cc|cc|cc|cc|cc}
\hline\hline
     &     &         & \multicolumn{6}{c|}{Empirical FDRs} & \multicolumn{6}{c}{Empirical Powers} \\
     &     &         & \multicolumn{2}{c|}{Problem (a)}& \multicolumn{2}{c|}{Problem (b)}& \multicolumn{2}{c|}{Problem (c)}& \multicolumn{2}{c|}{Problem (a)}& \multicolumn{2}{c|}{Problem (b)}& \multicolumn{2}{c}{Problem (c)}   \\
$G$  & $n$ & $V$     & $L=1$         & $L=3$        & $L=1$         &  $L=3$       &  $L=1$        & $L=3$  & $L=1$         & $L=3$        & $L=1$         &  $L=3$       &  $L=1$        & $L=3$ \\ \hline
  16 & 300 & $100^2$ & 2.29          & 2.62         & 3.47          & 4.70         & 7.20          & 2.16   & 99.95         & 99.65        & 29.09         & 56.11        & 22.07         & 32.05 \\
     &     & $200^2$ & 3.06          & 2.92         & 3.54          & 4.27         & 7.58          & 2.18   & 99.95         & 99.78        & 28.65         & 55.48        & 22.07         & 32.20 \\
     &     & $500^2$ & 2.45          & 2.82         & 3.34          & 4.57         & 6.02          & 3.30   & 99.98         & 99.83        & 29.29         & 55.60        & 19.35         & 29.89 \\
     & 600 & $100^2$ & 3.29          & 3.50         & 4.39          & 6.63         & 4.58          & 5.06   & 100.00        & 100.00       & 91.01         & 97.89        & 81.35         & 92.26 \\
     &     & $200^2$ & 2.92          & 3.14         & 4.61          & 6.51         & 4.54          & 5.10   & 100.00        & 100.00       & 91.03         & 97.98        & 81.45         & 92.53 \\
     &     & $500^2$ & 2.90          & 3.49         & 4.83          & 6.62         & 3.17          & 4.15   & 100.00        & 100.00       & 91.64         & 97.95        & 79.49         & 92.04 \\
     & 900 & $100^2$ & 2.96          & 3.53         & 5.06          & 7.32         & 4.96          & 6.37   & 100.00        & 100.00       & 99.57         & 99.87        & 99.18         & 99.81 \\
     &     & $200^2$ & 3.38          & 3.72         & 5.28          & 7.19         & 4.97          & 6.25   & 100.00        & 100.00       & 99.60         & 99.89        & 99.20         & 99.82 \\
     &     & $500^2$ & 3.47          & 3.82         & 5.13          & 7.02         & 4.19          & 5.55   & 100.00        & 100.00       & 99.61         & 99.89        & 99.16         & 99.66 \\ \hline
  25 & 300 & $100^2$ & 3.47          & 3.70         & 5.24          & 7.80         & 6.13          & 3.25   & 100.00        & 100.00       & 23.91         & 59.96        & 16.07         & 26.85 \\
     &     & $200^2$ & 3.84          & 3.71         & 5.87          & 7.98         & 8.58          & 3.77   & 99.95         & 99.78        & 24.51         & 61.25        & 15.30         & 26.88 \\
     &     & $500^2$ & 4.01          & 3.71         & 5.69          & 8.17         & 8.98          & 3.25   & 100.00        & 99.80        & 24.52         & 60.31        & 15.40         & 26.26 \\
     & 600 & $100^2$ & 4.06          & 4.68         & 5.49          & 8.66         & 4.53          & 5.18   & 100.00        & 100.00       & 90.72         & 96.98        & 84.18         & 93.35 \\
     &     & $200^2$ & 3.77          & 3.67         & 5.59          & 8.52         & 4.33          & 5.13   & 100.00        & 100.00       & 90.20         & 97.20        & 84.79         & 93.25 \\
     &     & $500^2$ & 3.63          & 4.07         & 5.64          & 8.46         & 4.11          & 5.29   & 100.00        & 100.00       & 91.00         & 97.19        & 85.04         & 93.40 \\
     & 900 & $100^2$ & 3.55          & 4.26         & 5.31          & 8.79         & 4.56          & 6.09   & 100.00        & 100.00       & 99.27         & 99.86        & 98.57         & 99.52 \\
     &     & $200^2$ & 3.95          & 4.71         & 5.89          & 9.04         & 4.35          & 6.09   & 100.00        & 100.00       & 99.24         & 99.83        & 98.68         & 99.58 \\
     &     & $500^2$ & 3.68          & 4.05         & 5.49          & 9.25         & 4.31          & 5.52   & 100.00        & 100.00       & 99.28         & 99.86        & 98.50         & 99.55 \\
\hline\hline
\end{tabular}
\end{threeparttable}
\end{table}

\begin{table}[htbp]
\renewcommand\arraystretch{1.2}
\centering
\begin{threeparttable}
\tiny
\caption{Empirical sizes and powers of the proposed global tests with $K=40$ for the testing problems (a)--(c) given in (1.1)--(1.3), respectively. All numbers reported below are multiplied by 100.}\label{tb:global_K40}
\begin{tabular}{cccc|ccc|ccc|ccc}
\hline\hline
&&&&\multicolumn{3}{c|}{Problem (a)}&\multicolumn{3}{c|}{Problem (b)}&\multicolumn{3}{c}{Problem (c)}\tabularnewline
&$G$&$n$&$V$&$L=1$&$L=3$&$L=5$&$L=1$&$L=3$&$L=5$&$L=1$&$L=3$&$L=5$\tabularnewline\hline
       & 16 & 300 & $100^2$ & 6.30   & 5.70   & 5.50   & 3.40   & 2.80   & 2.30   & 4.80   & 1.90   & 1.00   \\
       &    &     & $200^2$ & 4.70   & 4.00   & 4.10   & 3.30   & 2.50   & 1.80   & 3.80   & 2.00   & 0.80   \\
       &    &     & $500^2$ & 4.80   & 3.90   & 3.50   & 3.30   & 1.50   & 1.40   & 4.20   & 2.20   & 1.60   \\
       &    & 600 & $100^2$ & 3.40   & 3.00   & 2.60   & 4.50   & 2.20   & 1.50   & 4.10   & 2.40   & 2.10   \\
       &    &     & $200^2$ & 4.80   & 4.80   & 4.40   & 4.60   & 2.40   & 1.70   & 3.80   & 2.40   & 1.80   \\
       &    &     & $500^2$ & 4.50   & 4.50   & 4.30   & 4.20   & 2.50   & 1.80   & 2.40   & 1.90   & 1.60   \\
       &    & 900 & $100^2$ & 4.60   & 4.00   & 3.40   & 2.90   & 1.40   & 0.90   & 3.80   & 2.10   & 1.60   \\
       &    &     & $200^2$ & 4.00   & 4.70   & 4.20   & 3.40   & 2.30   & 1.90   & 3.30   & 2.80   & 2.50   \\
Global &    &     & $500^2$ & 4.40   & 4.10   & 4.00   & 4.90   & 3.60   & 2.60   & 3.30   & 1.90   & 1.30   \\ \cline{2-13}
Sizes  & 25 & 300 & $100^2$ & 5.90   & 5.90   & 4.90   & 4.50   & 2.50   & 1.20   & 3.50   & 2.10   & 0.90   \\
       &    &     & $200^2$ & 5.60   & 5.20   & 4.80   & 4.80   & 2.60   & 2.10   & 4.80   & 2.00   & 0.80   \\
       &    &     & $500^2$ & 5.50   & 5.00   & 3.70   & 5.00   & 3.60   & 1.80   & 5.60   & 3.30   & 2.30   \\
       &    & 600 & $100^2$ & 3.70   & 3.40   & 3.50   & 4.40   & 2.40   & 1.70   & 4.70   & 3.20   & 2.20   \\
       &    &     & $200^2$ & 3.50   & 3.00   & 2.60   & 4.80   & 2.10   & 1.60   & 3.90   & 2.30   & 1.80   \\
       &    &     & $500^2$ & 6.10   & 4.60   & 4.00   & 4.00   & 2.40   & 1.70   & 5.10   & 3.10   & 2.10   \\
       &    & 900 & $100^2$ & 3.60   & 3.50   & 3.10   & 4.30   & 1.80   & 1.60   & 2.30   & 1.20   & 0.70   \\
       &    &     & $200^2$ & 5.20   & 4.00   & 3.50   & 2.60   & 1.90   & 2.00   & 3.10   & 2.30   & 1.50   \\
       &    &     & $500^2$ & 4.60   & 4.30   & 4.00   & 3.80   & 2.30   & 1.90   & 3.80   & 1.80   & 0.70   \\ \hline
       & 16 & 300 & $100^2$ & 100.00 & 100.00 & 100.00 & 83.90  & 97.70  & 99.10  & 75.10  & 91.70  & 94.90  \\
       &    &     & $200^2$ & 100.00 & 100.00 & 100.00 & 84.70  & 97.40  & 99.20  & 77.00  & 92.30  & 96.10  \\
       &    &     & $500^2$ & 100.00 & 100.00 & 100.00 & 83.30  & 97.30  & 99.30  & 74.60  & 91.70  & 94.10  \\
       &    & 600 & $100^2$ & 100.00 & 100.00 & 100.00 & 100.00 & 100.00 & 100.00 & 100.00 & 100.00 & 100.00 \\
       &    &     & $200^2$ & 100.00 & 100.00 & 100.00 & 100.00 & 100.00 & 100.00 & 100.00 & 100.00 & 100.00 \\
       &    &     & $500^2$ & 100.00 & 100.00 & 100.00 & 100.00 & 100.00 & 100.00 & 100.00 & 100.00 & 100.00 \\
       &    & 900 & $100^2$ & 100.00 & 100.00 & 100.00 & 100.00 & 100.00 & 100.00 & 100.00 & 100.00 & 100.00 \\
       &    &     & $200^2$ & 100.00 & 100.00 & 100.00 & 100.00 & 100.00 & 100.00 & 100.00 & 100.00 & 100.00 \\
Global &    &     & $500^2$ & 100.00 & 100.00 & 100.00 & 100.00 & 100.00 & 100.00 & 100.00 & 100.00 & 100.00 \\ \cline{2-13}
Powers & 25 & 300 & $100^2$ & 100.00 & 100.00 & 100.00 & 82.30  & 96.10  & 98.50  & 70.10  & 89.70  & 94.80  \\
       &    &     & $200^2$ & 100.00 & 100.00 & 100.00 & 83.10  & 97.20  & 99.00  & 71.10  & 88.70  & 92.00  \\
       &    &     & $500^2$ & 100.00 & 100.00 & 100.00 & 83.20  & 97.60  & 98.80  & 69.40  & 86.70  & 92.40  \\
       &    & 600 & $100^2$ & 100.00 & 100.00 & 100.00 & 100.00 & 100.00 & 100.00 & 100.00 & 100.00 & 100.00 \\
       &    &     & $200^2$ & 100.00 & 100.00 & 100.00 & 100.00 & 100.00 & 100.00 & 100.00 & 100.00 & 100.00 \\
       &    &     & $500^2$ & 100.00 & 100.00 & 100.00 & 100.00 & 100.00 & 100.00 & 100.00 & 100.00 & 100.00 \\
       &    & 900 & $100^2$ & 100.00 & 100.00 & 100.00 & 100.00 & 100.00 & 100.00 & 100.00 & 100.00 & 100.00 \\
       &    &     & $200^2$ & 100.00 & 100.00 & 100.00 & 100.00 & 100.00 & 100.00 & 100.00 & 100.00 & 100.00 \\
       &    &     & $500^2$ & 100.00 & 100.00 & 100.00 & 100.00 & 100.00 & 100.00 & 100.00 & 100.00 & 100.00 \\
\hline\hline
\end{tabular}
\end{threeparttable}
\end{table}

\begin{table}[htbp]
\renewcommand\arraystretch{1.2}
\centering
\begin{threeparttable}
\tiny
\caption{Empirical FDRs and powers of the proposed multiple tests with $K=40$ for the testing problems (a)--(c) given in (1.1)--(1.3), respectively. All numbers reported below are multiplied by 100.}\label{tb:efdr_K40}
\begin{tabular}{ccc|cc|cc|cc|cc|cc|cc}
\hline\hline
     &     &         & \multicolumn{6}{c|}{Empirical FDRs} & \multicolumn{6}{c}{Empirical Powers} \\
     &     &         & \multicolumn{2}{c|}{Problem (a)}& \multicolumn{2}{c|}{Problem (b)}& \multicolumn{2}{c|}{Problem (c)}& \multicolumn{2}{c|}{Problem (a)}& \multicolumn{2}{c|}{Problem (b)}& \multicolumn{2}{c}{Problem (c)}   \\
$G$  & $n$ & $V$     & $L=1$         & $L=3$        & $L=1$         &  $L=3$       &  $L=1$        & $L=3$  & $L=1$         & $L=3$        & $L=1$         &  $L=3$       &  $L=1$        & $L=3$ \\ \hline
  16 & 300 & $100^2$ & 2.21          & 2.50         & 3.14          & 2.92         & 5.58          & 2.47   & 99.28         & 98.18        & 16.02         & 39.53        & 12.00         & 20.41 \\
     &     & $200^2$ & 2.40          & 2.18         & 2.95          & 2.54         & 5.70          & 2.68   & 99.60         & 99.05        & 16.29         & 39.91        & 11.54         & 20.54 \\
     &     & $500^2$ & 2.39          & 2.66         & 3.34          & 2.95         & 6.07          & 2.63   & 99.73         & 99.18        & 16.25         & 39.70        & 11.61         & 20.21 \\
     & 600 & $100^2$ & 2.41          & 2.53         & 2.65          & 4.25         & 1.58          & 1.78   & 100.00        & 100.00       & 80.49         & 95.47        & 59.57         & 81.41 \\
     &     & $200^2$ & 1.89          & 2.17         & 2.56          & 4.34         & 1.54          & 1.98   & 100.00        & 100.00       & 81.11         & 95.41        & 59.49         & 80.79 \\
     &     & $500^2$ & 2.36          & 2.39         & 2.77          & 3.96         & 1.60          & 2.15   & 100.00        & 100.00       & 79.79         & 95.34        & 58.75         & 82.34 \\
     & 900 & $100^2$ & 2.13          & 2.42         & 3.41          & 4.63         & 2.69          & 2.87   & 100.00        & 100.00       & 99.25         & 99.85        & 97.83         & 99.18 \\
     &     & $200^2$ & 2.21          & 2.23         & 3.09          & 4.18         & 2.63          & 2.82   & 100.00        & 100.00       & 99.06         & 99.82        & 97.91         & 99.27 \\
     &     & $500^2$ & 1.95          & 1.86         & 3.35          & 4.42         & 3.16          & 3.09   & 100.00        & 100.00       & 99.09         & 99.80        & 97.98         & 99.51 \\ \hline
  25 & 300 & $100^2$ & 2.50          & 3.11         & 3.18          & 3.58         & 5.60          & 4.29   & 99.88         & 99.58        & 10.46         & 36.50        & 6.75          & 13.74 \\
     &     & $200^2$ & 2.87          & 2.89         & 5.34          & 3.87         & 6.67          & 4.26   & 99.63         & 99.08        & 10.54         & 36.57        & 6.46          & 13.16 \\
     &     & $500^2$ & 2.39          & 2.70         & 4.28          & 4.06         & 5.29          & 4.21   & 99.73         & 99.08        & 10.33         & 36.62        & 6.44          & 13.14 \\
     & 600 & $100^2$ & 2.17          & 2.09         & 3.45          & 4.72         & 3.60          & 3.21   & 100.00        & 100.00       & 82.83         & 94.35        & 71.10         & 87.27 \\
     &     & $200^2$ & 2.29          & 2.56         & 3.61          & 4.71         & 3.51          & 2.76   & 100.00        & 100.00       & 83.18         & 94.46        & 73.20         & 87.73 \\
     &     & $500^2$ & 2.12          & 2.31         & 3.43          & 4.86         & 3.18          & 3.36   & 100.00        & 100.00       & 83.15         & 94.56        & 73.08         & 87.67 \\
     & 900 & $100^2$ & 2.36          & 2.48         & 3.66          & 5.11         & 2.77          & 3.06   & 100.00        & 100.00       & 98.74         & 99.65        & 97.50         & 99.25 \\
     &     & $200^2$ & 2.82          & 2.84         & 3.49          & 5.14         & 2.96          & 3.16   & 100.00        & 100.00       & 98.66         & 99.75        & 97.59         & 99.25 \\
     &     & $500^2$ & 2.49          & 2.85         & 3.62          & 5.03         & 2.87          & 3.18   & 100.00        & 100.00       & 98.75         & 99.71        & 97.25         & 98.98 \\
\hline\hline
\end{tabular}
\end{threeparttable}
\end{table}


\subsection{A New Simulation Setting}
We consider a new simulation setting to compare the performance of the proposed multiple method with those of the JdCov, dHSIC and GdCov methods for the testing problem (a). We first partition the $V$ voxels of the whole image template into $G$ predefined brain regions $\mathcal{R}_1,\ldots,\mathcal{R}_G$ with equal number of voxels $p_r=V/G$ in each $\mathcal{R}_g$, and then divide each region $\mathcal{R}_g$ into $S$ subregions $\mathcal{R}_g^{(1)}, \ldots, \mathcal{R}_g^{(S)}$ by the $k$-means clustering method with $S=p_r/10$. Let $v \in [V]$ denote the index of the voxels. The coordinates of the voxel $v$ are denoted by $h_{v}\in \mathbb{R}^2$, and the coordinates of the center in  $\mathcal{R}_g^{(s)}$ are denoted by $\bar{h}_g^{(s)}$. Let $Y^{(i)}_{j}(v)$ be the image signal of type $j$ at voxel $v$ for the subject $i$. For each $i\in[n]$, $j\in[J]$ and $v\in[V]$, assume that $Y_{j}^{(i)}(v)$ is generated by
\begin{align}\label{eq:simu}
Y^{(i)}_{j}(v) &= \sum_{j'=1}^J d_{j,j'}(v) f_{j,j'}\{\alpha^{(i)}_{j,j'}(v)\}\, ,
\end{align}
where $d_{j,j'}(v) =\sum_{g=1}^G \delta_{j,j',g} I(v \in \mathcal{R}_g)$ with $\delta_{j, j',g}\geq 0$ for any $j,j'\in [J]$. For any $i\in[n]$ and $(j,j')$ such that $j \leq j'$, $\{\alpha_{j,j'}^{(i)}(v)\}_{v \in \mathcal{R}}$ is independently generated from a Gaussian process using the Karhunen-Lo\`eve expansion approach with 31 basis functions and the covariance function $$\kappa_{\alpha}(v,v') =
\sum_{g=1}^G\sum_{s=1}^S\exp\{-0.001(|h_v-\bar{h}_g^{(s)}|_2^2+|h_{v'}-\bar{h}_g^{(s)}|_2^2)-0.01|h_v-h_{v'}|_2^2\}I(v, v'\in \mathcal{R}_g^{(s)})\,.$$ For $j > j'$, let $\alpha_{j,j'}^{(i)}(v) = \alpha_{j',j}^{(i)}(v)$. 
We set $J=3$, $G=25$ and $n\in\{300, 600, 900\}$. Letting the imaging size be $100,150,200,250$ and $300$, then $V =100^2,150^2,200^2,250^2$ and $300^2$, respectively. 
Consider the nonlinear dependence between modalities with $f_{j,j'}(x)=x^2$ for $(j,j')=(2,1)$ and $f_{j,j'}(x)=x$ otherwise. For any $j,j'=1,2,3$ and $g=1,\ldots,25$, let
\begin{align*}
  \delta_{j,j',g} & = \{I(j=1,j'=2)+I(j=3,j'=3)\}I(1\leq g \leq 25) +  \\
  & ~~~~~~~~~~ I(j=2,j'=1)I(1\leq g \leq 5)+ I(j=2,j'=2)I(6\leq g \leq 25),
\end{align*}
which indicates that the three modalities are not mutually independent only for $g \in \{1,\ldots,5\}$.

The empirical FDRs and powers of the proposed multiple tests and three comparing methods are reported in Table \ref{nestedMultimodal2}. The proposed test is implemented by the distributed algorithm with $K=30$ and $N=5000$. All simulation results are based on 1000 replications except that the results of JdCov are only based on 100 repetitions,
since it takes a too long computational time. For the testing problems (a), we have $(Q, Q_0) = (25, 20)$ in this setting and the FDR should be controlled by $\alpha Q_0/Q=4.00\%$ according to Theorem 3. It can be observed that, although the empirical FDRs of the proposed method are slightly higher than $4.00\%$ when $n=300$, they are well controlled around $4.00\%$ when $n$ is as large as 600 and 900. The GdCov and JdCov methods also have control on the empirical FDRs in most scenarios, and the dHSIC test yields very small empirical FDRs. However, the empirical powers
of the three competitors decrease with $V$. The proposed method does not suffer significant power loss when $V$ grows.

\begin{table}[!tbp]
\renewcommand\arraystretch{1.2}
\centering
\begin{threeparttable}
\scriptsize
\caption{Empirical FDRs and powers of the proposed multiple tests with $K=30$ and $L=1$, and three competitors (JdCov, dHSIC, GdCov) for the testing problem (a). All numbers reported below are multiplied by 100. All simulation results are based on 1000 replications except that the results of JdCov are only based on 100 repetitions. }\label{nestedMultimodal2}
\begin{tabular}{cc|cccc|cccc}
\hline\hline
    &      & \multicolumn{4}{c|}{Empirical FDRs}                    & \multicolumn{4}{c}{Empirical Powers}                   \\
$n$ & $V$     & Proposed      & JdCov   & dHSIC & GdCov & Proposed      & JdCov  & dHSIC  & GdCov  \\ \hline
300 & $100^2$ & 6.74          & 10.08   & 0.10  & 4.01  & 72.32         & 16.40   & 0.30   & 66.24  \\
    & $150^2$ & 6.96          & 8.33    & 0.00  & 5.30  & 76.26         & 4.40    & 0.00   & 45.82  \\
    & $200^2$ & 6.00          & 8.00    & 0.00  & 7.10  & 76.84         & 1.60    & 0.00   & 36.46  \\
    & $250^2$ & 5.60          & 10.00   & 0.00  & 8.44  & 78.70         & 1.00    & 0.00   & 33.50  \\
    & $300^2$ & 5.58          & 5.00    & 0.00  & 7.93  & 78.98         & 0.60    & 0.00   & 30.18  \\
600 & $100^2$ & 3.66          & 4.22    & 0.07  & 3.13  & 98.28         & 70.40   & 78.48  & 95.08  \\
    & $150^2$ & 4.24          & 6.42    & 0.10  & 4.90  & 99.02         & 13.20   & 0.02   & 68.46  \\
    & $200^2$ & 4.04          & 12.00   & 0.00  & 6.20  & 99.18         & 4.20    & 0.00   & 50.84  \\
    & $250^2$ & 3.80          & 9.33    & 0.00  & 6.32  & 99.36         & 1.60    & 0.00   & 41.56  \\
    & $300^2$ & 4.27          & 6.50    & 0.00  & 7.56  & 99.28         & 1.20    & 0.00   & 37.56  \\
900 & $100^2$ & 3.58          & 2.85    & 0.18  & 3.38  & 99.96         & 97.00   & 99.96  & 99.72  \\
    & $150^2$ & 3.39          & 7.68    & 0.00  & 3.72  & 100.00        & 36.80   & 8.28   & 85.66  \\
    & $200^2$ & 3.62          & 8.33    & 0.00  & 4.43  & 100.00        & 11.80   & 0.00   & 64.28  \\
    & $250^2$ & 3.65          & 11.33   & 0.00  & 5.63  & 100.00        & 5.00    & 0.00   & 51.64  \\
    & $300^2$ & 3.31          & 12.50   & 0.00  & 7.56  & 99.96         & 3.20    & 0.00   & 43.26  \\
\hline\hline
\end{tabular}
\end{threeparttable}
\end{table}

\setcounter{figure}{0}
\renewcommand{\thefigure}{S\arabic{figure}}
\section{Additional Details in Analysis of HCP Data}\label{sec:suppl_app}
\begin{figure}[htbp]
  \centering
  \includegraphics[width=15cm]{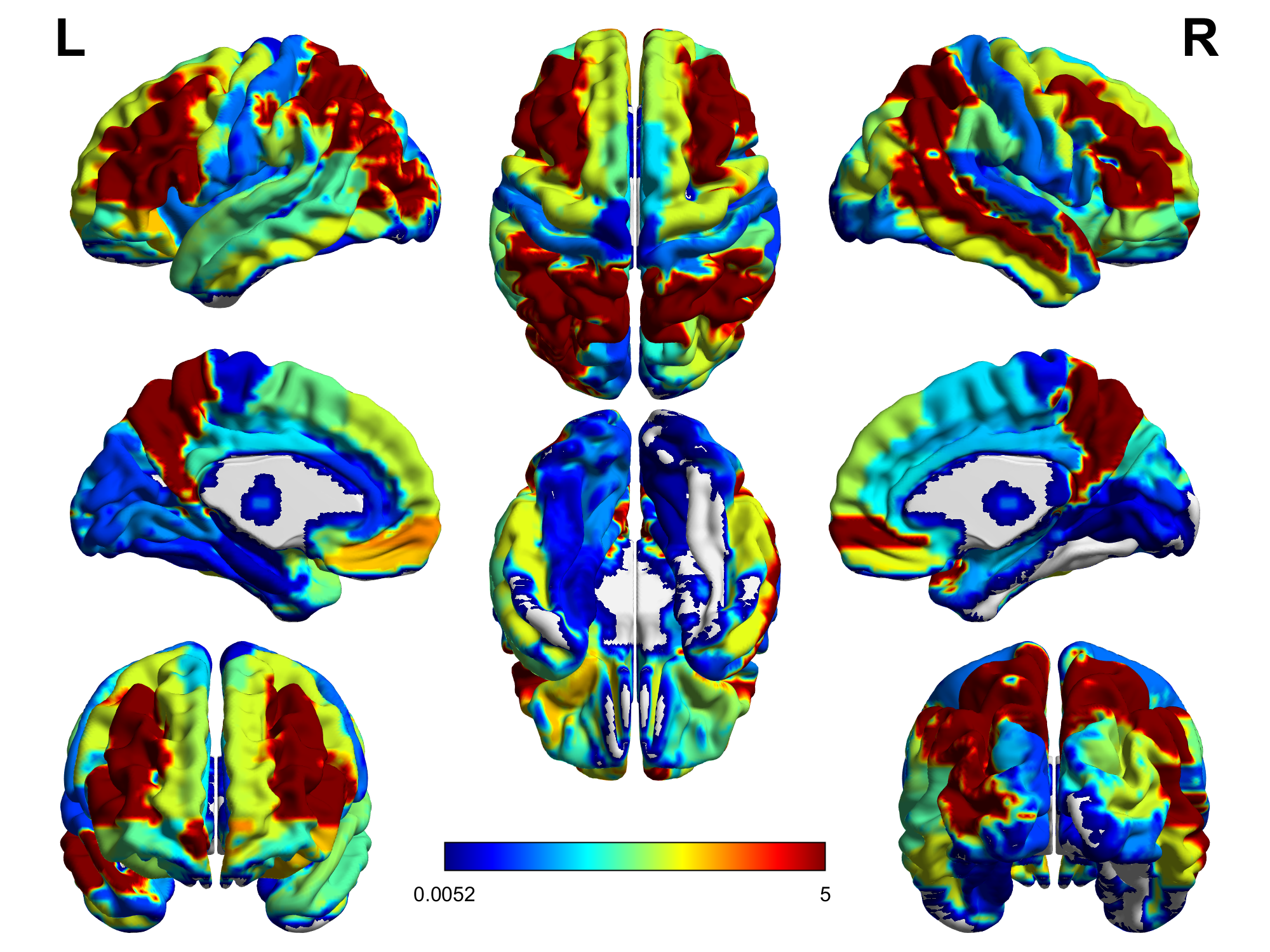}\\
  \caption{ Regionwise ``$-\log_{10}$(p-values)" for testing whether the working memory (2bk-0bk), language processing (story-math) and social cognition (mental) brain activities are dependent in each brain region. Set $-\log_{10}$(p-value) = 5 if $\mbox{p-value} < 0.0001$. The grey regions are not considered in the analysis. }\label{fig1}
\end{figure}
\paragraph{Testing independence among modalities over regions} For the multiple tests, the region-wise negative $\log_{10}$(p-values) were shown in Figure \ref{fig1}, where the value was set to $5$ if the p-value $< 0.0001$.  Based on our analysis, the language processing, working memory and social cognition brain activities are strongly dependent with statistical significance in most parts of the inferior frontal gyrus (Frontal\_Inf\_Oper\_L, Frontal\_Inf\_Tri\_L, Frontal\_Inf\_Tri\_R), while the dependence of the three types of brain activities is not statistically significant in the most part of cingulate cortex (Cingulum\_Ant\_L, Cingulum\_Ant\_R,  Cingulum\_Mid\_L,  Cingulum\_Mid\_R, Cingulum\_Post\_L,  Cingulum\_Post\_R), indicating that  there are more individual differences in the three fMRI task activities in cingulate cortex. It is well-known that inferior frontal gyrus~\citep{becker2013inferior} and anterior cingulate cortex~\citep{lenartowicz2005role} are associated with language processing and working memory tasks~\citep{barch2013function} respectively.

\section{An Auxiliary Lemma}\label{se:lemma}
\begin{la}\label{la:2}If $B\{\log (dB)\}^2=o(n)$, then it holds that $|\hat{\Sigma}_n-{\Sigma}_{n}|_{\infty}=O_{\p}[n^{-1/2}B^{3/2}\{\log(dB)\}^{1/2}]+O_{\p}[n^{-1}B^{2}\{\log(dB)\}^{2}]$.
\end{la}
\noindent{\it Proof.} Note that $\mathbb{E}(U_m)$ is a constant vector for any $m\in[M]$. Write $\mu=\mathbb{E}(U_m)$ and define
$H_j=M^{-1}\sum_{m=j+1}^{M}\mathbb{E}\{(U_{m}-\mu)(U_{m-j}-\mu)^{\T}\}$ for $j \geq 0$ and $H_j=M^{-1}\sum_{m=-j+1}^{M}\mathbb{E}\{(U_{m+j}-\mu)(U_{m}-\mu)^{\T}\}$ for $j < 0$.
Then $\Sigma_n=\sum_{j=-B+1}^{B-1}H_j$. Write $H_j=(H_{j,k,l})_{k,l\in[d]}$, $\hat{H}_j=(\hat{H}_{j,k,l})_{k,l\in[d]}$, $\bar{U}=(\bar{U}_1,\ldots,\bar{U}_d)^{\T}$ and $\mu=(\mu_{1},\ldots,\mu_{d})^{\T}$.
For any $j\in \{0\}\cup[B-1]$ and $k,l\in [d]$, it holds that
\begin{align*}
\hat{H}_{j,k,l}-H_{j,k,l} =&~\dfrac{1}{M}\sum_{m=j+1}^{M}\{U_{m,k}U_{m-j,l}-\mathbb{\mathbb{E}}(U_{m,k}U_{m-j,l})\}\\
&-\dfrac{M-j}{M}\big[\bar{U}_{l}\{\bar{U}_{k,(j+1)}-\mu_{k}\} + \bar{U}_{k}\{\bar{U}_{l}^{(M-j)}-\mu_{l}\} \big] \\
&+\dfrac{M-j}{M}(\bar{U}_{l}-\mu_{l})( \bar{U}_{k} - \mu_{k})
\end{align*}
with $\bar{U}_{k,(j+1)} = (M-j)^{-1}\sum_{m=j+1}^{M}U_{m,k}$ and $\bar{U}_{l}^{(M-j)} = (M-j)^{-1}\sum_{m=j+1}^{M}U_{m-j,l}$. Notice that  $\max_{m\in[M]}\max_{j\in[d]}|U_{m,j}|\le C$. We then have that
\begin{align}\label{eq:var_mb}
|\hat{H}_j-H_j|_{\infty}=&~\max_{k,l\in[d]}|\hat{H}_{j,k,l}-H_{j,k,l}| \notag\\
\lesssim&~\max_{k,l\in[d]}\bigg|\dfrac{1}{M-j}\sum_{m=j+1}^{M}\{U_{m,k}U_{m-j,l}-\mathbb{E}(U_{m,k}U_{m-j,l})\}\bigg|\notag\\
&+\max_{k\in[d]}|\bar{U}_{k,(j+1)}-\mu_{k}|+\max_{l\in[d]}|\bar{U}_l^{(M-j)}-\mu_l|+\max_{l\in [d]}|\bar{U}_{l}-\mu_{l}|^2 \notag\\
=&: I_{1,j} + I_{2,j}+I_{3,j}+I_4\,.
\end{align}
We next bound $I_{1,j}$, $I_{2,j}$, $I_{3,j}$ and $I_4$ in \eqref{eq:var_mb}, respectively. Notice that $\{U_{m}-\mu_m\}$ is a $(B-1)$-dependent sequence with uniformly bounded components. Recall $M\asymp n$. Applying Lemma 7 of \cite{ChangChenWu_2021} with $(\tilde{B}_{\tilde{n}},\tilde{L}_{\tilde{n}},\tilde{j}_{\tilde{n}},r_1,r_2)=(1,B-1,0,2,\infty)$, it holds that
	\begin{align*}
	\mathbb{P}\bigg(\max_{l\in[d]}|\bar{U}_{l}-\mu_{l}|\ge z\bigg)\le&~\sum_{l=1}^{d}\mathbb{P}\big(|\bar{U}_{l}-\mu_{l}|\ge z\big)\notag\\
	\lesssim&~d\exp(-CB^{-1}n z^2)+d\exp(-CB^{-1/2}n ^{1/2}z^{1/2})
	\end{align*}
	for any $z>0$, which implies
	\begin{align}\label{eq:i4tail}
	\mathbb{P}(I_{4}\ge z)&\lesssim d\exp(-CB^{-1}n z)+d\exp(-CB^{-1/2}n ^{1/2}z^{1/4})
	\end{align}
	for any $z>0$. Analogously, due to $B=o(n)$, we also have
	\begin{align*}
	\mathbb{P}\bigg(\max_{j\in\{0\}\cup[B-1]}I_{1,j}\ge z\bigg)&\lesssim d^2B\exp(-CB^{-1}n z^2)+d^2B\exp(-CB^{-1/2}n ^{1/2}z^{1/2})\,,\\
	\mathbb{P}\bigg(\max_{j\in\{0\}\cup[B-1]}I_{2,j}\ge z\bigg)&\lesssim dB\exp(-CB^{-1}n z^2)+dB\exp(-CB^{-1/2}n ^{1/2}z^{1/2})\,,\\
	\mathbb{P}\bigg(\max_{j\in\{0\}\cup[B-1]}I_{3,j}\ge z\bigg)&\lesssim dB\exp(-CB^{-1}n z^2)+dB\exp(-CB^{-1/2}n ^{1/2}z^{1/2})
	\end{align*}
	for any $z>0$. Together with \eqref{eq:i4tail}, \eqref{eq:var_mb} yields
	\begin{align*}
	\mathbb{P}\bigg(\max_{j\in\{0\}\cup[B-1]}|\hat{H}_j-H_j|_\infty \ge z\bigg)\lesssim d^2B\exp(-CB^{-1}n z^2)+d^2B\exp(-CB^{-1/2}n ^{1/2}z^{1/2})
	\end{align*}
	for any $z=o(1)$.
	Similarly, we also have
	\begin{align*}
	\mathbb{P}\bigg(\max_{-j\in[B-1]}|\hat{H}_j-H_j|_\infty \ge z\bigg)\lesssim d^2B\exp(-CB^{-1}n z^2)+d^2B\exp(-CB^{-1/2}n ^{1/2}z^{1/2})
	\end{align*}
	for any $z=o(1)$. Notice that $\hat{\Sigma}_n-\Sigma_n=\sum_{j=-B+1}^{B-1}(\hat{H}_j-H_j)$. Thus,
	\begin{align}\label{eq:covestbd1}
	\mathbb{P}(|\hat{\Sigma}_n-\Sigma_n|_\infty\geq z)\lesssim d^2B\exp(-CB^{-3}n z^2)+d^2B\exp(-CB^{-1}n ^{1/2}z^{1/2})
	\end{align}
	for any $z=o(B)$, which implies $|\hat{\Sigma}_n-{\Sigma}_{n}|_{\infty}=O_{\p}[n^{-1/2}B^{3/2}\{\log(dB)\}^{1/2}]+O_{\p}[n^{-1}B^{2}\{\log(dB)\}^{2}]$ provided that $B\{\log (dB)\}^2=o(n)$. $\hfill\Box$

\section{Proof of Proposition 1}\label{se:pfpn1}
Define $\mathring{T}_{n}=\sqrt{M}D_n^{-1/2}\bar{U}$ and $\mathring{W}_{n,L}=f_L(\mathring{T}_n)$. Recall $\mathcal{F} = \{v=(v_1,\ldots,v_d)^{\T}: v_j \in \{ 0,1 \}~\textrm{for each}~j \in [d],~\textrm{and}~|v|_{0}= L \}$.
Due to the countability of the set $\mathcal{F}$, we can index all the elements in $\mathcal{F}$ as $\{\tilde{v}_{1},\ldots,\tilde{v}_{|\mathcal{F}|}\}$ in a certain order, and
define a new sequence $\{U_{m}^{\dag}\}_{m=1}^{M}$ with $U_{m}^{\dag}=(U_{m,1}^{\dag},\ldots,U_{m,|\mathcal{F}|}^{\dag})^{\T}:=(\tilde{v}^{\T}_{1}D_n^{-1/2}U_{m},\ldots,\tilde{v}^{\T}_{|\mathcal{F}|}D_n^{-1/2}U_{m})^{\T}$. Write $S_{U^\dag}=M^{-1/2}\sum_{m=1}^MU_m^\dag=(S_{U^\dag,1},\ldots,S_{U^\dag,|\mathcal{F}|})^{\T}$. We have
\begin{align}\label{eq:Wnlexp}
\{\mathring{W}_{n,L}\le z\}=\bigg\{\max_{v\in \mathcal{F}}\sqrt{M}v^{\T}D_n^{-1/2}\bar {U}\le z\bigg\}=\{{S}_{U^\dag}\le z\cdot 1_{|\mathcal{F}|}\}\,,
\end{align}
where $1_{|\mathcal{F}|}$ is a $|\mathcal{F}|$-dimensional vector with all the components being $1$. Recall $R_n=D_n^{-1/2}\Sigma_n D_n^{-1/2}$. Let $\xi=(\xi_{1},\ldots,\xi_{d})^{\T}\sim \mathcal{N}(0,R_n)$ and $g=V^{\T}\xi$ with $V=(\tilde{v}_{1},\ldots,\tilde{v}_{|\mathcal{F}|})$. Then $\mathrm{Cov}( S_{U^\dag})=V^{\T}R_n V=\mathrm{Cov}(g)$. Identical to \eqref{eq:Wnlexp}, we also have
\begin{align}\label{eq:fxiexp}
\{f_L(\xi)\le z\}=\bigg\{\max_{j\in[|\mathcal{F}|]}\tilde{v}_{j}^{\T}\xi\le z\bigg\}=\{g\le z\cdot 1_{|\mathcal{F}|} \}\,.
\end{align}
Recall $\{U_{m}^{\dag}\}_{m=1}^{M}$ is also a $(B-1)$-dependent sequence and $\max_{m\in[M],j\in[|\mathcal{F}|]}|U_{m,j}^{\dag}|\le CL$. Notice that $\mathrm{Var}( S_{U^\dag,j})=\tilde{v}_{j}^{\T}R_n\tilde{v}_{j}\ge\min_{v\in\mathcal{F}}v^{\T}R_n v\ge C$
holds uniformly over $j\in|\mathcal{F}|$ with $|\mathcal{F}|=C_d^L=d!/\{L!(d-L)!\}$. Applying Corollary 1 of \cite{ChangChenWu_2021} with $B_n=CL$ and $m=B-1$, we have
	\begin{align}\label{eq:gaussbd}
	\sup_{y\in \mathbb{R}^{|\mathcal{F}|}}\big|\mathbb{P}( S_{U^\dag}\le y) - \mathbb P(g\le y)\big|\lesssim \frac{B^{2/3}L(\log |\mathcal{F}|)^{7/6}}{n^{1/6}}\lesssim\frac{B^{2/3}L^{13/6}(\log d)^{7/6}}{n^{1/6}}\,.
	\end{align}
Together with \eqref{eq:Wnlexp} and \eqref{eq:fxiexp}, we have
	\begin{align}\label{eq:gb1}
	\sup_{z\in \mathbb{R}}\big|\mathbb{P}\big(\mathring{W}_{n,L}\le z\big) - \mathbb P\{f_L(\xi)\le z\}\big|\lesssim \frac{B^{2/3}L^{13/6}(\log d)^{7/6}}{n^{1/6}}\,.
	\end{align}
For any $z\in\mathbb{R}$ and $\epsilon>0$, it holds that
\begin{align*}
\mathbb{P}(W_{n,L}> z)&=\mathbb{P}(W_{n,L}> z,|W_{n,L}-\mathring{W}_{n,L}|>\epsilon)+\mathbb{P}(W_{n,L}> z,|W_{n,L}-\mathring{W}_{n,L}|\le\epsilon)\\
&\le\mathbb{P}(|W_{n,L}-\mathring{W}_{n,L}|> \epsilon) + \mathbb{P}(\mathring{W}_{n,L}> z-\epsilon)\,.
\end{align*}
Thus, by \eqref{eq:gb1} and the Nazarov's inequality \cite[Lemma A.1]{CCK2015}, we have
	\begin{align}\label{eq:Ts1}
	\mathbb{P}(W_{n,L}> z)-\mathbb P\{f_L(\xi)> z\}\le&~\mathbb{P}(|W_{n,L}-\mathring{W}_{n,L}| >\epsilon)+\P\{z-\epsilon< f_L(\xi)\le z\}\notag\\
	&+\mathbb{P}(\mathring{W}_{n,L} > z-\epsilon)- \P\{f_L(\xi)>z-\epsilon\}\notag\\
	\lesssim&~\mathbb{P}(|W_{n,L}-\mathring{W}_{n,L}| >\epsilon)+\frac{B^{2/3}L^{13/6}(\log d)^{7/6}}{n^{1/6}}+\epsilon(L\log d)^{1/2}
	\end{align}
for any $z\in\mathbb{R}$ and $\epsilon>0$.
On the other hand, it can be derived that
\begin{align*}
\mathbb{P}(W_{n,L}> z)&\geq\P(W_{n,L}> z, \mathring{W}_{n,L}> z+\epsilon)\\
&= \P(\mathring{W}_{n,L}> z+\epsilon)-\P(W_{n,L}\leq z, \mathring{W}_{n,L}> z+\epsilon)\\
&\geq\P(\mathring{W}_{n,L}> z+\epsilon)-\P(|W_{n,L}-\mathring{W}_{n,L}| > \epsilon)\,.
\end{align*}
Applying the Nazarov's inequality and \eqref{eq:gb1} again, we have
	\begin{align*}
	\mathbb{P}(W_{n,L}> z)-\mathbb P\{f_L(\xi)> z\}\gtrsim-\mathbb{P}(|W_{n,L}-\mathring{W}_{n,L}|>\epsilon)-\frac{B^{2/3}L^{13/6}(\log d)^{7/6}}{n^{1/6}}-\epsilon(L\log d)^{1/2}
	\end{align*}
for any $z\in\mathbb{R}$ and $\epsilon>0$. Together with \eqref{eq:Ts1}, it holds that
	\begin{align}\label{eq:gabd1}
	&\sup_{z\in \mathbb{R}}|\mathbb{P}(W_{n,L}>z)-\mathbb P\{f_L(\xi)>z\}|\notag\\
	&~~~~~~~~~\lesssim \mathbb{P}(|W_{n,L}-\mathring{W}_{n,L}| >\epsilon)+\frac{B^{2/3}L^{13/6}(\log d)^{7/6}}{n^{1/6}}+\epsilon(L\log d)^{1/2}\,.
	\end{align}

Recall $D_n= \diag(\sigma_1^2, \ldots, \sigma_d^2)$ and $\hat{D}_{n}= \diag(\hat\sigma_1^2, \ldots, \hat\sigma_d^2)$. It holds that $
|T_n-\mathring{T}_n|_\infty\leq \sqrt{M}|\hat{D}_n^{-1/2}-D_n^{-1/2}|_\infty|\bar{U}|_\infty$,
which implies that
$
|W_{n,L}-\mathring{W}_{n,L}|\leq L|T_n-\mathring{T}_n|_\infty\leq L\sqrt{M}|\hat{D}_n^{-1/2}-D_n^{-1/2}|_\infty|\bar{U}|_\infty$. To bound $\mathbb{P}(|W_{n,L}-\mathring{W}_{n,L}| >\epsilon)$ for some $\epsilon=o(1)$, it suffices to specify the upper bounds of $\mathbb{P}(|\hat{D}_n^{-1/2}-D_n^{-1/2}|_\infty>z)$ and $\mathbb{P}(|\bar{U}|_\infty>z)$ for any $z=o(1)$, respectively. Notice that $\mathbb{E}(\bar{U})=0$ under $H_0$. As we have shown in the proof of Lemma \ref{la:2},  it holds under $H_0$ that
\begin{align*}
	\mathbb{P}(|\bar{U}|_{\infty}> z)\lesssim~d\exp(-CB^{-1}n z^2)+d\exp(-CB^{-1/2}n ^{1/2}z^{1/2})
\end{align*}
for any $z>0$. On the other hand, since $\max_{l\in[d]}|\hat{\sigma}_{l}^2-\sigma_{l}^2|\le |\hat{\Sigma}_n-\Sigma_n|_\infty$, it follows from \eqref{eq:covestbd1} that
	\begin{align*}
	\mathbb{P}\bigg(\max_{l\in[d]}|\hat{\sigma}_{l}^2-\sigma_{l}^2| \ge z\bigg)\lesssim d^2B\exp(-CB^{-3}n z^2)+d^2B\exp(-CB^{-1}n ^{1/2}z^{1/2})
	\end{align*}
for any $z=o(B)$.
Due to $\min_{l\in[d]}\sigma_{l}^2\ge c_{1}$, then we have
	\begin{align}
	\mathbb{P}\bigg(\min_{l\in[d]}\hat{\sigma}_{l}^2 \le \frac{c_{1}}{2}\bigg)\le&~\mathbb{P}\bigg\{\min_{l\in[d]}(\hat{\sigma}_{l}^2-\sigma_{l}^2)+\min_{l\in[d]}\sigma_{l}^2\le \frac{c_{1}}{2}\bigg\}\label{eq:tm3p4}\\
	\leq&~\mathbb{P}\bigg(\max_{l\in[d]}|\hat{\sigma}_{l}^2-\sigma_{l}^2|\ge\frac{c_1}{2}\bigg)\lesssim d^2B\exp(-CB^{-3}n)+d^2B\exp(-CB^{-1}n ^{1/2})\,.\notag
	\end{align}
If $B^2\log(dB)\cdot\max\{B,\log(dB)\}=o(n)$, we know the event $\mathcal{E}=\{\min_{l\in[d]}\hat{\sigma}_l^2>c_1/2\}$ occurs with probability approaching one.
Restricted on the event $\mathcal{E}$, we have $\max_{l\in[d]}|\hat{\sigma}_{l}^{-1}-\sigma_{l}^{-1}|\leq \tilde{c}\max_{l\in[d]}|\hat{\sigma}_l^2-\sigma_l^2|$ for some universal constant $\tilde{c}>0$. Therefore,
	\begin{align}
	\mathbb{P}\bigg(\max_{l\in[d]}|\hat{\sigma}_{l}^{-1}-\sigma_{l}^{-1}|\ge z\bigg)&\leq\mathbb{P}\bigg(\max_{l\in[d]}|\hat{\sigma}_{l}^2-\sigma_{l}^2| \ge \tilde{c}^{-1}z\bigg)+\mathbb{P}\bigg(\min_{l\in[d]}\hat{\sigma}_{l}^2 \le \frac{c_{1}}{2}\bigg)\notag\\
	&\lesssim d^2B\exp(-CB^{-3}n z^2)+d^2B\exp(-CB^{-1}n ^{1/2}z^{1/2}) \label{eq:tm3p5}
	\end{align}
for any $z=o(1)$. Due to $|W_{n,L}-\mathring{W}_{n,L}|\leq L M^{1/2}|\hat{D}_n^{-1/2}-D_n^{-1/2}|_\infty|\bar{U}|_\infty\le  Ln^{1/2}|\bar{U}|_\infty\max_{l\in[d]}|\hat{\sigma}_{l}^{-1}-\sigma_{l}^{-1}|$, we have
	\begin{align*}
	\mathbb{P}\big(|W_{n,L}-\mathring{W}_{n,L}|> \epsilon\big)&\leq \mathbb{P}\bigg(\max_{l\in[d]}|\hat{\sigma}_{l}^{-1}-\sigma_{l}^{-1}|> B^{1/2}L^{-1/2}n^{-1/4}\epsilon^{1/2}\bigg)\\
	&~~~~~~~+\mathbb{P}(|\bar{U}|_{\infty}>B^{-1/2}L^{-1/2}n^{-1/4}\epsilon^{1/2})\\
	&\lesssim d^2B\exp(-CB^{-2}L^{-1}n^{1/2}\epsilon)+d^2B\exp(-CB^{-3/4}L^{-1/4}n^{3/8}\epsilon^{1/4})
	\end{align*}
for any $\epsilon=o(1)$. By \eqref{eq:gabd1}, it holds that
	\begin{align*}
	\sup_{z\in \mathbb{R}}|\mathbb{P}(W_{n,L}>z)-\mathbb P\{f_L(\xi)>z\}|\lesssim&~d^{2}B\exp\bigg(-\frac{Cn^{1/2}\epsilon}{B^{2}L}\bigg)+d^2B\exp\bigg(-\frac{Cn^{3/8}\epsilon^{1/4}}{B^{3/4}L^{1/4}}\bigg)\\
	&+\frac{B^{2/3}L^{13/6}(\log d)^{7/6}}{n^{1/6}}+\epsilon(L\log d)^{1/2}\,.
	\end{align*}
With selecting $\epsilon=n^{-1/6}B^{2/3}L^{5/3}(\log d)^{2/3}$, we have
	\begin{align}\label{eq:f_L_1}
	\sup_{z\in \mathbb{R}}|\mathbb{P}(W_{n,L}>z)-\mathbb P\{f_L(\xi)>z\}|\lesssim&~d^{2}B\exp\{-CB^{-4/3}n^{1/3}L^{2/3}(\log d)^{2/3}\}\notag\\
	&+d^2B\exp\{-CB^{-7/12}n^{1/3}L^{1/6}(\log d)^{1/6}\}\\
	&+\frac{B^{2/3}L^{13/6}(\log d)^{7/6}}{n^{1/6}}\,.\notag
\end{align}
Hence,
$
\sup_{z\in \mathbb{R}}|\mathbb{P}(W_{n,L}>z)-\mathbb P\{f_L(\xi)>z \}|=o(1)$
provided that $B^4 L^{13}(\log d)^7=o(n)$, $B^4L^{-2}(\log B)^3\\ \cdot(\log d)^{-2}=o(n)$ and $B^{7/4}L^{-1/2}(\log B)^3(\log d)^{-1/2}=o(n)$. We complete the proof of part (i) of Proposition 1.

Write $\mathcal{X}_n=\{X_1,\ldots,X_n\}$. To prove part (ii) of Proposition 1, it suffices to show $\sup_{z\in \mathbb{R}}|\mathbb P\{f_L(\xi)> z\}-\mathbb P\{f_L(\hat{\xi})>z\,|\,\mathcal{X}_n\}|=o_\p(1)$. Recall $\xi\sim\mathcal{N}(0,R_n)$ and $\hat{\xi}\,|\,\mathcal{X}_n\sim\mathcal{N}(0,\hat{R}_n)$. We define ${\Delta}_{n}= \max_{\check{v}_{1},\check{v}_{2}\in\mathcal{F}}|\check{v}_{1}^{\T}(R_n-\hat{R}_{n})\check{v}_{2}|$, and write $\Gamma_n=D_n^{-1/2}$ and $\hat{\Gamma}_{n}= \hat{D}_n^{-1/2}$. For any $\check{v}_{1},\check{v}_{2}\in\mathcal{F}$, it follows from the triangle inequality that
\begin{align*}
|\check{v}_{1}^{\T}(R_n-\hat{R}_{n})\check{v}_{2}| & = |\check{v}_{1}^{\T}(\Gamma_n\Sigma_n\Gamma_n-\hat{\Gamma}_n\hat{\Sigma}_n\hat{\Gamma}_n)\check{v}_{2}|\\
& \le |\check{v}_{1}^{\T}\Gamma_n\Sigma_n(\Gamma_n-\hat{\Gamma}_n)\check{v}_{2}| + |\check{v}_{1}^{\T}\Gamma_n(\Sigma_n-\hat{\Sigma}_n)\hat{\Gamma}_n\check{v}_{2}| + |\check{v}_{1}^{\T}(\Gamma_n-\hat{\Gamma}_n)\hat{\Sigma}_n\hat{\Gamma}_n\check{v}_{2}|\,,
\end{align*}
which implies
\begin{align}
\Delta_{n}\le&\max_{\check{v}_{1},\check{v}_{2}\in\mathcal{F}}|\check{v}_{1}^{\T}\Gamma_n\Sigma_n(\Gamma_n-\hat{\Gamma}_n)\check{v}_{2}| + \max_{\check{v}_{1},\check{v}_{2}\in\mathcal{F}}|\check{v}_{1}^{\T}\Gamma_n(\Sigma_n-\hat{\Sigma}_n)\hat{\Gamma}_n\check{v}_{2}| + \max_{\check{v}_{1},\check{v}_{2}\in\mathcal{F}}|\check{v}_{1}^{\T}(\Gamma_n-\hat{\Gamma}_n)\hat{\Sigma}_n\hat{\Gamma}_n\check{v}_{2}|\notag\\
=&:I_{1}+I_{2}+I_{3}\,.\label{eq:tm3pf3}
\end{align}
Note that
$I_{1}\le|\Sigma_n|_\infty\max_{\check{v}_{1}\in\mathcal{F}}|\Gamma_n\check{v}_{1}|_{1}\max_{\check{v}_{2}\in\mathcal{F}}|(\Gamma_n-\hat{\Gamma}_n)\check{v}_{2}|_{1}\lesssim BL^2|\Gamma_n-\hat{\Gamma}_n|_\infty$. By \eqref{eq:tm3p5}, we have $I_1=O_{\p}[n^{-1/2}B^{5/2}L^2\{\log(dB)\}^{1/2}]+O_{\p}[n^{-1}B^{3}L^2\{\log(dB)\}^{2}]$
	provided that $B^{2}\log (dB)\max\{B,\log(dB)\}=o(n)$. Similarly, we also have $I_{2}=O_{\p}[n^{-1/2}B^{3/2}L^2\{\log(dB)\}^{1/2}]+O_{\p}[n^{-1}B^{2}L^2\{\log(dB)\}^{2}]$ and $I_{3}=O_{\p}[n^{-1/2}B^{5/2}L^2\{\log(dB)\}^{1/2}]+O_{\p}[n^{-1}B^{3}L^2\{\log(dB)\}^{2}]$. Hence, $\Delta_n=O_{\p}[n^{-1/2}B^{5/2}L^2\{\log(dB)\}^{1/2}]+O_{\p}[n^{-1}B^{3}L^2\{\log(dB)\}^{2}]=o_\p(1)$ if $B^3L^2\log(dB)\cdot\max\{B^2L^2,\log(dB)\}=o(n)$. It follows from Lemma 3.1 of \cite{CCK2013} that
\begin{align}\label{eq:f_L_2}
&\sup_{z\in \mathbb{R}}|\mathbb P\{f_L(\xi)>z\}-\mathbb P\{f_L(\hat{\xi})> z\,|\,\mathcal{X}_n\}|\notag\\
&~~~~~~~~~~= \sup_{z\in \mathbb{R}}\bigg|\P\bigg(\max_{v\in \mathcal{F}}v^{\T}\xi\le z\bigg)-\P\bigg(\max_{v\in \mathcal{F}}v^{\T}\hat{\xi}\le z\,|\,\mathcal{X}_n\bigg)\bigg|\\
&~~~~~~~~~~\lesssim \Delta_{n}^{1/3}\{1\vee \log(d^{L}/\Delta_{n})\}^{2/3}=o_\p(1)\notag
\end{align}
provided that $B^5L^8(\log d)^4\log(dB)=o(n)$. $\hfill\Box$

\section{Proof of Theorem 1}

Notice that $\hat{{\rm cv}}_{L,\alpha}$ is the $(1-\alpha)$-quantile of the conditional distribution of $f_L(\hat{\xi})$ given $\{X_1,\ldots,X_n\}$, where $\hat{\xi}\,|\,X_1,\ldots,X_n\sim \mathcal{N}(0,\hat{R}_n)$. By Part (ii) of Proposition 1, we have
$
\mathbb{P}_{H_0}(W_{n,L}>\hat{{\rm cv}}_{L,\alpha})=\alpha+o(1)$, which completes the proof of Theorem 1.$\hfill\Box$

\section{Proof of Theorem 2}
Write $\mathcal{X}_{n}=\{ X_1,\ldots,X_n \}$. Recall $V=(\tilde{v}_{1},\ldots,\tilde{v}_{|\mathcal{F}|})\in\mathbb{R}^{d\times|\mathcal{F}|}$ with $\mathcal{F}=\{\tilde{v}_{1},\ldots,\tilde{v}_{|\mathcal{F}|}\}$. Let $\hat{g}=V^{\T}\hat{\xi}$ with $\hat{\xi}\,|\,\mathcal{X}_n\sim\mathcal{N}(0,\hat{R}_n)$. Then $\hat{g}\,|\,\mathcal{X}_n\sim\mathcal{N}(0,V^{\T}\hat{R}_nV)$ and
\begin{align}\label{eq:tailbd1}
\mathbb{E}\big(|\hat{g}|_{\infty} \, | \, \mathcal{X}_n\big) \le \{1+(2\log |\mathcal{F}|)^{-1}\}(2\log |\mathcal{F}|)^{1/2}\max_{j\in [|\mathcal{F}|]}(\tilde{v}_j^{\T}\hat{R}_n\tilde{v}_j)^{1/2}\,.
\end{align}
As shown in \cite{Borell_1975}, for any $u>0$, it holds that
\begin{align}\label{eq:tailbd}
\P\big\{|\hat{g}|_{\infty}\ge \mathbb{E}\big(|\hat{g}|_{\infty} \, | \,\mathcal{X}_n\big)+u \, | \, \mathcal{X}_n\big\}\le \exp\bigg(-\frac{u^{2}}{2\max_{j\in[|\mathcal{F}|]}\tilde{v}_j^{\T}\hat{R}_n\tilde{v}_j}\bigg)\,.
\end{align}

Recall $D_n={\rm diag}(\Sigma_n)={\rm diag}(\sigma_1^2,\ldots,\sigma_d^2)$ and $\hat{D}_n={\rm diag}(\hat{\Sigma}_n)={\rm diag}(\hat{\sigma}_1^2,\ldots,\hat{\sigma}_d^2)$. For any $v>0$, define
\begin{align*}
\mathcal{E}(v)=\bigg\{\max_{j\in[|\mathcal{F}|]}\bigg|\frac{(\tilde{v}_j^{\T}\hat{R}_n\tilde{v}_j)^{1/2}}{(\tilde{v}_j^{\T}R_n\tilde{v}_j)^{1/2}}-1\bigg|\le v,\, \max_{k\in[d]}\bigg|\frac{\hat{\sigma}_{k}}{\sigma_{k}}-1\bigg|\le v\bigg\}\,.
\end{align*}
We have shown in Section \ref{se:pfpn1} that $\max_{j\in[|\mathcal{F}|]}|\tilde{v}_j^{\T}(\hat{R}_n-R_n)\tilde{v}_j|=O_{\p}[n^{-1/2}B^{5/2}L^2\{\log(dB)\}^{1/2}]+O_{\p}[n^{-1}B^{3}L^2\{\log(dB)\}^{2}]$. Due to $\min_{j\in[|\mathcal{F}|]}\tilde{v}_j^{\T}R_n\tilde{v}_j\geq C$, we have
	\begin{align*}
	\max_{j\in[|\mathcal{F}|]}\bigg|\frac{(\tilde{v}_j^{\T}\hat{R}_n\tilde{v}_j)^{1/2}}{(\tilde{v}_j^{\T}R_n\tilde{v}_j)^{1/2}}-1\bigg|=O_{\p}[n^{-1/2}B^{5/2}L^2\{\log(dB)\}^{1/2}]+O_{\p}[n^{-1}B^{3}L^2\{\log(dB)\}^{2}]\,.
	\end{align*}
On the other hand, it follows from Lemma \ref{la:2} that $\max_{k\in[d]}|\hat{\sigma}_k^2-\sigma_k^2|=O_{\p}[n^{-1/2}B^{3/2}\{\log(dB)\}^{1/2}]+O_{\p}[n^{-1}B^{2}\{\log(dB)\}^{2}]$. Due to $\min_{k\in[d]}\sigma_k^2\geq C$, we have
	\begin{align*}
	\max_{k\in[d]}\bigg|\frac{\hat{\sigma}_{k}}{\sigma_{k}}-1\bigg|=O_{\p}[n^{-1/2}B^{3/2}\{\log(dB)\}^{1/2}]+O_{\p}[n^{-1}B^{2}\{\log(dB)\}^{2}]\,.
	\end{align*}
With selecting $v\asymp \eta_n[n^{-1/2}B^{5/2}L^2\{\log(dB)\}^{1/2}+n^{-1}B^{3}L^2\{\log(dB)\}^{2}]\rightarrow0$ for some $\eta_n\rightarrow\infty$, we have $\mathbb{P}\{\mathcal{E}(v)\}\rightarrow1$ as $n\rightarrow\infty$. Restricted on $\mathcal{E}(v)$, \eqref{eq:tailbd1} and \eqref{eq:tailbd} yield
\begin{align}\label{eq:P2}
\hat{\mathrm{cv}}_{L,\alpha}\le&~(1+v)[\{1+(2\log |\mathcal{F}|)^{-1}\}(2\log |\mathcal{F}|)^{1/2}+\{2\log(1/\alpha)\}^{1/2}]\max_{j\in[|\mathcal{F}|]}(\tilde{v}_j^{\T}R_n\tilde{v}_j)^{1/2}\notag\\
\le&~\{1+(\log |\mathcal{F}|)^{-1}\}\lambda(|\mathcal{F}|,\alpha)\max_{j\in[|\mathcal{F}|]}(\tilde{v}_j^{\T}R_n\tilde{v}_j)^{1/2}
\end{align}
with $\lambda(|\mathcal{F}|,\alpha)=(2\log |\mathcal{F}|)^{1/2}+\{2\log(1/\alpha)\}^{1/2}$, provided that $B^5L^4\log(dB)(\log|\mathcal{F}|)^2 = o(n)$ and $B^3L^2\{\log(dB)\}^2\log|\mathcal{F}| = o(n)$.

Let $\tilde{U}=M^{-1}\sum_{m=1}^{M}(U_{m}-\mu)=:(\tilde{U}_1,\ldots,\tilde{U}_d)^{\T}$ with $\mu=\mathbb{E}(U_{m})=(\mu_{1},\ldots,\mu_{d})^{\T}$.
We write $\tilde{T}_{n,j} = \mu_j/\sigma_j$ for any $j\in[d]$, and sort $\{\tilde{T}_{n,j}\}_{j=1}^d$ in the decreasing order as $\tilde{T}_{n,j_1^*} \ge \cdots \ge \tilde{T}_{n,j_d^*}$. Restricted on $\mathcal{E}(v)$, we have
\begin{align*}
W_{n,L}\ge\sum_{\ell=1}^{L}T_{n,j_\ell^*}=&~\sqrt{M}\sum_{\ell=1}^{L}\frac{\tilde{U}_{{j_\ell^*}}}{\hat{\sigma}_{j_\ell^*}}+\sqrt{M}\sum_{\ell=1}^L\frac{\mu_{j_\ell^*}}{\hat{\sigma}_{j_\ell^*}} \\
\ge&~\sqrt{M}\sum_{\ell=1}^{L}\frac{\tilde{U}_{{j_\ell^*}}}{\hat{\sigma}_{j_\ell^*}}+ \sqrt{M}\sum_{\ell=1}^{L}\frac{\mu_{j_\ell^*}}{(1+v)\sigma_{j_\ell^*}}\,.
\end{align*}
Given $\epsilon_{n}>0$ such that $\epsilon_{n}\rightarrow 0$ and $\epsilon_{n}^2\log |\mathcal{F}|\rightarrow\infty$, we select $u>0$ satisfying that $(1+v)\{1+(\log |\mathcal{F}|)^{-1}+u\}(M^{-1}n)^{1/2}=1+\epsilon_{n}$. Then $u\sim\epsilon_n$.
Due to $\max_{1\le j_{1}< j_{2}<\cdots<j_{L}\le d}\sum_{l=1}^{L}\mu_{j_{l}}/\sigma_{j_{l}}\ge  (1+\epsilon_{n})n^{-1/2}\lambda(|\mathcal{F}|,\alpha)\max_{j\in[|\mathcal{F}|]}(\tilde{v}_j^{\T}R_n\tilde{v}_j)^{1/2}$, we know
\begin{align}\label{eq:P3}
\sqrt{M}\sum_{\ell=1}^{L}\frac{\mu_{j_\ell^*}}{(1+v)\sigma_{j_\ell^*}}\ge \{1+(\log |\mathcal{F}|)^{-1}+u\}\lambda(|\mathcal{F}|,\alpha)\max_{j\in[|\mathcal{F}|]}(\tilde{v}_j^{\T}R_n\tilde{v}_j)^{1/2}\,.
\end{align}
Write $\aleph=\max_{j\in[|\mathcal{F}|]}(\tilde{v}_j^{\T}R_n\tilde{v}_j)^{1/2}$. Therefore, \eqref{eq:P2} and \eqref{eq:P3} yield that
\begin{align}\label{eq:power}
&\P_{H_1}(W_{n,L}>\hat{\mathrm{cv}}_{L,\alpha}) \notag\\
&~~~\ge \P_{H_1}\bigg\{ \sqrt{M}\sum_{\ell=1}^{L}\frac{\tilde{U}_{j_\ell^*}}{\hat{\sigma}_{j_\ell^*}} + \sqrt{M}\sum_{\ell=1}^{L}\frac{\mu_{j_\ell^*}}{(1+v)\sigma_{j_\ell^*}} >\{1+(\log |\mathcal{F}|)^{-1}\}\lambda(|\mathcal{F}|,\alpha)\aleph,~\mathcal{E}(v)\bigg\}\notag\\
&~~~\ge \P\bigg\{\sqrt{M}\sum_{\ell=1}^{L}\frac{\tilde{U}_{{j_\ell^*}}}{\hat{\sigma}_{j_\ell^*}} > -u\lambda(|\mathcal{F}|,\alpha)\aleph,~\mathcal{E}(v)\bigg\}\notag\\
&~~~\ge 1- \P\bigg\{ \sqrt{M}\sum_{\ell=1}^{L}\frac{\tilde{U}_{j_\ell^*}}{\hat{\sigma}_{j_\ell^*}} \le -u\lambda(|\mathcal{F}|,\alpha)\aleph\bigg \} - \P\{\mathcal{E}(v)^{c}\}\,.
\end{align}
For each $m\in[M]$, we define $\hat{U}_{m}^{\dag}:=\{\tilde{v}^{\T}_{1}\hat{D}_n^{-1/2}(U_{m}-\mu),\ldots,\tilde{v}^{\T}_{|\mathcal{F}|}\hat{D}_n^{-1/2}(U_{m}-\mu)\}^{\T}$. Write $S_{\hat{U}^\dag}=M^{-1/2}\sum_{m=1}^M\hat{U}_m^\dag=:(S_{\hat{U}^\dag,1},\ldots,\,S_{\hat{U}^\dag,|\mathcal{F}|})^{\T}$. Based on the definition of $(j_1^*,\ldots,j_d^*)$, there exists some $j_*\in[|\mathcal{F}|]$ such that $\sqrt{M}\sum_{\ell=1}^L\tilde{U}_{j_\ell^*}/\hat{\sigma}_{j_\ell^*}=S_{\hat{U}^\dag,j_*}$. Since $(\tilde{v}_{j_*}^{\T}R_n\tilde{v}_{j_*})^{-1/2}S_{\hat{U}^\dag,j_*}\rightarrow\mathcal{N}(0,1)$ in distribution and $u\lambda(|\mathcal{F}|,\alpha)\aleph(\tilde{v}_{j_*}^{\T}R_n\tilde{v}_{j_*})^{-1/2}\gtrsim \epsilon_n(\log |\mathcal{F}|)^{1/2}\rightarrow\infty$, then
\begin{align*}
\P\bigg\{ \sqrt{M}\sum_{\ell=1}^{L}\frac{\tilde{U}_{j_\ell^*}}{\hat{\sigma}_{j_\ell^*}} \le -u\lambda(|\mathcal{F}|,\alpha)\aleph\bigg \}=o(1)\,.
\end{align*}
Together with \eqref{eq:power}, we have $\P_{H_1}(W_{n,L}>\hat{\mathrm{cv}}_{L,\alpha})\rightarrow1$ as $n\rightarrow\infty$. We complete the proof of Theorem 2. $\hfill\Box$

\section{Proof of Theorem 3}

Let $\xi^{(q)} = \{\xi_1^{(q)},\ldots,\xi_{|\mathcal{K}_q|}^{(q)}\}^{\T} \sim \mathcal{N}\{0,R_n^{(q)}\}$ and write $\mathcal{X}_{n}=\{ X_1,\ldots,X_n \}$. Recall $V_{n,L}^{(q)}=\Phi^{-1}\{1-\mathrm{pv}_{L}^{(q)}\}$ with $\mathrm{pv}_{L}^{(q)} = \P[ f_{L}\{\hat{\xi}^{(q)}\} \ge W_{n,L}^{(q)}\, |\, \mathcal{X}_n ]$.
Let $F_q(z) = \P\{W_{n,L}^{(q)}\leq z\}$ for any $z\in \mathbb{R}$.

We first show that $\P\{V_{n,L}^{(q)} \ge t\} = 1-\Phi(t) + o(1)$ for any $q\in\mathcal{H}_0$. Identical to \eqref{eq:f_L_1}, we have
	\begin{align*}
	\sup_{z\in\mathbb{R}}\big|\mathbb{P}\{W_{n,L}^{(q)}>z\}-\mathbb{P}[f_L\{\xi^{(q)}\}>z]\big|\leq&~C_*|\mathcal{K}_q|^{2}B\exp\{-CB^{-4/3}n^{1/3}L^{2/3}(\log |\mathcal{K}_q|)^{2/3}\}\notag\\
	&+C_*|\mathcal{K}_q|^2B\exp\{-CB^{-7/12}n^{1/3}L^{1/6}(\log |\mathcal{K}_q|)^{1/6}\}\\
	&+\frac{C_*B^{2/3}L^{13/6}(\log |\mathcal{K}_q|)^{7/6}}{n^{1/6}}\notag
	\end{align*}
for any $q\in\mathcal{H}_0$, where $C_*$ is independent of $q$.
Given $W_{n,L}^{(q)}$ with $q\in\mathcal{H}_0$, define the event
\begin{align*}
\mathcal{E}_{n,1}^{(q)}=\big\{ \big|1-F_q\{W_{n,L}^{(q)}\}-\mathbb P\big[f_L\{\hat{\xi}^{(q)}\}> W_{n,L}^{(q)}\,|\,\mathcal{X}_n\big]\big| \le \bar{C}_1\mathcal{I}_{n}^{(q)} \big\}
\end{align*}
for some sufficiently large constant $\bar{C}_1>0$ independent of $q$, where
	\begin{align*}
	\mathcal{I}_{n}^{(q)}&= \Delta_{n,q}^{1/3}\{1\vee \log(|\mathcal{K}_q|^{L}\Delta_{n,q}^{-1})\}^{2/3}+|\mathcal{K}_q|^{2}B\exp\{-CB^{-4/3}n^{1/3}L^{2/3}(\log |\mathcal{K}_q|)^{2/3}\}\\
	&~~~~~~+|\mathcal{K}_q|^{2}B\exp\{-CB^{-7/12}n^{1/3}L^{1/6}(\log |\mathcal{K}_q|)^{1/6}\}+\frac{B^{2/3}L^{13/6}(\log |\mathcal{K}_q|)^{7/6}}{n^{1/6}}
	\end{align*}
and $\Delta_{n,q}$ is defined in the same manner of $\Delta_n$ appeared in the proof of Proposition 1 but with replacing $\mathcal{F}$, $R_n$ and $\hat{R}_n$ by $\mathcal{F}_q$, $R_n^{(q)}$ and $\hat{R}_n^{(q)}$, respectively. Write $\Gamma_n^{(q)}=\{D_n^{(q)}\}^{-1/2}$ and $\hat{\Gamma}_{n}^{(q)}= \{\hat{D}_n^{(q)}\}^{-1/2}$. Identical to \eqref{eq:tm3pf3}, it holds that
\begin{align*}
\Delta_{n,q}\le&~\max_{\check{v}_{1},\check{v}_{2}\in\mathcal{F}_q}|\check{v}_{1}^{\T}\Gamma_n^{(q)}\Sigma_n^{(q)}\{\Gamma_n^{(q)}-\hat{\Gamma}_n^{(q)}\}\check{v}_{2}| + \max_{\check{v}_{1},\check{v}_{2}\in\mathcal{F}_q}|\check{v}_{1}^{\T}\Gamma_n^{(q)}\{\Sigma_n^{(q)}-\hat{\Sigma}_n^{(q)}\}\hat{\Gamma}_n^{(q)}\check{v}_{2}| \\
&+ \max_{\check{v}_{1},\check{v}_{2}\in\mathcal{F}_q}|\check{v}_{1}^{\T}\{\Gamma_n^{(q)}-\hat{\Gamma}_n^{(q)}\}\hat{\Sigma}_n^{(q)}\hat{\Gamma}_n^{(q)}\check{v}_{2}|\\
\leq&~\max_{\check{v}_1\in\mathcal{F}_q}|\Gamma_n^{(q)}\check{v}_1|_1\cdot|\Sigma_n^{(q)}|_\infty\cdot\max_{\check{v}_2\in\mathcal{F}_q}|\{\Gamma_n^{(q)}-\hat{\Gamma}_n^{(q)}\}\check{v}_{2}|_1\\
&+\max_{\check{v}_1\in\mathcal{F}_q}|\Gamma_n^{(q)}\check{v}_1|_1\cdot|\hat{\Sigma}_n^{(q)}-\Sigma_n^{(q)}|_\infty\cdot\max_{\check{v}_2\in\mathcal{F}_q}|\hat{\Gamma}_n^{(q)}\check{v}_{2}|_1\\
&+\max_{\check{v}_1\in\mathcal{F}_q}|\{\hat{\Gamma}_n^{(q)}-\Gamma_n^{(q)}\}\check{v}_1|_1\cdot|\hat{\Sigma}_n^{(q)}|_\infty\cdot\max_{\check{v}_2\in\mathcal{F}_q}|\hat{\Gamma}_n^{(q)}\check{v}_{2}|_1\,.
\end{align*}
Define $\mathcal{E}=\{\min_{l\in[d]}\hat{\sigma}_l>c_1/2,|\hat{\Sigma}_n-\Sigma_n|_\infty\leq1\}$. Restricted on $\mathcal{E}$, we have $\Delta_{n,q}\leq \bar{C}_2BL^2|\hat{\Gamma}_n^{(q)}-\Gamma_n^{(q)}|_\infty+\bar{C}_2L^2|\hat{\Sigma}_n^{(q)}-\Sigma_n^{(q)}|_\infty$ for some universal constant $\bar{C}_2>0$ independent of $q$. By \eqref{eq:covestbd1} and \eqref{eq:tm3p4},
	\begin{align*}
	\mathbb{P}\bigg(\max_{q\in\mathcal{H}_0}\Delta_{n,q}>\epsilon\bigg)\leq&~\mathbb{P}(\mathcal{E}^c)+\sum_{q\in\mathcal{H}_0}\mathbb{P}\big\{\bar{C}_2BL^2|\hat{\Gamma}_n^{(q)}-\Gamma_n^{(q)}|_\infty+\bar{C}_2L^2|\hat{\Sigma}_n^{(q)}-\Sigma_n^{(q)}|_\infty>\epsilon\big\}\\
	\lesssim&~\sum_{q\in\mathcal{H}_0}\mathbb{P}\bigg\{\bar{C}_2BL^2|\hat{\Gamma}_n^{(q)}-\Gamma_n^{(q)}|_\infty>\frac{\epsilon}{2}\bigg\}+\sum_{q\in\mathcal{H}_0}\mathbb{P}\bigg\{\bar{C}_2L^2|\hat{\Sigma}_n^{(q)}-\Sigma_n^{(q)}|_\infty>\frac{\epsilon}{2}\bigg\}\\
	&~+d^2B\exp(-CB^{-3}n)+d^2B\exp(-CB^{-1}n ^{1/2})
	\end{align*}
for any $\epsilon>0$. Write $K_{\max}=\max_{q\in\mathcal{H}_0}|\mathcal{K}_q|$. Identical to \eqref{eq:covestbd1} and \eqref{eq:tm3p5}, we have
	\begin{align*}
	\max_{q\in\mathcal{H}_0}\mathbb{P}\bigg\{\bar{C}_2BL^2|\hat{\Gamma}_n^{(q)}-\Gamma_n^{(q)}|_\infty>\frac{\epsilon}{2}\bigg\}\lesssim&~K_{\max}^2B\exp(-CB^{-5}L^{-4}n \epsilon^2)\\
	&+K_{\max}^2B\exp(-CB^{-3/2}L^{-1}n ^{1/2}\epsilon^{1/2})\,,\\
	\max_{q\in\mathcal{H}_0}\mathbb{P}\bigg\{\bar{C}_2L^2|\hat{\Sigma}_n^{(q)}-\Sigma_n^{(q)}|_\infty>\epsilon\bigg\}\lesssim&~K_{\max}^2B\exp(-CB^{-3}L^{-4}n \epsilon^2)\\
	&+K_{\max}^2B\exp(-CB^{-1}L^{-1}n ^{1/2}\epsilon^{1/2})
	\end{align*}
for any $\epsilon=o(1)$. Hence,
	\begin{align*}
	\mathbb{P}\bigg(\max_{q\in\mathcal{H}_0}\Delta_{n,q}>\epsilon\bigg)&\lesssim d^2B\exp(-CB^{-3}n)+d^2B\exp(-CB^{-1}n ^{1/2})\\
	&~~~~+Q_0K_{\max}^2B\exp(-CB^{-5}L^{-4}n \epsilon^2)+Q_0K_{\max}^2B\exp(-CB^{-3/2}L^{-1}n ^{1/2}\epsilon^{1/2})
	\end{align*}
for any $\epsilon=o(1)$.
Identical to \eqref{eq:f_L_2}, we have $\P\{\mathcal{E}_{n,1}^{(q)}\}=1$ for any $q\in\mathcal{H}_0$. Thus, it holds that
\begin{align} \label{eq:Normal}
\P\big\{V_{n,L}^{(q)} \ge t \big\} =&~\P\Big( \P\big[ f_{L}\{\hat{\xi}^{(q)}\} \ge W_{n,L}^{(q)} \,|\, \mathcal{X}_n \big]  \le 1-\Phi(t) \Big)\notag\\
=&~ \P\Big(\P\big[ f_{L}\{\hat{\xi}^{(q)}\} \ge W_{n,L}^{(q)} \,|\, \mathcal{X}_n \big]  \le 1-\Phi(t), \mathcal{E}_{n,1}^{(q)} \Big) \\
\le&~ \P\Big[1-F_q\{W_{n,L}^{(q)}\}  \le 1-\Phi(t) + \bar{C}_1\mathcal{I}_{n}^{(q)} \Big] \notag\\
=&~  \P\big[  W_{n,L}^{(q)} \ge F_q^{-1}\{\Phi(t) - \bar{C}_1\mathcal{I}_{n}^{(q)} \} \big]\,.\notag
\end{align}
Given $\epsilon=o(1)$, define $\mathcal{E}_{n,2} = \{ \max_{q\in\mathcal{H}_0}\Delta_{n,q} \leq\epsilon \}$.  Recall $\mathcal{I}_{n}^{(q)} = \Delta_{n,q}^{1/3}\{1\vee \log(|\mathcal{K}_q|^{L}\Delta_{n,q}^{-1})\}^{2/3}+|\mathcal{K}_q|^{2}B\exp\{-CB^{-4/3}n^{1/3}L^{2/3}(\log |\mathcal{K}_q|)^{2/3}\}+|\mathcal{K}_q|^{2}B\exp\{-CB^{-7/12}n^{1/3}L^{1/6}(\log |\mathcal{K}_q|)^{1/6}\}+n^{-1/6}\cdot\\B^{2/3}L^{13/6}(\log |\mathcal{K}_q|)^{7/6}$. Under $\mathcal{E}_{n,2}$, we have
	\begin{align*}
	\mathcal{I}_n^{(q)}&\leq\epsilon^{1/3}\{\log(|\mathcal{K}_q|^{L}\epsilon^{-1})\}^{2/3}+|\mathcal{K}_q|^{2}B\exp\{-CB^{-4/3}n^{1/3}L^{2/3}(\log |\mathcal{K}_q|)^{2/3}\}\\
	&~~~~~~+|\mathcal{K}_q|^{2}B\exp\{-CB^{-7/12}n^{1/3}L^{1/6}(\log |\mathcal{K}_q|)^{1/6}\}+\frac{B^{2/3}L^{13/6}(\log |\mathcal{K}_q|)^{7/6}}{n^{1/6}}\\
	&=:\Omega_q\,.
	\end{align*}
Then
	\begin{align} \label{eq:Normal2}
	&\P\big[  W_{n,L}^{(q)} \ge F_q^{-1}\{\Phi(t) - \bar{C}_1\mathcal{I}_{n}^{(q)} \} \big]\notag\\
	&~~~~\le\P\big[  W_{n,L}^{(q)} \ge F_q^{-1}\{\Phi(t) - \bar{C}_1\mathcal{I}_{n}^{(q)} \}, \mathcal{E}_{n,2} \big] + \P(\mathcal{E}_{n,2}^c) \notag\\
	&~~~~\le\P\big[  W_{n,L}^{(q)} \ge F_q^{-1}\{\Phi(t) - \bar{C}_1\Omega_q \} \big]\notag+\bar{C}_3d^2B\exp(-CB^{-3}n)+\bar{C}_3d^2B\exp(-CB^{-1}n ^{1/2})\notag\\
	&~~~~~~~+\bar{C}_3Q_0K_{\max}^2B\exp(-CB^{-5}L^{-4}n \epsilon^2)+\bar{C}_3Q_0K_{\max}^2B\exp(-CB^{-3/2}L^{-1}n ^{1/2}\epsilon^{1/2})\\
	&~~~~\leq1-\Phi(t)+\bar{C}_1\Omega_q+\bar{C}_3d^2B\exp(-CB^{-3}n)+\bar{C}_3d^2B\exp(-CB^{-1}n ^{1/2})\notag\\ &~~~~~~~+\bar{C}_3Q_0K_{\max}^2B\exp(-CB^{-5}L^{-4}n \epsilon^2)+\bar{C}_3Q_0K_{\max}^2B\exp(-CB^{-3/2}L^{-1}n ^{1/2}\epsilon^{1/2})\notag
	\end{align}
where $\bar{C}_3>0$ is a universal constant independent of $q$ and $t$.
Likewise, we also have
	\begin{align*}
	\P\big\{ V_{n,L}^{(q)}\geq t\big\}=&~ \P\Big(\P\big[ f_{L}\{\hat{\xi}^{(q)}\} \ge W_{n,L}^{(q)} \,|\, \mathcal{X}_n \big]  \le 1-\Phi(t), \mathcal{E}_{n,1}^{(q)} \Big) \\
	\geq&~\P\Big( 1-F_q\{W_{n,L}^{(q)}\}  \le 1-\Phi(t) -  \bar{C}_1\mathcal{I}_{n}^{(q)} \Big)\\
	=&~\P\big[W_{n,L}^{(q)}\geq F_q^{-1}\{\Phi(t) + \bar{C}_1\mathcal{I}_n^{(q)}\} \big] \\
	\ge&~1-\Phi(t)-\bar{C}_1\Omega_q-\bar{C}_3d^2B\exp(-CB^{-3}n)-\bar{C}_3d^2B\exp(-CB^{-1}n ^{1/2})\notag\\ &-\bar{C}_3Q_0K_{\max}^2B\exp(-CB^{-5}L^{-4}n \epsilon^2)-\bar{C}_3Q_0K_{\max}^2B\exp(-CB^{-3/2}L^{-1}n ^{1/2}\epsilon^{1/2})\notag\,.
	\end{align*}
	Thus, if $B^4L^{-2}\log K_{\max}=o(n)$ and $B^{7/4}L^{-1/2}(\log K_{\max})^{5/2}=o(n)$, we have
	\begin{align}\label{eq:tm3delta}
	&\sup_{t\in\mathbb{R
	}}\max_{q\in\mathcal{H}_0}|\mathbb{P}\{V_{n,L}^{(q)}\geq t\}-\{1-\Phi(t)\}|\notag\\
	&~~~~\lesssim\epsilon^{1/3}\{\log(K_{\max}^{L}\epsilon^{-1})\}^{2/3}+B\exp(-CB^{-4/3}n^{1/3}L^{2/3})\notag\\
	&~~~~~~~+B\exp(-CB^{-7/12}n^{1/3}L^{1/6})+n^{-1/6}B^{2/3}L^{13/6}(\log K_{\max})^{7/6}\\
	&~~~~~~~+\bar{C}_3d^2B\exp(-CB^{-3}n)+\bar{C}_3d^2B\exp(-CB^{-1}n ^{1/2})\notag\\ &~~~~~~~+\bar{C}_3QK_{\max}^2B\exp(-CB^{-5}L^{-4}n \epsilon^2)+\bar{C}_3QK_{\max}^2B\exp(-CB^{-3/2}L^{-1}n ^{1/2}\epsilon^{1/2})\notag\\
	&~~~~=:\delta(\epsilon)\notag
	\end{align}
for any $\epsilon=o(1)$.

Let $G(t) = 1-\Phi(t)$ and $t_{\max} = (2\log Q - 2\log\log Q)^{1/2}$. We consider two cases: (i) there exists $t\in [0,t_{\max}]$ such that $\widehat{{\rm FDP}}(t)\leq\alpha$, and (ii) $\widehat{{\rm FDP}}(t)>\alpha$ for any $t\in [0,t_{\max}]$. For Case (i), by the definition of $\hat{t}$, it holds that $\widehat{\mathrm{FDP}}(t) > \alpha$ for any $t < \hat{t}$. Notice that $I\{V_{n,L}^{(q)} > t \} \ge I\{V_{n,L}^{(q)} > \hat{t}\}$ for $t < \hat{t}$.
Thus,
\begin{align*}
\frac{Q G(t)}{1 \vee \sum_{q\in [Q]}I\{V_{n,L}^{(q)} \ge \hat{t}\}} \geq \frac{Q G(t)}{1 \vee \sum_{q\in [Q]}I\{V_{n,L}^{(q)} \ge t\}}=\widehat{\mathrm{FDP}}(t)> \alpha\,. 
\end{align*}
Letting $t \uparrow \hat{t}$ in the numerator of the first term in above inequality, we have $\widehat{\mathrm{FDP}}(\hat{t}) \ge \alpha$. By the definition of $\hat{t}$, there exists a sequence $\{ t_i\}$ with $t_i \ge \hat{t}$ and $t_i \to \hat{t}$ such that $\widehat{\mathrm{FDP}}(t_i) \le \alpha$. Due to the fact that $I\{V_{n,L}^{(q)} > \hat{t}\} \ge I\{V_{n,L}^{(q)} > t_i\}$, it implies that
\begin{align*}
\frac{Q G(t_i)}{1 \vee \sum_{q\in [Q]}I\{V_{n,L}^{(q)} > \hat{t}\}}\leq \frac{Q G(t_i)}{1 \vee \sum_{q\in [Q]}I\{V_{n,L}^{(q)} > t_i\}} \le \alpha\,. 
\end{align*}
Letting $t_i \downarrow \hat{t}$ in the numerator of the first term in above inequality, we have $\widehat{\mathrm{FDP}}(\hat{t}) \le \alpha$. Thus, it holds that $\widehat{\mathrm{FDP}}(\hat{t}) = \alpha$ in Case (i). For Case (ii), by \eqref{eq:tm3delta}, we have $\hat{t} = (2\log Q)^{1/2}$ and
\begin{align*}
\P\bigg[ \sum_{q \in \mathcal{H}_0}I\{ V_{n,L}^{(q)} \ge \hat{t} \}\geq1 \bigg] \le&~ Q_0 \max_{q \in \mathcal{H}_0}\P\{ V_{n,L}^{(q)} \ge \hat{t} \} \\
\le&~ Q_0 G\{(2\log Q)^{1/2} \} +Q_0\delta(\epsilon)\\
\lesssim&~o(1)+Q\delta(\epsilon)
\end{align*}
as $Q\rightarrow\infty$, where $\delta(\epsilon)$ is given in \eqref{eq:tm3delta}. 
If $Q^{12}B^5L^4\log(nBK_{\max})\max\{(L\log K_{\max})^4,(\log n)^4\}=o(n)$ and $Q^{12}B^4L^{13}(\log K_{\max})^7=o(n)$, we can select suitable $\epsilon=o(1)$ such that $Q^{2}\delta(\epsilon)\rightarrow0$ as $n\rightarrow\infty$, which implies that $\mathbb{P}\{{\rm FDP}(\hat{t})=0\}\rightarrow1$ in Case (ii).  In the sequel, we will assume $Q^{2}\delta(\epsilon)\rightarrow0$ as $n\rightarrow\infty$. Then \begin{align}
\mathbb{P}\bigg\{{\rm FDP}(\hat{t})\leq\frac{\alpha Q_0}{Q}+\varepsilon\bigg\}=&~\mathbb{P}\bigg\{{\rm FDP}(\hat{t})\leq\frac{\alpha Q_0}{Q}+\varepsilon,~{\rm Case~(i)~holds}\bigg\}\notag\\
&+\mathbb{P}\bigg\{{\rm FDP}(\hat{t})\leq\frac{\alpha Q_0}{Q}+\varepsilon,~{\rm Case~(ii)~holds}\bigg\}\notag\\
=&~\mathbb{P}\big\{{\rm Case~(i)~holds}\big\}+\mathbb{P}\big\{{\rm Case~(ii)~holds}\big\}\label{eq:tm3p1}\\
&-\mathbb{P}\bigg\{{\rm FDP}(\hat{t})>\frac{\alpha Q_0}{Q}+\varepsilon,~{\rm Case~(i)~holds}\bigg\}\notag\\
=&~1-\mathbb{P}\bigg\{{\rm FDP}(\hat{t})>\frac{\alpha Q_0}{Q}+\varepsilon,~{\rm Case~(i)~holds}\bigg\}\,.\notag
\end{align}
Notice that $\widehat{{\rm FDP}}(\hat{t})=\alpha$ and $\hat{t}\in [0,t_{\max}]$ under Case (i).
If we can show
\begin{align}\label{eq:tm3p2}
\sup_{t \in [0,t_{\max}]}\bigg| \frac{ \mathrm{FDP}(t)}{\widehat{\mathrm{FDP}}(t)} - \frac{Q_0}{Q} \bigg| \to 0~~\mbox{in probability}
\end{align}
as $n,Q\rightarrow\infty$, then we have $\lim_{n,Q\rightarrow\infty}\mathbb{P}\{{\rm FDP}(\hat{t})\leq\alpha Q_0/Q+\varepsilon\}=1$ by \eqref{eq:tm3p1}.

In the sequel, we will show \eqref{eq:tm3p2}. Due to
\begin{align*}
\sup_{t \in [0,t_{\max}]}\bigg| \frac{ \mathrm{FDP}(t) }{\widehat{\mathrm{FDP}}(t)} - \frac{Q_0}{Q} \bigg| = \sup_{t \in [0,t_{\max}]}\bigg| \frac{ \sum_{q\in \mathcal{H}_0} I\{V_{n,L}^{(q)} > t\} - Q_0 G(t)  }{Q G(t)} \bigg|\,,
\end{align*}
\eqref{eq:tm3p2} is equivalent to
\begin{align*}
\sup_{t \in [0,t_{\max}]}\Bigg| \frac{ \sum_{q\in \mathcal{H}_0} [I\{V_{n,L}^{(q)} > t\} - G(t)] }{Q G(t)} \Bigg|  \to 0~~\mbox{in probability}
\end{align*}
as $n,Q\rightarrow\infty$. Let $0 = t_0 < t_1 < \cdots < t_s = t_{\max}$, where $t_i - t_{i-1} = v_*$ for $1 \le i \le s-1$ and $t_s - t_{s-1} \le v_*$ with $v_* = \{(\log Q)(\log\log Q)^{1/2}\}^{-1/2}$ and $s \sim t_{\max}/v_*$. For any $t\in[t_{i-1},t_i]$, we have
\begin{align*}
\frac{ \sum_{q\in \mathcal{H}_0} I\{V_{n,L}^{(q)} > t_{i}\} }{Q G(t_{i})}\frac{G(t_{i})}{G(t_{i-1})}  \le \frac{ \sum_{q\in \mathcal{H}_0} I\{V_{n,L}^{(q)} > t\} }{Q G(t)} \le \frac{ \sum_{q\in \mathcal{H}_0} I\{V_{n,L}^{(q)} > t_{i-1}\} }{Q G(t_{i-1})}\frac{G(t_{i-1})}{G(t_i)}\,.
\end{align*}
Notice that there exists a universal constant $\bar{C}_4>0$ such that $e^{-t^2/2}\leq \max(\bar{C}_4,2t)\int_t^\infty e^{-x^2/2}\,{\rm d}x$ for any $t>0$. Then
\begin{align*}
0<1-\frac{G(t_i)}{G(t_{i-1})}=\frac{\int_{t_{i-1}}^{t_i}e^{-x^2/2}\,{\rm d}x}{\int_{t_{i-1}}^\infty e^{-x^2/2}\,{\rm d}x}\leq\frac{v_*e^{-t_{i-1}^2/2}}{\int_{t_{i-1}}^\infty e^{-x^2/2}\,{\rm d}x}\leq v_*\max(\bar{C}_4,2t_{i-1})\,,
\end{align*}
which implies that
$
\max_{1\leq i\leq s}|1-G(t_i)/G(t_{i-1})|\leq 2v_*t_{\max}\rightarrow0$
as $Q\rightarrow\infty$. Thus, to prove \eqref{eq:tm3p2}, it suffices to show that
\begin{align*}
\max_{0 \le i \le s} \Bigg| \frac{ \sum_{q\in \mathcal{H}_0} [I\{V_{n,L}^{(q)} > t_i\} - G(t_i)] }{Q G(t_i)} \Bigg|  \to 0 ~~\mbox{in probability}
\end{align*}
as $n,Q\rightarrow\infty$. For any $\varepsilon > 0$, by the Bonferroni inequality and the Markov's inequality, we have
\begin{align}\label{eq:tm3p3}
& \P\bigg(\max_{0 \le i \le s} \bigg| \frac{ \sum_{q\in \mathcal{H}_0} [I\{V_{n,L}^{(q)} > t_i\} - G(t_i)] }{Q G(t_i)} \bigg| \ge \varepsilon \bigg)\notag \\
&~~~~~~~~~~\le \sum_{i=0}^{s}\P \bigg( \bigg| \frac{ \sum_{q\in \mathcal{H}_0} [I\{V_{n,L}^{(q)} > t_i\} - G(t_i)] }{Q G(t_i)} \bigg| \ge \varepsilon \bigg)\\
&~~~~~~~~~~\le \frac{\bar{C}_5}{v_*}\int_{0}^{t_{\max}} \P \bigg(\bigg| \frac{ \sum_{q\in \mathcal{H}_0} [I\{V_{n,L}^{(q)} > t\} - G(t)] }{Q G(t)} \bigg| \ge \varepsilon \bigg)\,\mathrm{d}t \notag\\
&~~~~~~~~~~\leq \frac{\bar{C}_5}{v_*\varepsilon^2}\int_{0}^{t_{\max}}\mathbb{E}\bigg(\bigg| \frac{ \sum_{q\in \mathcal{H}_0} [I\{V_{n,L}^{(q)} > t\} - G(t)] }{Q G(t)} \bigg|^2\bigg)\,{\rm d}t \notag \end{align}
for some constant $\bar{C}_5>0$, where the second step is based on the relationship between integration and its associated Riemann sum.
Notice that
\begin{align}\label{eq:FDP2}
&\mathbb{E}\bigg(\bigg| \frac{ \sum_{q\in \mathcal{H}_0} [I\{V_{n,L}^{(q)} > t\} - G(t)] }{Q G(t)} \bigg|^2\bigg)\notag\\
&~~~~~~~~~\leq2\mathbb{E}\bigg(\bigg| \frac{ \sum_{q\in \mathcal{H}_0} [I\{V_{n,L}^{(q)} > t\} - \mathbb{P}\{V_{n,L}^{(q)} > t\}] }{Q G(t)} \bigg|^2\bigg)\notag\\
&~~~~~~~~~~~~~+2\mathbb{E}\bigg(\bigg| \frac{ \sum_{q\in \mathcal{H}_0} [\mathbb{P}\{V_{n,L}^{(q)} > t\}-G(t)] }{Q G(t)} \bigg|^2\bigg)\\
&~~~~~~~~~= 2\sum_{q,q' \in \mathcal{H}_0} \frac{\P\{ V_{n,L}^{(q)} > t, V_{n,L}^{(q')} > t\} - \P\{ V_{n,L}^{(q)} > t\}\P\{ V_{n,L}^{(q')} > t\}}{Q^2G^2(t)}\notag\\
&~~~~~~~~~~~~~+2\mathbb{E}\bigg(\bigg| \frac{ \sum_{q\in \mathcal{H}_0} [\mathbb{P}\{V_{n,L}^{(q)} > t\}-G(t)] }{Q G(t)} \bigg|^2\bigg)\,.\notag
\end{align}
By \eqref{eq:tm3delta}, we have
\begin{align*}
\mathbb{E}\bigg(\bigg| \frac{ \sum_{q\in \mathcal{H}_0} [\mathbb{P}\{V_{n,L}^{(q)} > t\}-G(t)] }{Q G(t)} \bigg|^2\bigg) \lesssim \bigg\{\frac{Q_0\delta(\epsilon)}{QG(t)}\bigg\}^2\leq\frac{\delta^2(\epsilon)}{G^2(t)}\,.
\end{align*}
Notice that $\int_0^{t_{\max}}\{G(t)\}^{-2}\,{\rm d}t\lesssim t_{\max}\exp(t_{\max}^{2})\lesssim Q^2(\log Q)^{-3/2}$. Recall $v_* = \{(\log Q)(\log\log Q)^{1/2}\}^{-1/2}$. Then
\begin{align*}
\int_0^{t_{\max}}\mathbb{E}\bigg(\bigg| \frac{ \sum_{q\in \mathcal{H}_0} [\mathbb{P}\{V_{n,L}^{(q)} > t\}-G(t)] }{Q G(t)} \bigg|^2\bigg) \,{\rm d}t \lesssim \int_0^{t_{\max}}\frac{\delta^2(\epsilon)}{G^2(t)}\,{\rm d}t \lesssim \frac{Q^2\delta^2(\epsilon)}{(\log Q)^{3/2}} = o(v_*)\,.
\end{align*}
By \eqref{eq:tm3p3} and \eqref{eq:FDP2}, to prove \eqref{eq:tm3p2}, it suffices to show that
\begin{align}\label{eq:toprove5}
\int_0^{t_{\max}}\sum_{q,q' \in \mathcal{H}_0} \frac{\P\{ V_{n,L}^{(q)} > t, V_{n,L}^{(q')} > t\} - \P\{ V_{n,L}^{(q)} > t\}\P\{ V_{n,L}^{(q')} > t\}}{Q^2G^2(t)} \,{\rm d}t  = o(v_*)\,.
\end{align}
Define $\mathcal{H}_{01} = \{ (q, q'): q,q'\in \mathcal{H}_0, q = q' \}$, $\mathcal{H}_{02} = \{ (q,q'):q, q' \in \mathcal{H}_0, q \neq q', q \in \mathcal{A}_{q'}(\gamma)$ or $q' \in \mathcal{A}_{q}(\gamma) \}$ and $\mathcal{H}_{03} = \{(q,q'):q,q'\in\mathcal{H}_{0}\}\backslash (\mathcal{H}_{01} \cup \mathcal{H}_{02})$. Then $\sum_{q,q'\in\mathcal{H}_0}=\sum_{(q,q')\in\mathcal{H}_{01}}+\sum_{(q,q')\in\mathcal{H}_{02}}+\sum_{(q,q')\in\mathcal{H}_{03}}$.

When $(q,q')\in\mathcal{H}_{01}$, $\P\{ V_{n,L}^{(q)} > t, V_{n,L}^{(q')} > t\} - \P\{ V_{n,L}^{(q)} > t\}\P\{ V_{n,L}^{(q')} > t\}=\P\{ V_{n,L}^{(q)} > t\} [1- \P\{ V_{n,L}^{(q)} > t\}]\leq \P\{ V_{n,L}^{(q)} > t\}$. By \eqref{eq:tm3delta},
\begin{align*}
\sum_{(q,q')\in \mathcal{H}_{01}}\frac{\P\{ V_{n,L}^{(q)} > t, V_{n,L}^{(q')} > t\} - \P\{ V_{n,L}^{(q)} > t\}\P\{ V_{n,L}^{(q')} > t\} }{Q^2G^2(t)} \lesssim \frac{1}{Q G(t)}+\frac{\delta(\epsilon)}{QG^2(t)}\,.
\end{align*}
Due to $\int_0^{t_{\max}}\{G(t)\}^{-1}\,{\rm d}t\lesssim \exp(2^{-1}t_{\max}^2)=Q(\log Q)^{-1}$ and $\int_0^{t_{\max}}\{G(t)\}^{-2}\,{\rm d}t\lesssim t_{\max}\exp(t_{\max}^{2})\lesssim Q^2(\log Q)^{-3/2}$, we have
\begin{align}
&\int_0^{t_{\max}}\sum_{(q,q')\in \mathcal{H}_{01}}\frac{\P\{ V_{n,L}^{(q)} > t, V_{n,L}^{(q')} > t\} - \P\{ V_{n,L}^{(q)} > t\}\P\{ V_{n,L}^{(q')} > t\} }{Q^2G^2(t)}\,{\rm d}t\notag\\
&~~~~~~~~~~\lesssim \int_0^{t_{\max}}\frac{{\rm d}t}{QG(t)}+\delta(\epsilon) \int_0^{t_{\max}}\frac{{\rm d}t}{QG^2(t)}\lesssim \frac{1}{\log Q}+\frac{Q\delta(\epsilon)}{(\log Q)^{3/2}}=o(v_*)\,.\label{eq:toprove51}
\end{align}

Analogous to \eqref{eq:Normal} and \eqref{eq:Normal2}, it holds that
	\begin{align}\label{eq:Nquantile}
	\P\big\{ V_{n,L}^{(q)} > t,  V_{n,L}^{(q')} > t \big\} \leq&~\P\big[  W_{n,L}^{(q)} \ge F_q^{-1}\{\Phi(t) - \bar{C}_1\mathcal{I}_{n}^{(q)} \} , W_{n,L}^{(q')} \ge F_{q'}^{-1}\{\Phi(t) - \bar{C}_1\mathcal{I}_{n}^{(q')} \} \big]\notag\\
	\leq&~\P\big[  W_{n,L}^{(q)} \ge F_q^{-1}\{\Phi(t) - \bar{C}_1\Omega_q \} , W_{n,L}^{(q')} \ge F_{q'}^{-1}\{\Phi(t) - \bar{C}_1\Omega_{q'} \} \big]\notag\\
	&~+\bar{C}_3d^2B\exp(-CB^{-3}n)+\bar{C}_3 d^2B\exp(-CB^{-1}n ^{1/2})\notag\\ &~+\bar{C}_3 Q_0K_{\max}^2B\exp(-CB^{-5}L^{-4}n \epsilon^2)\notag\\
	&~+\bar{C}_3Q_0K_{\max}^2B\exp(-CB^{-3/2}L^{-1}n ^{1/2}\epsilon^{1/2})\\
	\leq &~\P\big[  F_q\{W_{n,L}^{(q)}\} \ge \Phi(t) , F_{q'}\{W_{n,L}^{(q')}\} \ge \Phi(t) \big]+\bar{C}_1\Omega_q+\bar{C}_1\Omega_{q'}\notag\\
	&~+\bar{C}_3 d^2B\exp(-CB^{-3}n)+\bar{C}_3d^2B\exp(-CB^{-1}n ^{1/2})\notag\\ &~+\bar{C}_3 Q_0K_{\max}^2B\exp(-CB^{-5}L^{-4}n\epsilon^2)\notag\\
	&~+\bar{C}_3Q_0K_{\max}^2B\exp(-CB^{-3/2}L^{-1}n ^{1/2}\epsilon^{1/2})\notag\\
	\leq&~ \P\big[  F_q\{W_{n,L}^{(q)}\} \ge \Phi(t) , F_{q'}\{W_{n,L}^{(q')}\} \ge \Phi(t) \big]+\delta(\epsilon) \,,\notag
	\end{align}
where $\delta(\epsilon)$ is given in \eqref{eq:tm3delta}, and the third step is  due to the fact $F_q\{W_{n,L}^{(q)}\}$ and $F_{q'}\{W_{n,L}^{(q')}\}$ are uniformly distributed on $[0,1]$.  Notice that $\Phi^{-1}[F_q\{W_{n,L}^{(q)}\}]\sim \mathcal{N}(0,1)$ and $\Phi^{-1}[F_{q'}\{W_{n,L}^{(q')}\}]\sim\mathcal{N}(0,1)$. Define $\eta^{(q)} = \Phi^{-1}[F_{q}\{W_{n,L}^{(q)}\}]$ and $\eta^{(q')} = \Phi^{-1}[F_{q'}\{W_{n,L}^{(q')}\}]$. When $(q,q')\in\mathcal{H}_{02}$, since $\max_{q,q' \in [Q]}|\mathrm{Corr}\{\eta^{(q)}, \eta^{(q')}\}| \le r<1$,   Lemma 2 in \cite{Berman_1962} implies that $\P\{ \eta^{(q)}\ge t , \eta^{(q')} \ge t \}\lesssim t^{-2}\exp\{-t^2/(1+r)\}$ for any $t>\bar{C}_6$, where $\bar{C}_6>0$ is a universal constant. As we have mentioned above,  there exists a universal constant $\bar{C}_4>0$ such that $e^{-t^2/2}\leq \max(\bar{C}_4,2t)\int_t^\infty e^{-x^2/2}\,{\rm d}x$ for any $t>0$. Therefore, by \eqref{eq:Nquantile},
\begin{align*}
\P\{ V_{n,L}^{(q)} > t, V_{n,L}^{(q')} > t\} \lesssim&~\frac{1}{t^2}\exp\bigg(-\frac{t^2}{1+r}\bigg)+\delta(\epsilon)\\
\lesssim&~ t^{-2}\{\max(\bar{C}_4,2t)\}^{2/(1+r)}\{G(t)\}^{2/(1+r)}+\delta(\epsilon)\,.
\end{align*}
for any $t>\max\{\bar{C}_6,\bar{C}_4/2\}$.
Since $\max_{q \in [Q]}|\mathcal{A}_q(\gamma)| = o(Q^{\nu})$, we have $|\mathcal{H}_{02}| = O(Q^{1+\nu})$. Note that $\P\{ V_{n,L}^{(q)} > t, V_{n,L}^{(q')} > t\} \leq1$ for any $0<t<\bar{C}_4/2$. Together with $\nu<(1-r)/(1+r)$, we have
\begin{align}
&\int_0^{t_{\max}}\sum_{(q,q')\in\mathcal{H}_{02}}\frac{\P\{V_{n,L}^{(q)}>t,V_{n,L}^{(q')}>t\}-\P\{V_{n,L}^{(q)}>t\}\P\{V_{n,L}^{(q')}>t\}}{Q^2G^2(t)}\,{\rm d}t\notag \\ &~~~~~~~~~~ \lesssim  \int_0^{\bar{C}_4/2} \frac{{\rm d}t}{Q^{1-\nu}G^2(t)}  + \int_{\bar{C}_4/2}^{t_{\max}} \frac{{\rm d}t}{Q^{1-\nu}\{G(t)\}^{2r/(1+r)}} +\delta(\epsilon)\int_0^{t_{\max}}\frac{{\rm d}t}{Q^{1-\nu}G^2(t)}\notag\\
&~~~~~~~~~~ \lesssim \frac{1}{Q^{1-\nu}} + \frac{Q^{(r-1)/(1+r)+\nu}}{(\log Q)^{(1+3r)/(2+2r)}}+\frac{Q^{1+\nu}\delta(\epsilon)}{(\log Q)^{3/2}} =o(v_*)\,.\label{eq:toprove52}
\end{align}

Define $\rho_{q,q'}={\rm Corr}\{\eta^{(q)},\eta^{(q')}\}$. Then $|\rho_{q,q'}|\leq (\log Q)^{-2-\gamma}$ for any $(q,q') \in \mathcal{H}_{03}$. By Theorem 2.1.e of \cite{LinBai_2010},
\begin{equation}\label{eq:tm3p11}
\P\big\{ \eta^{(q)}>t, \eta^{(q')}>t \big\} \le \left\{\begin{aligned} G(t)G\bigg\{\frac{(1-\rho_{q,q'})t}{(1-\rho_{q,q'}^2)^{1/2}}\bigg\}\,, ~~~~~~~~~~&\mbox{if} ~ -1 < \rho_{q,q'} \le 0\,; \\
(1+\rho_{q,q'})G(t)G\bigg\{\frac{(1-\rho_{q,q'})t}{(1-\rho_{q,q'}^2)^{1/2}}\bigg\}\,, ~~~~&\mbox{if} ~ 0 \le \rho_{q,q'} < 1\,.\end{aligned}\right.
\end{equation}
Note that $|\rho_{q,q'}|\leq(\log Q)^{-2-\gamma}$ for any $(q,q')\in\mathcal{H}_{03}$. When $-(\log Q)^{-2-\gamma} \le \rho_{q,q'} \le 0$, due to $(1-\rho_{q,q'})/(1-\rho_{q,q'}^2)^{1/2} \ge 1$, we have $G\{(1-\rho_{q,q'})t/(1-\rho_{q,q'}^2)^{1/2}\} \le G(t)$, which implies that $\P\{ \eta^{(q)}>t, \eta^{(q')}>t \} \le G^2(t) \le \{1+(\log Q)^{-1-\gamma}\}G^2(t)$. When $0 <\rho_{q,q'} \le (\log Q)^{-2-\gamma}$, by the Taylor expansion, it holds that
\begin{align*}
G\bigg\{\frac{(1-\rho_{q,q'})t}{(1-\rho_{q,q'}^2)^{1/2}}\bigg\} = G(t)+\phi(\tilde{t})\bigg\{t-\frac{(1-\rho_{q,q'})t}{(1-\rho_{q,q'}^2)^{1/2}} \bigg\}
\end{align*}
for $(1-\rho_{q,q'})t/(1-\rho_{q,q'}^2)^{1/2}<\tilde{t}<t$, where $\phi(\cdot)$ is the density function of the standard normal distribution $\mathcal{N}(0,1)$. Then, if $\rho_{q,q'}>0$,
\begin{align*}
G\bigg\{\frac{(1-\rho_{q,q'})t}{(1-\rho_{q,q'}^2)^{1/2}}\bigg\}\{G(t)\}^{-1} = &~ 1+\frac{\phi(\tilde{t})}{G(t)}\bigg\{t-\frac{(1-\rho_{q,q'})t}{(1-\rho_{q,q'}^2)^{1/2}} \bigg\} \\
\leq &~ 1+\frac{t\phi(\tilde{t})}{t\phi(t)/(1+t^2)}\bigg\{1-\frac{(1-\rho_{q,q'})}{(1-\rho_{q,q'}^2)^{1/2}}\bigg\} \\
\leq &~ 1+\frac{\phi(\tilde{t})}{\phi(t)}\cdot2\rho_{q,q'}(1+t^2)
\end{align*}
for any $t>0$. For $0<t<t_{\max}$, it holds that
\begin{align*}
\frac{\phi(\tilde{t})}{\phi(t)}=\exp\bigg\{\frac{1}{2}(t-\tilde{t})(t+\tilde{t})\bigg\}\leq&~\exp\bigg\{t_{\max}^2\bigg(1-\sqrt{\frac{1-\rho_{q,q'}}{1+\rho_{q,q'}}}\bigg)
\bigg\}\\
\leq&~\exp(2\rho_{q,q'}t_{\max}^2)\leq \exp\{4(\log Q)^{-1-\gamma}\}\lesssim 1\,,
\end{align*}
which implies that $G\{(1-\rho_{q,q'})t/(1-\rho_{q,q'}^2)^{1/2}\} \le G(t)[1+O\{(\log Q)^{-1-\gamma}\}]$ for any $t\in[0,t_{\max}]$ if $0<\rho_{q,q'}\leq(\log Q)^{-2-\gamma}$, where the term $O\{(\log Q)^{-1-\gamma}\}$ holds uniformly over $t\in[0,t_{\max}]$. Then $(1+\rho_{q,q'})G(t)G\{(1-\rho_{q,q'})t/(1-\rho_{q,q'}^2)^{1/2}\}\leq G^2(t)[1+O\{(\log Q)^{-1-\gamma}\}]$ for any $t\in[0,t_{\max}]$ if $0<\rho_{q,q'}\leq(\log Q)^{-2-\gamma}$. Hence, by \eqref{eq:tm3p11}, when $(q,q')\in\mathcal{H}_{03}$, $\P\{ \eta^{(q)}>t, \eta^{(q')}>t \} \le [1+O\{(\log Q)^{-1-\gamma}\}]G^2(t)$ for $t \in [0,t_{\max}]$, which implies that
\begin{align*}
\max_{(q,q') \in \mathcal{H}_{03}}\P\{ V_{n,L}^{(q)} > t, V_{n,L}^{(q')} > t\} \le [1+O\{(\log Q)^{-1-\gamma}\}]G^2(t) + \delta(\epsilon)
\end{align*}
for $t \in [0,t_{\max}]$. Due to $\delta(\epsilon)\int_0^{t_{\max}}\{G(t)\}^{-2}\,{\rm d}t\lesssim Q^2\delta(\epsilon)(\log Q)^{-3/2}=o(v_*)$, it then holds that
\begin{align*}
&\int_0^{t_{\max}}\sum_{(q,q')\in \mathcal{H}_{03}}\frac{\P\{V_{n,L}^{(q)}>t,V_{n,L}^{(q')}>t\}-\P\{V_{n,L}^{(q)}>t\}\P\{V_{n,L}^{(q')}>t\} }{Q^2G^2(t)}\,{\rm d}t \\
&~~~~~~~~~~~~~ \lesssim O\{(\log Q)^{-1-\gamma}\}\cdot\int_0^{t_{\max}}\,{\rm d}t  + \int_0^{t_{\max}} \frac{\delta(\epsilon)}{G^2(t)}\,{\rm d}t =o(v_*)\,.
\end{align*}
Together with \eqref{eq:toprove51} and \eqref{eq:toprove52}, we know \eqref{eq:toprove5} holds. Then $\lim_{n,Q\rightarrow\infty}\mathbb{P}\{{\rm FDP}(\hat{t})\leq\alpha Q_0/Q+\varepsilon\}=1$. Since ${\rm FDR}(t)={\E}\{{\rm FDP}(t)\}$, it holds that $\limsup_{n,Q\to\infty}{\rm FDR}(\hat{t})\leq\alpha Q_0/Q$. We complete the proof of Theorem 3. $\hfill\Box$

\section{Proof of Proposition 2}
Let $n_{{\rm dist}}=\sum_{k=1}^{K}M_{k}$. Define $\Sigma_{n}^{{\rm dist}}=\mathrm{Cov}(n_{{\rm dist}}^{1/2}\bar{U}^{{\rm dist}})$, $D_{n}^{{\rm dist}}=\diag(\Sigma_{n}^{{\rm dist}})=\diag\{(\sigma_{1}^{{\rm dist}})^2,\ldots,(\sigma_{d}^{{\rm dist}})^2\}$, $\mathring{T}_{n}^{{\rm dist}}=n_{{\rm dist}}^{1/2}(D_{n}^{{\rm dist}})^{-1/2}\bar{U}^{{\rm dist}}$ and $\mathring{W}_{n,L}^{{\rm dist}}=f_{L}(\mathring{T}_{n}^{{\rm dist}})$. Recall $\mathcal{F} = \{v=(v_1,\ldots,v_d)^{\T}: v_j \in \{ 0,1 \}~\textrm{for each}~j \in [d],~\textrm{and}~|v|_{0}= L \}$.
Due to the countability of the set $\mathcal{F}$, we can index all the elements in $\mathcal{F}$ as $\{\tilde{v}_{1},\ldots,\tilde{v}_{|\mathcal{F}|}\}$ in a certain order. Since $U_{1}^{(1),{\rm dist}},\ldots,U_{M_1}^{(1),{\rm dist}},\ldots,U_1^{(K),{\rm dist}},\ldots,U_{M_K}^{(K),{\rm dist}}$ is a $(B-1)$-dependent sequence,
using the same arguments in the proof of Proposition 1, for any $\epsilon>0$, we have
	\begin{align}\label{eq:gaussbd_distributed}
	&\sup_{z\in \mathbb{R}}\big|\mathbb{P}(W_{n,L}^{{\rm dist}}\le z)-\mathbb P(g\le z\cdot 1_{|\mathcal{F}|})\big|\notag\\
	&~~~~~~~~~\lesssim \mathbb{P}\big(|W_{n,L}^{{\rm dist}}-\mathring{W}_{n,L}^{{\rm dist}}| >\epsilon\big)+\frac{B^{2/3}L^{13/6}(\log d)^{7/6}}{n^{1/6}}+\epsilon(L\log d)^{1/2}\,,
	\end{align}
where $g\sim \mathcal{N}(0, V^{\T}R_{n}^{{\rm dist}}V)$ with $R_{n}^{{\rm dist}}=(D_{n}^{{\rm dist}})^{-1/2}\Sigma_{n}^{{\rm dist}}(D_{n}^{{\rm dist}})^{-1/2}$ and $V=(\tilde{v}_{1},\ldots,\tilde{v}_{|\mathcal{F}|})$, provided that $BK=o(n)$.

Next, we give an upper bound for $\mathbb{P}(|W_{n,L}^{{\rm dist}}-\mathring{W}_{n,L}^{{\rm dist}}| >\epsilon)$.
Let
$
\hat{D}_{n}^{{\rm dist}}=\diag[n_{{\rm dist}}^{-1}\sum_{k=1}^{K}M_k\{\hat{\sigma}_{1}^{(k),{\rm dist}}\}^2,$ $\ldots,n_{{\rm dist}}^{-1}\sum_{k=1}^{K}M_k\{\hat{\sigma}_{d}^{(k),{\rm dist}}\}^2]=\diag\{(\hat{\sigma}_{1}^{{\rm dist}})^2,\ldots,(\hat{\sigma}_{d}^{\rm dist})^2\}$.
Define $T_{n}^{{\rm dist}}=n_{{\rm dist}}^{1/2}(\hat D_{n}^{{\rm dist}})^{-1/2}\bar{U}^{{\rm dist}}$. Then we have $W_{n,L}^{{\rm dist}}=f_{L}(T_{n}^{{\rm dist}})$.
Identical to the arguments used in the proof of Lemma \ref{la:2} for deriving the upper bound of $\mathbb{P}(|\bar{U}|_\infty>z)$, it holds under $H_{0}$ that $\mathbb{P}(|\bar{U}^{{\rm dist}}|_{\infty}>z)\lesssim d\exp(-CB^{-1}n z^2)+d\exp(-CB^{-1/2}n ^{1/2}z^{1/2})$ for any $z>0$.
Write $\{\sigma_{j}^{(k),{\rm dist}}\}^2={\rm Var}\{\sqrt{M_{k}}\bar{U}^{(k),{\rm dist}}_{j}\}$ for any $k\in [K]$ and $j\in [d]$. Then we have $(\sigma_{j}^{\rm dist})^2=n_{\rm dist}^{-1}\sum_{k=1}^{K}M_{k}\{\sigma_{j}^{(k),{\rm dist}}\}^2$ for $j\in [d]$. Due to $|(\hat{\sigma}_{j}^{\rm dist})^2-(\sigma_{j}^{\rm dist})^2|=|n_{\rm dist}^{-1}\sum_{k=1}^{K}M_{k}\{\hat{\sigma}_{j}^{(k),{\rm dist}}\}^2-n_{\rm dist}^{-1}\sum_{k=1}^{K}M_{k}\{\sigma_{j}^{(k),{\rm dist}}\}^2|\le \max_{k\in [K]}|\{\hat{\sigma}_{j}^{(k),{\rm dist}}\}^2-\{\sigma_{j}^{(k),{\rm dist}}\}^2|$ for all $j\in [d]$, we have
	\begin{align}\label{eq:sigma_dist_bound}
	\mathbb{P}\bigg\{\max_{j\in [d]}|(\hat{\sigma}_{j}^{\rm dist})^2-(\sigma_{j}^{\rm dist})^2|\ge z\bigg\}&\le 	\mathbb{P}\bigg\{\max_{j\in [d]}\max_{k\in [K]}|\{\hat{\sigma}_{j}^{(k),{\rm dist}}\}^2-\{\sigma_{j}^{(k),{\rm dist}}\}^2|\ge z\bigg\}\notag\\
	&\le \sum_{k=1}^{K}\mathbb{P}\bigg[\max_{j\in [d]}|\{\hat{\sigma}_{j}^{(k),{\rm dist}}\}^2-\{\sigma_{j}^{(k),{\rm dist}}\}^2|\ge z\bigg]\\
	&\lesssim d^2B\sum_{k=1}^{K}\exp(-CB^{-3}n_k z^2)+d^2B\sum_{k=1}^{K}\exp(-CB^{-1}n_k^{1/2}z^{1/2})\notag\\
	&\lesssim d^2BK\exp(-CB^{-3}K^{-1}n z^2)+d^2BK\exp(-CB^{-1}K^{-1/2}n^{1/2}z^{1/2})\notag
	\end{align}
for any $z=o(B)$, provided that $BK=o(n)$, where the third step is same as \eqref{eq:covestbd1}. Identical to \eqref{eq:tm3p5}, we have
	\begin{align}\label{eq:tail_inverse}
	&\mathbb{P}\bigg\{\max_{j\in[d]}|(\hat{\sigma}_{j}^{\rm dist})^{-1}-(\sigma_{j}^{\rm dist})^{-1}|\ge z\bigg\}\\&~~~~~~\lesssim d^2BK\exp(-CB^{-3}K^{-1}n z^2)+d^2BK\exp(-CB^{-1}K^{-1/2}n^{1/2}z^{1/2})\notag
	\end{align}
for any $z=o(1)$. Due to $BK=o(n)$, we know $n_{\rm dist}\asymp n$. Notice that $|W_{n,L}^{{\rm dist}}-\mathring{W}_{n,L}^{{\rm dist}}|\le L|T_{n}^{{\rm dist}}-\mathring{T}_{n}^{{\rm dist}}|_{\infty}\le Ln_{{\rm dist}}^{1/2}|\bar{U}^{{\rm dist}}|_{\infty}\max_{j\in[d]}|(\hat{\sigma}_{j}^{\rm dist})^{-1}-(\sigma_{j}^{\rm dist})^{-1}|$.  It then holds that
	\begin{align*}
	\mathbb{P}\big(|W_{n,L}^{{\rm dist}}-\mathring{W}_{n,L}^{{\rm dist}}|> \epsilon\big)&\leq \mathbb{P}\bigg\{\max_{j\in[d]}|(\hat{\sigma}_{j}^{\rm dist})^{-1}-(\sigma_{j}^{\rm dist})^{-1}|> \epsilon^{1/2}L^{-1/2}K^{1/4}n^{-1/4}B^{1/2}\bigg\}\\
	&~~~+\mathbb{P}(|\bar{U}^{\rm dist}|_{\infty}>\epsilon^{1/2} L^{-1/2}K^{-1/4}n^{-1/4}B^{-1/2})\\
	&\lesssim d^2 BK\exp(-CB^{-2}K^{-1/2}L^{-1}n^{1/2}\epsilon)\\\
	&~~~+d^2 BK\exp(-CB^{-3/4}K^{-3/8}L^{-1/4}n^{3/8}\epsilon^{1/4})
	\end{align*}
	for any $\epsilon=o(1)$, provided that $B^{2}KL^{-2}=o(n)$. Together with \eqref{eq:gaussbd_distributed}, we have
	\begin{align*}
	\sup_{z\in \mathbb{R}}\big|\mathbb{P}(W_{n,L}^{{\rm dist}}\le z)-\mathbb P(g\le z\cdot 1_{|\mathcal{F}|})\big|\lesssim&~d^2 BK\exp(-CB^{-2}K^{-1/2}L^{-1}n^{1/2}\epsilon)\notag\\
	&+d^2 BK\exp(-CB^{-3/4}K^{-3/8}L^{-1/4}n^{3/8}\epsilon^{1/4})\\
	&+\frac{B^{2/3}L^{13/6}(\log d)^{7/6}}{n^{1/6}}+\epsilon(L\log d)^{1/2}
	\end{align*}
with $g\sim \mathcal{N}(0, V^{\T}R_{n}^{{\rm dist}}V)$. Selecting $\epsilon= n^{-1/6}B^{2/3}L^{5/3}(\log d)^{2/3}=o(1)$, we have
	$
	\sup_{z\in \mathbb{R}}|\mathbb{P}(W_{n,L}^{{\rm dist}}\le z)-\mathbb P(g\le z\cdot 1_{|\mathcal{F}|})|\lesssim d^2 BK\exp\{-CB^{-4/3}K^{-1/2}L^{2/3}n^{1/3}(\log d)^{2/3}\}+n^{-1/6}B^{2/3}L^{13/6}(\log d)^{7/6}+d^2BK\exp\{-CB^{-7/12}K^{-3/8}L^{1/6}n^{1/3}(\log d)^{1/6}\}$,
which implies
\begin{align}\label{eq:gaussbd_distributed_1}
\sup_{z\in \mathbb{R}}\big|\mathbb{P}(W_{n,L}^{{\rm dist}}\le z)-\mathbb P(g\le z\cdot 1_{|\mathcal{F}|})\big|=o(1)
\end{align}
provided that $B^{4}K^{3/2}L^{-2}\{\log(dBK)\}^3 (\log d)^{-2}=o(n)$, $B^{7/4}K^{9/8}L^{-1/2}\{\log(dBK)\}^{3} (\log d)^{-1/2}=o(n)$ and $B^{4}L^{13}(\log d)^{7}=o(n)$.

Recall $\tilde{D}_n^{{\rm dist}}={\rm diag}[n_{\rm dist}^{-1}\sum_{k=1}^KM_k\{\tilde{\sigma}_1^{(k),{\rm dist}}\}^2,\ldots,n_{\rm dist}^{-1}\sum_{k=1}^KM_k\{\tilde{\sigma}_d^{(k),{\rm dist}}\}^2]$. For each $k\in[K]$, define
$
Y_{k}=\varepsilon_{k}K^{1/2}M_kn_{\rm dist}^{-1/2}(\tilde{D}_{n}^{\rm dist})^{-1/2}\{\bar{U}^{(k),{\rm dist}}-\bar{U}^{\rm dist}\}$.
Then $\mathring{\xi}=(\tilde{D}_n^{\rm dist})^{-1/2}
\sum_{k=1}^K\varepsilon_kn_{\rm dist}^{-1/2}M_k\{\bar{U}^{(k),{\rm dist}}-\bar{U}^{{\rm dist}}\}=K^{-1/2}\sum_{k=1}^{K}Y_{k}$.
Since $\varepsilon_{1},\ldots,\varepsilon_K$ is an independent Rademacher sequence, we know $Y_1,\ldots,Y_K$ is also an independent sequence with mean zero given $\mathcal{X}_{n}=\{X_{1},\ldots,X_{n}\}$. Let $M_{\max}=\max_{k\in[K]}M_k$. Conditional on $\mathcal{X}$, due to $M_{\max}\asymp nK^{-1}$ and $n_{\rm dist}\asymp n$, we have $\max_{v\in\mathcal{F}}\max_{k\in[K]}|v^{\T}Y_{k}|\leq \max_{v\in\mathcal{F}}|v|_1\cdot\max_{k\in[K]}|Y_k|_\infty\lesssim Ln^{1/2}K^{-1/2}|(\tilde{D}_{n}^{\rm dist})^{-1/2}|_{\infty}\max_{k\in [K]}|\bar{U}^{(k),{\rm dist}}-\bar{U}^{\rm dist}|_{\infty}$. Define $\hat{R}_{n}^{\rm dist}={\rm Cov}(K^{-1/2}\sum_{k=1}^{K}Y_{k}\,|\,\mathcal{X}_n)$. We will show later that $\max_{v_{1},v_{2}\in\mathcal{F}}|v_{1}(R_{n}^{\rm dist}-\hat{R}_{n}^{\rm dist})v_{2}|=o_{\p}(1)$.
Due to $\min_{v\in\mathcal{F}}v^{\T}R_{n}^{{\rm dist}}v\ge c_{2}$, we have
$\min_{v\in\mathcal{F}}v^{\T}\hat{R}_{n}^{{\rm dist}}v\ge c_{2}/2$ with probability approaching to one. Recall $V=(\tilde{v}_{1},\ldots,\tilde{v}_{|\mathcal{F}|})$ with $\mathcal{F}=\{\tilde{v}_{1},\ldots,\tilde{v}_{|\mathcal{F}|}\}$. Identical to \eqref{eq:gaussbd}, it holds with probability approaching one that
\begin{align}\label{eq:GA_dist_2}
&\sup_{y\in\mathbb{R}^{|\mathcal{F}|}} \bigg|\mathbb{P}\bigg(\frac{1}{\sqrt{K}}\sum_{k=1}^{K}V^{\T}Y_{k} \le y\,\bigg|\,\mathcal{X}_{n}\bigg)-\mathbb{P}(\hat{g}\le y\,|\,\mathcal{X}_{n})\bigg|\notag\\
&\quad\quad\quad\lesssim \frac{L^{13/6}n^{1/2}(\log d)^{7/6}}{K^{2/3}}|(\tilde{D}_{n}^{\rm dist})^{-1/2}|_{\infty}\max_{k\in [K]}|\bar{U}^{(k),{\rm dist}}-\bar{U}^{\rm dist}|_{\infty}
\end{align}
where $\hat{g}\,|\,\mathcal{X}_n\sim \mathcal{N}(0,V^{\T}\hat{R}_{n}^{\rm dist}V)$.
As we have shown that $\mathbb{P}(|\bar{U}^{{\rm dist}}|_{\infty}>z)\lesssim d\exp(-CB^{-1}n z^2)+d\exp(-CB^{-1/2}n ^{1/2}z^{1/2})$ for any $z>0$, we have $|\bar{U}^{\rm dist}|_{\infty}=O_{\p}\{n^{-1/2}B^{1/2}(\log d)^{1/2}\}$.
Analogously, we can also show $\max_{k\in[K]}\mathbb{P}\{|\bar{U}^{(k),{\rm dist}}|_{\infty}> z\}\lesssim d\exp(-CB^{-1}K^{-1}n z^2)+d\exp(-CB^{-1/2}K^{-1/2}n ^{1/2}z^{1/2})$ for any $z>0$, which implies
	\begin{align*}
	\mathbb{P}\bigg\{\max_{k\in [K]}|\bar{U}^{(k),{\rm dist}}|_{\infty}> z\bigg\}\le&\, \sum_{k=1}^{K}\mathbb{P}\{|\bar{U}^{(k),{\rm dist}}|_{\infty}> z\}\\
	\lesssim&\, dK\exp(-CB^{-1}K^{-1}n z^2)+dK\exp(-CB^{-1/2}K^{-1/2}n ^{1/2}z^{1/2})
	\end{align*}
	for any $z>0$. Then $\max_{k\in [K]}|\bar{U}^{(k),{\rm dist}}|_{\infty}=O_{\p}[n^{-1/2}K^{1/2}B^{1/2}\{\log (dK)\}^{1/2}]+O_{\p}\{n^{-1}KB(\log dK)^{2}\}$. Thus, $\max_{k\in[K]}|\bar{U}^{(k),{\rm dist}}-\bar{U}^{\rm dist}|_{\infty}\le \max_{k\in [K]}|\bar{U}^{(k),{\rm dist}}|_{\infty}+|\bar{U}^{\rm dist}|_{\infty}=O_{\p}[n^{-1/2}K^{1/2}B^{1/2}\{\log (dK)\}^{1/2}]+O_{\p}\{n^{-1}KB(\log dK)^{2}\}$. Notice that $\varepsilon_k^2=1$ for any $k\in[K]$. By the definition of $\tilde{\sigma}_j^{(k),{\rm dist}}$, we know $\tilde{\sigma}_j^{(k),{\rm dist}}=\hat{\sigma}_j^{(k),{\rm dist}}$ for any $k\in[K]$ and $j\in[d]$. As we have shown in \eqref{eq:sigma_dist_bound} that $\max_{k\in[K]}\mathbb{P}[\max_{j\in [d]}|\{\hat{\sigma}_{j}^{(k),{\rm dist}}\}^2-\{\sigma_{j}^{(k),{\rm dist}}\}^2|\ge z]\lesssim d^2B\exp(-CB^{-3}K^{-1}n z^2)+d^2B\exp(-CB^{-1}K^{-1/2}\\\cdot n^{1/2}z^{1/2})$ for any $z=o(B)$, we have $\max_{j\in [d]}\max_{k\in [K]}|\{\hat{\sigma}_{j}^{(k),{\rm dist}}\}^2-\{\sigma_{j}^{(k),{\rm dist}}\}^2|=o_{\p}(1)$ provided that $B^3 K\log (dBK)=o(n)$ and $B^2K(\log dBK)^{2}=o(n)$. Due to $\min_{j\in [d]}\min_{k\in [K]}\{\sigma_{j}^{(k),{\rm dist}}\}^2\ge c_{2}$, we have $\min_{j\in [d]}\min_{k\in [K]}\{\tilde{\sigma}_{j}^{(k),{\rm dist}}\}^2\ge c_{2}/2$ with probability approaching one, which implies that $\min_{j\in[d]}n_{\rm dist}^{-1}\sum_{k=1}^{K}M_{k}\{\tilde{\sigma}^{(k),\rm dist}_{j}\}^2\ge c_{2}/2$. Hence, $|(\tilde{D}_{n}^{\rm dist})^{-1/2}|_{\infty}=[\min_{j\in[d]}n_{\rm dist}^{-1}\sum_{k=1}^{K}M_{k}\{\tilde{\sigma}^{(k),\rm dist}_{j}\}^2]^{-1/2}=O_{\p}(1)$. By \eqref{eq:GA_dist_2}, it holds that
	\begin{align*}
	&\sup_{y\in\mathbb{R}^{|\mathcal{F}|}} \bigg|\mathbb{P}\bigg(\frac{1}{\sqrt{K}}\sum_{k=1}^{K}V^{\T}Y_{k} \le y\,\bigg|\,\mathcal{X}_{n}\bigg)-\mathbb{P}(\hat{g}\le y\,|\,\mathcal{X}_{n})\bigg|\\
	&~~~~~~=O_{\p}\bigg[\frac{B^{1/2}L^{13/6}\{\log (dK)\}^{1/2}(\log d)^{7/6}}{K^{1/6}}\bigg]+O_{\p}\bigg[\frac{BK^{1/3}L^{13/6}(\log dK)^{2}(\log d)^{7/6}}{n^{1/2}}\bigg]\,.
	\end{align*}
Together with the fact
$
\{f_L(\mathring{\xi})\le z\}=\{\max_{j\in[|\mathcal{F}|]}\tilde{v}_{j}^{\T}\mathring{\xi}\le z\}=\{K^{-1/2}\sum_{k=1}^{K}V^{\T}Y_{k}\le z\cdot 1_{|\mathcal{F}}|\}
$
for any $z\in\mathbb{R}$, we have
\begin{align}\label{eq:rdbd}
\sup_{z\in\mathbb{R}} \big|\mathbb{P}\{f_{L}(\mathring{\xi})\le z\,|\,\mathcal{X}_{n}\}-\mathbb{P}(\hat{g}\le z \cdot1_{|\mathcal{F}|}\,|\,\mathcal{X}_{n})\big|=o_{\p}(1)
\end{align}
provided that $B^{3}L^{13}\{\log(dK)\}^3(\log d)^7 =o(K)$ and $B^2K^{2/3}L^{13/3}(\log dK)^4(\log d)^{7/3}=o(n)$.

Define $\Delta_{n}^{\rm dist}=\max_{v_{1},v_{2}\in\mathcal{F}}|v_{1}^{\T}(R_{n}^{\rm dist}-\hat{R}_{n}^{\rm dist})v_{2}|$. Recall $\Sigma_{n}^{{\rm dist}}=\mathrm{Cov}(n_{{\rm dist}}^{1/2}\bar{U}^{{\rm dist}})$. We write $\Gamma_n^{\rm dist}=(D_n^{\rm dist})^{-1/2}$,  $\hat{\Gamma}_{n}^{\rm dist}= (\tilde{D}_n^{\rm dist})^{-1/2}$ and $\hat{\Sigma}_{n}^{\rm dist}=n_{\rm dist}^{-1}\sum_{k=1}^{K}M_{k}^2\{\bar{U}^{(k),{\rm dist}}-\bar{U}^{{\rm dist}}\}\{\bar{U}^{(k),{\rm dist}}-\bar{U}^{{\rm dist}}\}^{\T}$. Then $R_n^{\rm dist}=\Gamma_n^{{\rm dist}}\Sigma_n^{\rm dist}\Gamma_n^{{\rm dist}}$ and $\hat{R}_{n}^{\rm dist}={\rm Cov}(K^{-1/2}\sum_{k=1}^{K}Y_{k}\,|\,\mathcal{X}_n)=\hat{\Gamma}_{n}^{\rm dist}\hat{\Sigma}_{n}^{\rm dist}\hat{\Gamma}_{n}^{\rm dist}$.
For any $\check{v}_{1},\check{v}_{2}\in\mathcal{F}$, it follows from the triangle inequality that
\begin{align*}
|\check{v}_{1}^{\T}(R_n^{\rm dist}-\hat{R}_{n}^{\rm dist})\check{v}_{2}| &\le |\check{v}_{1}^{\T}\Gamma_n^{\rm dist}\Sigma_n^{\rm dist}(\Gamma_n^{\rm dist}-\hat{\Gamma}_n^{\rm dist})\check{v}_{2}| + |\check{v}_{1}^{\T}\Gamma_n^{\rm dist}(\Sigma_n^{\rm dist}-\hat{\Sigma}_n^{\rm dist})\hat{\Gamma}_n^{\rm dist}\check{v}_{2}|\\ &\quad\quad+ |\check{v}_{1}^{\T}(\Gamma_n^{\rm dist}-\hat{\Gamma}_n^{\rm dist})\hat{\Sigma}_n^{\rm dist}\hat{\Gamma}_n^{\rm dist}\check{v}_{2}|\\
&=: I_{1}^{\rm dist}+I_{2}^{\rm dist}+I_{3}^{\rm dist}\,.
\end{align*}
Notice that
$I_{1}^{\rm dist}\le|\Sigma_n^{\rm dist}|_\infty\max_{\check{v}_{1}\in\mathcal{F}}|\Gamma_n^{\rm dist}\check{v}_{1}|_{1}\max_{\check{v}_{2}\in\mathcal{F}}|(\Gamma_n^{\rm dist}-\hat{\Gamma}_n^{\rm dist})\check{v}_{2}|_{1}\lesssim BL^2|\Gamma_n^{\rm dist}-\hat{\Gamma}_n^{\rm dist}|_\infty$ and $\tilde{D}_n^{{\rm dist}}=\hat{D}_n^{\rm dist}$. By \eqref{eq:tail_inverse}, we have $|\Gamma_n^{\rm dist}-\hat{\Gamma}_n^{\rm dist}|_{\infty}=O_{\p}[n^{-1/2}B^{3/2}K^{1/2}\{\log(dBK)\}^{1/2}]+O_{\p}[n^{-1}B^{2}K\{\log(dBK)\}^{2}]$ provided that $B^{3}K\log(dBK)=o(n)$ and $B^{2}K\{\log(dBK)\}^2=o(n)$. Then
	\begin{align}\label{eq:I1dist}
	I_{1}^{\rm dist}= O_{\p}[n^{-1/2}B^{5/2}L^2K^{1/2}\{\log(dBK)\}^{1/2}]+O_{\p}[n^{-1}B^{3}L^2K\{\log(dBK)\}^{2}]\,.
	\end{align}
Note that $\Sigma_n^{\rm dist}=n_{\rm dist}^{-1}\sum_{k=1}^KM_k^2\mathbb{E}[\bar{U}^{(k),{\rm dist}}\{\bar{U}^{(k),{\rm dist}}\}^{\T}]$. For any $i,j\in [d]$, by the triangle inequality, we have
\begin{align}\label{eq:J123}
&\bigg|\frac{1}{n_{\rm dist}}\sum_{k=1}^{K}M_{k}^{2}\mathbb{E}\{\bar{U}^{(k),{\rm dist}}_{i}\bar{U}^{(k),{\rm dist}}_{j}\}-\frac{1}{n_{\rm dist}}\sum_{k=1}^{K}M_{k}^{2}\{\bar{U}^{(k),{\rm dist}}_{i}-\bar{U}_{i}^{{\rm dist}}\}\{\bar{U}^{(k),{\rm dist}}_{j}-\bar{U}_{j}^{{\rm dist}}\}\bigg|\notag\\
&~~~~~\le \bigg|\frac{1}{n_{\rm dist}}\sum_{k=1}^{K}M_{k}^{2}[\bar{U}^{(k),{\rm dist}}_{i}\bar{U}^{(k),{\rm dist}}_{j}-\mathbb{E}\{\bar{U}^{(k),{\rm dist}}_{i}\bar{U}^{(k),{\rm dist}}_{j}\}]\bigg|+|\bar{U}_{j}^{{\rm dist}}|\bigg|\frac{1}{n_{\rm dist}}\sum_{k=1}^{K}M_{k}^{2}\bar{U}_{i}^{(k),{\rm dist}}\bigg|\notag\\
&~~~~~~~~~~~+|\bar{U}_{i}^{{\rm dist}}|\bigg|\frac{1}{n_{{\rm dist}}}\sum_{k=1}^{K}M_{k}^{2}\bar{U}_{j}^{(k),{\rm dist}}\bigg|+|\bar{U}_{i}^{{\rm dist}}||\bar{U}_{j}^{{\rm dist}}|\bigg(\frac{1}{n_{\rm dist}}\sum_{k=1}^{K}M_{k}^{2}\bigg)\\
&~~~~~\le \max_{i,j\in [d]}\bigg|\frac{1}{n_{\rm dist}}\sum_{k=1}^{K}M_{k}^{2}[\bar{U}^{(k),{\rm dist}}_{i}\bar{U}^{(k),{\rm dist}}_{j}-\mathbb{E}\{\bar{U}^{(k),{\rm dist}}_{i}\bar{U}^{(k),{\rm dist}}_{j}\}]\bigg|\notag\\
&~~~~~~~~~~~+2\max_{j\in [d]}|\bar{U}_{j}^{{\rm dist}}|\cdot\max_{j\in [d]}\bigg|\frac{1}{n_{{\rm dist}}}\sum_{k=1}^{K}M_{k}^{2}\bar{U}_{j}^{(k),{\rm dist}}\notag\bigg|+\bigg(\frac{1}{n_{\rm dist}}\sum_{k=1}^{K}M_{k}^{2}\bigg)\max_{j\in [d]}|\bar{U}_{j}^{{\rm dist}}|^2\notag\\
&~~~~~=: J_{1}+J_{2}+J_{3}\,.\notag
\end{align}
Notice that $\max_{k\in[K]}\max_{i,j\in [d]}|M_{k}^{2}\bar{U}^{(k),{\rm dist}}_{i}\bar{U}^{(k),{\rm dist}}_{j}|=O_{\p}\{K^{-1}nB\log (dK)\}+O_{\p}[B^2\{\log (dK)\}^4]$. For any $\delta_{n}\rightarrow \infty$ as $n\rightarrow\infty$, we have $\max_{k\in[K]}\max_{i,j\in [d]}|M_{k}^{2}\bar{U}^{(k),{\rm dist}}_{i}\bar{U}^{(k),{\rm dist}}_{j}|\leq  K^{-1}nB\delta_n\log (dK)+B^2\delta_n\{\log (dK)\}^4$ with probability approaching one. Let $Z_{i,j}^{(k)}=M_{k}^{2}\bar{U}^{(k),{\rm dist}}_{i}\bar{U}^{(k),{\rm dist}}_{j}I[|M_{k}^{2}\bar{U}^{(k),{\rm dist}}_{i}\bar{U}^{(k),{\rm dist}}_{j}|\le K^{-1}nB\delta_n\log (dK)+B^2\delta_n\{\log (dK)\}^4]$ and $\tilde{Z}_{i,j}^{(k)}=M_{k}^{2}\bar{U}^{(k),{\rm dist}}_{i}\bar{U}^{(k),{\rm dist}}_{j}I[|M_{k}^{2}\bar{U}^{(k),{\rm dist}}_{i}\bar{U}^{(k),{\rm dist}}_{j}|> K^{-1}n\\\cdot B\delta_n\log (dK)+B^2\delta_n\{\log (dK)\}^4]$. Since $\max_{k\in[K]}\mathbb{P}\{|\bar{U}^{(k),{\rm dist}}|_{\infty}> z\}\lesssim d\exp(-CB^{-1}K^{-1}n z^2)+d\exp(-CB^{-1/2}K^{-1/2}n ^{1/2}z^{1/2})$ for any $z>0$, we have $\max_{i,j\in [d]}|n_{\rm dist}^{-1}\sum_{k=1}^{K}\mathbb{E}\{\tilde{Z}_{i,j}^{(k)}\}|\lesssim K^{-1}B+n^{-1}B^2$ provided that $BK\{\log (dK)\}^3=o(n)$.
	By the Hoeffding's inequality, we have
	\begin{align*}
	\mathbb{P}(|J_{1}|> z)&\le \mathbb{P}\bigg[ \max_{i,j\in [d]}\bigg|\frac{1}{n_{\rm dist}}\sum_{k=1}^{K}[Z_{i,j}^{(k)}-\mathbb{E}\{Z_{i,j}^{(k)}\}]\bigg|> \frac{z}{2}\bigg]\\
	&~~~~~~+\mathbb{P}\bigg[ \max_{i,j\in [d]}\bigg|\frac{1}{n_{\rm dist}}\sum_{k=1}^{K}[\tilde{Z}_{i,j}^{(k)}-\mathbb{E}\{\tilde{Z}_{i,j}^{(k)}\}]\bigg|> \frac{z}{2}\bigg]\\
	&\le d^{2}\max_{i,j\in [d]}\mathbb{P}\bigg[\bigg|\sum_{k=1}^{K}[Z_{i,j}^{(k)}-\mathbb{E}\{Z_{i,j}^{(k)}\}]\bigg|> \frac{n_{\rm dist}z}{2}\bigg]\\
	&~~~~~~+\mathbb{P}\bigg[ \max_{k\in[K]}\max_{i,j\in [d]}|M_{k}^2\bar{U}_{i}^{(k),\rm dist}\bar{U}_{j}^{(k),\rm dist}|>K^{-1}nB\delta_n\log (dK)+B^2\delta_n\{\log (dK)\}^4\bigg]\\
	&\le 2d^2\exp\bigg[-\frac{C Kn^2z^2}{n^2B^2\delta_{n}^{2}\log^2 (dK)+K^2B^4\delta_n^2\{\log (dK)\}^8}\bigg]+o(1)
	\end{align*}
	for any $z\ge C'(K^{-1}B+n^{-1}B^2)$, where $C'>0$ is some sufficiently large constant. Hence, it holds that  $|J_{1}|=O_{\p}\{K^{-1/2}B\delta_n(\log d)^{1/2}\log(dK)\}+O_{\p}[n^{-1}K^{1/2}B^2\delta_n(\log d)^{1/2}\{\log (dK)\}^4]$. Noting that we can select arbitrary slow diverging $\delta_n$, following a standard result from probability theory, we have $|J_{1}|=O_{\p}\{K^{-1/2}B(\log d)^{1/2}\log(dK)\}+O_{\p}[n^{-1}K^{1/2}B^2(\log d)^{1/2}\{\log (dK)\}^4]$.
	Using the same arguments for deriving the upper bound of $\mathbb{P}(|\bar{U}^{{\rm dist}}|_{\infty}>z)$, we have
	\begin{align*}
	\mathbb{P}\bigg\{\max_{j\in [d]}\bigg|\frac{1}{n_{\rm dist}}\sum_{k=1}^{K}M_{k}^{2}\bar{U}_{j}^{(k),\rm dist}\bigg|> z\bigg\}&=	\mathbb{P}\bigg\{\max_{j\in [d]}\bigg|\frac{1}{n_{\rm dist}}\sum_{k=1}^{K}\frac{M_{k}}{M_{\max}}M_{k}\bar{U}_{j}^{(k),\rm dist}\bigg|> \frac{z}{M_{\max}}\bigg\}\\
	&\lesssim d\exp(-CB^{-1}n^{-1}K^2 z^2)+d\exp(-CB^{-1/2}K ^{1/2}z^{1/2})
	\end{align*}
	for any $z>0$, which implies $\max_{j\in [d]}|n_{\rm dist}^{-1}\sum_{k=1}^{K}M_{k}^{2}\bar{U}_{j}^{(k),\rm dist}|=O_{\p}\{B^{1/2}K^{-1}n^{1/2}(\log d)^{1/2}\}$. Due to $|\bar{U}^{\rm dist}|_{\infty}=O_{\p}\{n^{-1/2}B^{1/2}(\log d)^{1/2}\}$, we have $|J_{2}|=O_{\p}(BK^{-1}\log d)$ and $|J_{3}|=O_{\p}(BK^{-1}\log d)$. \eqref{eq:J123} yields $|\Sigma_{n}^{\rm dist}-\hat{\Sigma}_{n}^{\rm dist}|_{\infty}=O_{\p}\{K^{-1/2}B(\log d)^{1/2}\log(dK)\}+O_{\p}[n^{-1}K^{1/2}B^2(\log d)^{1/2}\{\log (dK)\}^4]$, which implies
	\begin{align}\label{eq:I2dist}
	I_2^{\rm dist}&\le |\Sigma_n^{\rm dist}-\hat{\Sigma}_{n}^{\rm dist}|_{\infty}\max_{\check{v}_{1}\in\mathcal{F}}|\Gamma_{n}^{\rm dist}\check{v}_{1}|_{1}\max_{\check{v}_{2}\in\mathcal{F}}|\hat{\Gamma}_{n}^{\rm dist}\check{v}_{2}|_{1}\\
	&=O_{\p}\{K^{-1/2}BL^2(\log d)^{1/2}\log(dK)\}+O_{\p}[n^{-1}K^{1/2}B^2L^2(\log d)^{1/2}\{\log (dK)\}^4]\,,\notag
	\end{align}
	Using the same arguments for deriving the convergence rate of $I_{1}^{\rm dist}$, we have
	\begin{align*}
	I_{3}^{\rm dist}=O_{\p}[n^{-1/2}B^{5/2}L^2K^{1/2}\{\log(dBK)\}^{1/2}]+O_{\p}[n^{-1}B^{3}L^2K\{\log(dBK)\}^{2}]
	\end{align*}
	provided that $\{\log(dK)\}^2\log d=o(K)$ and $BK^{1/2}(\log d)^{1/2}\{\log (dK)\}^4=o(n)$. Together with \eqref{eq:I1dist} and \eqref{eq:I2dist}, it holds that
	\begin{align*}
	\Delta_{n}^{\rm dist}\le I_{1}^{\rm dist}+I_{2}^{\rm dist}+I_{3}^{\rm dist}=o_{\p}\{L^{-2}(\log d)^{-2}\}
	\end{align*}
	provided that $B^2L^8\{\log(dK)\}^{2}(\log d)^5=o(K)$, $B^3L^4K\{\log(dBK)\}^{2}(\log d)^2=o(n)$, $B^2L^4K^{1/2} (\log d)^{5/2}\\\cdot\{\log (dK)\}^4=o(n)$ and $B^5L^8K\log (dBK)(\log d)^{4}=o(n)$. Identical to \eqref{eq:f_L_2}, we have
\begin{align*}
\sup_{y\in\mathbb{R}^{|\mathcal{F}|}}\big|\mathbb{P}(g\le y)-\mathbb{P}(\hat{g}\le y\,|\,\mathcal{X}_{n})\big|\lesssim (\Delta_{n}^{\rm dist})^{1/3}\{1\vee \log(d^{L}/\Delta_{n}^{\rm dist})\}^{2/3}=o_{\p}(1)\,.
\end{align*}
Together with \eqref{eq:gaussbd_distributed_1} and \eqref{eq:rdbd}, it holds that
\begin{align*}
\sup_{z\in\mathbb{R}}\big|\mathbb{P}(W_{n,L}^{{\rm dist}}>z)-\mathbb{P}\{f_L(\mathring{\xi})>z\,|\,\mathcal{X}_{n}\}\big|\le&~\sup_{z\in\mathbb{R}}\big|\mathbb{P}(W_{n,L}^{{\rm dist}}\le z)-\mathbb{P}(g\le z\cdot 1_{|\mathcal{F}|})\big|\\
&+ \sup_{z\in\mathbb{R}}\big|\mathbb{P}(g\le z\cdot 1_{|\mathcal{F}|})-\mathbb{P}(\hat{g}\le z\cdot 1_{|\mathcal{F}|}\,|\,\mathcal{X}_{n})\big|\\
&+\sup_{z\in\mathbb{R}}\big |\mathbb{P}\{f_{L}(\mathring{\xi})\le z|\mathcal{X}_{n}\}-\mathbb{P}(\hat{g}\le z \cdot 1_{|\mathcal{F}|}\,|\,\mathcal{X}_{n})\big|\\
=&~o_{\p}(1)\,.
\end{align*}
We complete the proof of Proposition 2.$\hfill\Box$


\end{document}